\documentclass[twocolumn,aps,prd]{revtex4-1}
\usepackage{dcolumn}
\usepackage{enumerate}
\usepackage{amsmath}
\usepackage{amssymb}
\usepackage{graphicx}
\usepackage{graphics}

\begin{document}
\title{From helical to standard magnetorotational instability: predictions for upcoming liquid sodium experiments}

\author{Ashish Mishra}
\email{a.mishra@hzdr.de}
\affiliation{Helmholtz-Zentrum Dresden-Rossendorf, Bautzner Landstr.
400, D-01328 Dresden, Germany}

\author{George Mamatsashvili}
\affiliation{Helmholtz-Zentrum Dresden-Rossendorf, Bautzner Landstr.
400, D-01328 Dresden, Germany}
\affiliation{Abastumani Astrophysical Observatory, Abastumani 0301, Georgia}
\affiliation{Institute of Geophysics, Tbilisi State University, Tbilisi 0193, Georgia}

\author{Frank Stefani}
\affiliation{Helmholtz-Zentrum~Dresden-Rossendorf, Bautzner Landstr.
400, D-01328 Dresden, Germany}

\begin{abstract}
We conduct a linear analysis of axisymmetric magnetorotational instability (MRI) in a magnetized cylindrical Taylor-Couette (TC) flow for its standard version (SMRI) with a purely axial background magnetic field and two further types -- helically modified SMRI (H-SMRI) and helical MRI (HMRI) -- in the presence of combined axial and azimuthal magnetic fields. This study is intended as preparatory for upcoming large-scale liquid sodium MRI experiments planned within the DRESDYN project at Helmholtz-Zentrum Dresden-Rossendorf, so we explore these instability types for typical values of the main parameters: the magnetic Reynolds number, the Lundquist number and the ratio of the angular velocities of the cylinders, which are attainable in these experiments. In contrast to previous attempts at detecting MRI in the lab, our results demonstrate that SMRI and its helically modified version can in principle be detected in the DRESDYN-TC device for the  range of the above parameters, including the astrophysically most important Keplerian rotation, despite the extremely small magnetic Prandtl number of liquid sodium. Since in the experiments we plan to approach (H-)SMRI from the previously studied  HMRI regime, we characterise the continuous and monotonous transition between the both regimes. We show that H-SMRI, like HMRI, represents an overstability (travelling wave) with non-zero frequency linearly increasing with azimuthal field. Because of its relevance to finite size flow systems in experiments, we also analyse the absolute form of H-SMRI and compare its growth rate and onset criterion with the convective one. 
\end{abstract}	
	
\maketitle
\section{Introduction}

The magnetorotational instability (MRI) is of key importance for cosmic structure formation. Originally discovered by Velikhov in 1959 \cite{Velikhov_1959}, and then ``forgotten'' for nearly three decades, in 1991 it was ``rediscovered'' and successfully applied to the long-standing problem of turbulence and angular momentum transport in accretion disks around protostars and black holes \cite{Balbus_Hawley_1991}. While extensively studied analytically and numerically over the last three decades (see recent reviews \cite{Ruediger_etal_2018_PhysRepo, lesur_2021} and references therein), standard form of MRI (SMRI) in a classical setup --  rotating cylindrical flow of a conducting fluid threaded by a purely axial magnetic field -- where it was originally discovered theoretically, has yet eluded any clear experimental confirmation, despite great efforts and encouraging first results \cite{Sisan_etal2004,Nornberg_Ji_etal_2010_PhysRevLett,Roach_etal2012PhRvL, Hung_etal2019CmPhy}. This is mainly due to the fact that for the onset and efficient growth of SMRI both magnetic Reynolds ($Rm$) and Lundquist ($Lu$) numbers should be high enough $O(10)$, which makes dedicated liquid metal experiments challenging, because the extremely low  magnetic Prandtl numbers $Pm=\nu/\eta=10^{-6}-10^{-5}$ of strongly resistive liquid metals ($\nu$ is viscosity and $\eta$ resistivity) in turn requires very high Reynolds numbers $Re=Rm/Pm \gtrsim 10^6$.  

\begin{figure*}
\centering
\includegraphics[width=0.3\textwidth, height=6cm]{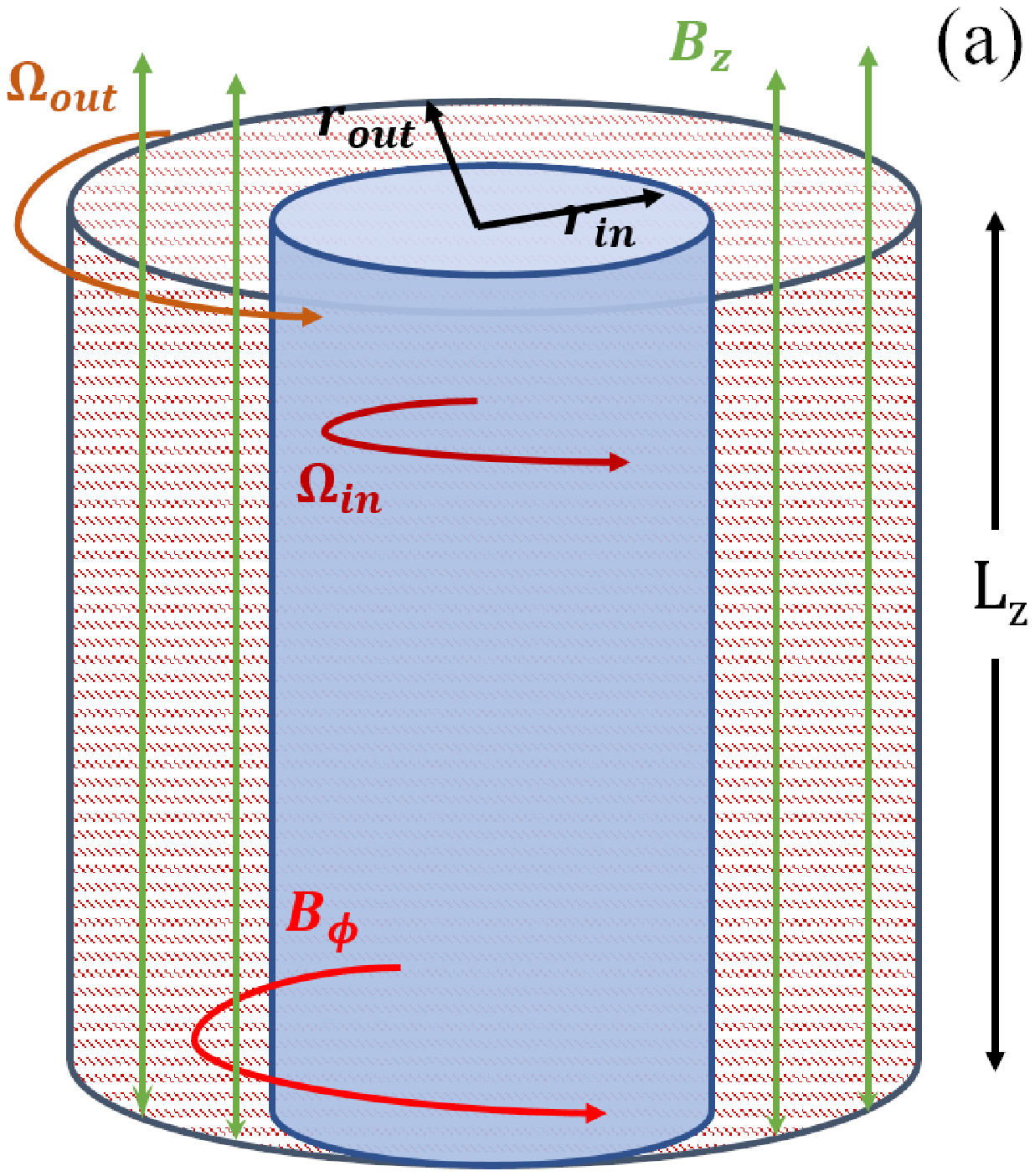}
\hspace{2em}
\includegraphics[width=0.4\textwidth, height=6cm]{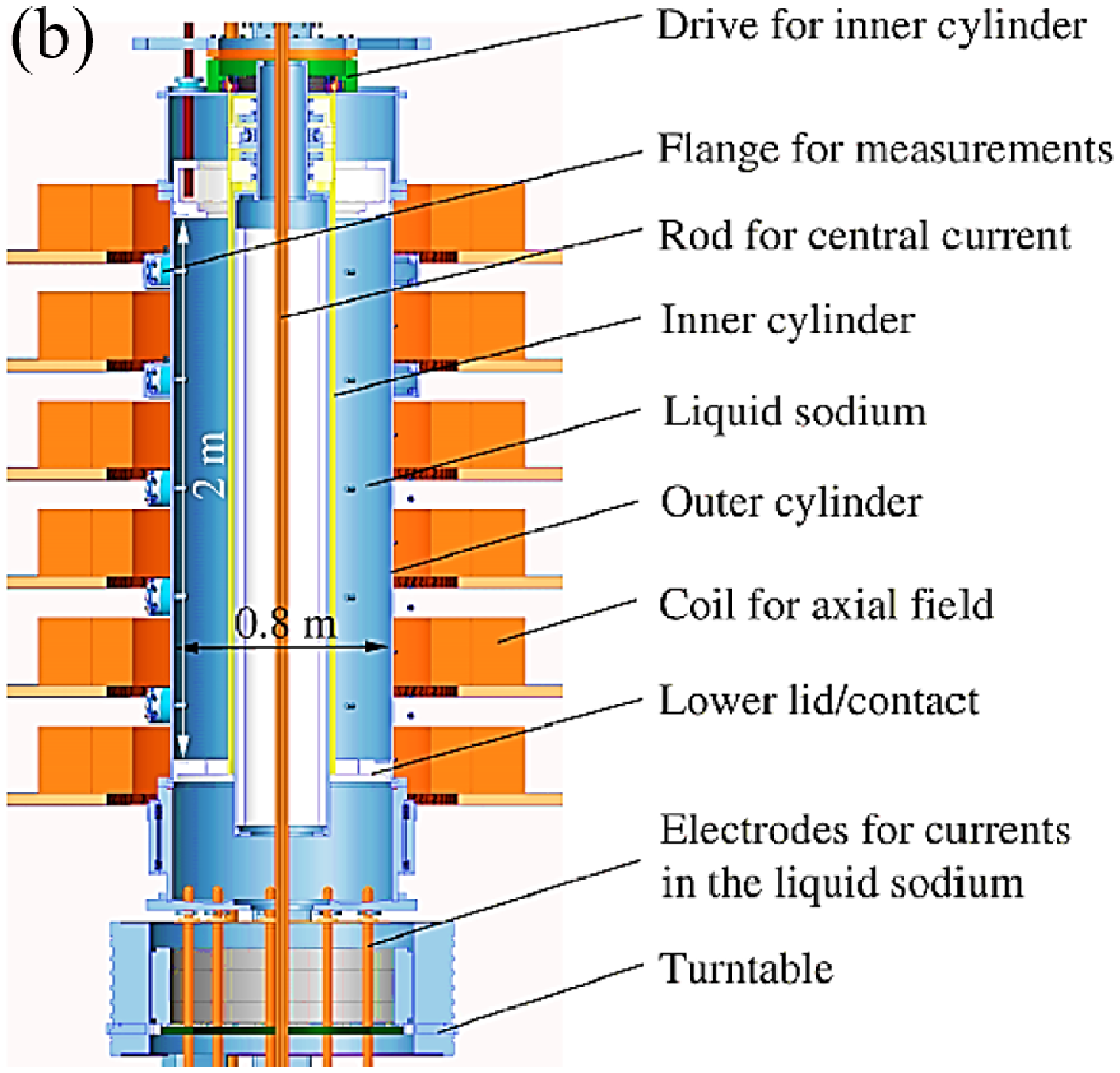}
\caption{ (a) A simplified Taylor-Couette flow setup with an imposed helical magnetic field, (b) a schematic view of the new DRESDYN-MRI machine.} \label{Fig1: TC_Setup}
\end{figure*}

It came, therefore, as a big surprise when Hollerbach and R\"udiger
found in 2005 \cite{Hollerbach_Rudiger_2005} that the addition
of an azimuthal magnetic field can drastically reduce the effort
for experimental realization of MRI, at least for rotational profiles
slightly steeper than the Keplerian one. This new version of MRI under the influence of a helical magnetic field, now known as  helical MRI (HMRI), was shown to be destabilized inertial oscillations \cite{Liu_Goodman_Herron_Ji_2006PhRvE}, governed by the {\it hydrodynamic} Reynolds number ($Re$) and Hartmann number ($Ha$), in contrast to SMRI which is a destabilized magneto-Coriolis wave. Many features of HMRI have since been revealed, in particular, its monotonous transition to SMRI \cite{Hollerbach_Rudiger_2005}, including, though, 
a spectral exceptional point \cite{Kirillov_Stefani_2010ApJ}, 
its restriction to rather steep profiles in terms of the the lower and upper Liu limits \cite{Liu_Goodman_Herron_Ji_2006PhRvE,Kirillov_etal2014},
the continuous connection of these two limits when allowing non-current free azimuthal magnetic fields \cite{Liu_Goodman_Herron_Ji_2006PhRvE,Kirillov_Stefani_PRL_2013} 
and the close connection of its growth rate to the growth 
factor of the underlying non-modal instability of the 
non-magnetic rotational flow \cite{Mamatsashvili_Stefani2016}.
Moreover, for purely or dominantly azimuthal 
magnetic fields, a non-axisymmetric ``sibling'' of HMRI,
called azimuthal MRI (AMRI), was discovered  \cite{Hollerbach_Rudiger_2010,Seilmayer_etal2014}, with otherwise very similar properties as HMRI.

Meanwhile, both HMRI and AMRI were found and characterized in the liquid metal experiment PROMISE at Helmholtz-Zentrum Dresden-Rossendorf (HZDR) \cite{Stefani_Gundrum_Gerbeth_etal_2006PhRvL,Stefani_Gerbeth_Gundrum_Etal_2009PhysRevE,Seilmayer_etal2014}. Being sufficient to provide some $Re$ of the order of $10^3$  and $Ha$ of the order of $10-10^2$ (rather than significant values of these numbers required for SMRI), PROMISE  is a medium-size Taylor-Coutte experiment with slow rotation rates (below 1\,Hz), using the very safe eutectic alloy GaInSn, with the current supply for some kiloAmps through the central rod  being the most expensive part. While the experimental results on HMRI were 
in very good agreement with numerical predictions \cite{Stefani_Gerbeth_Gundrum_Etal_2009PhysRevE},
the investigation  of the details of the "butterfly diagram" 
of AMRI is still ongoing \cite{Mishra_etal2021}.
Though its constructional restrictions to $Re \lesssim 10^4$ make 
PROMISE unsuitable for reaching very high $Re \sim 10^6$ necessary for SMRI to operate at very low magnetic Prandtl of liquid metals, the clear identification of HMRI in these experiments has strongly encouraged to investigate the monotonous transition to SMRI in a significantly bigger machine. 

Such a new experiment (Fig. \ref{Fig1: TC_Setup}b) is presently under construction in the framework of the DREsden Sodium facility for DYNamo and thermohydraulic studies
(DRESDYN) at HZDR. Actually, this machine is not limited to the detection of SMRI only but is also designed to study various combinations of MRI with the current-driven Tayler instability \cite{Seilmayer_etal2012PhRvL}, as well as the recently discovered ``Super-HMRI'' in rotating flows with positive shear \cite{Mamatsashvili_Stefani_Hollerbach_Rudiger_2019}. Yet, the focus of this paper is exclusively on the classical aspects of HMRI and SMRI. As a preparatory step towards large-scale new cutting-edge MRI experiments within DRESDYN project, we study these instabilities and the connection between them specifically for those ranges of the characteristic parameters of the magnetized Taylor-Couette flow that are achievable in these experiments. We aim, in particular, at a detailed analysis of the axisymmetric MRI when going from HMRI to SMRI regimes. Our work relies on, but extends significantly, the previous works on pure SMRI \cite{goodman_ji_2002,Rudiger_Schultz_Shalybkov_2003PhRvE}, on the monotonous transition between HMRI and SMRI \cite{Hollerbach_Rudiger_2005} and the connection between HMRI and SMRI \cite{Kirillov_Stefani_2010ApJ}. In the latter work, it was shown using a local WKB analysis, that HMRI arises from the exchange of instabilities of SMRI and inertial waves through the reconnection of these two modes at a spectral exceptional point as an imposed azimuthal magnetic field increases.

The paper is organized as follows. The basic equations and the formulation of the linear stability problem are given in Sec. II, the main results on SMRI, helically modified SMRI and HMRI are presented in Sec. III for the convective form of these instabilities and in Sec. IV for the absolute form. A summary and conclusions are presented in Sec. V.

\section{Mathematical Setting}
The basic equations of non-ideal MHD governing motion of an incompressible conducting fluid are
\begin{equation}
\frac{\partial {\bf U}}{\partial t}+({\bf U}\cdot \nabla) {\bf U}=-\frac{1}{\rho}\nabla P + \frac{{\bf J}\times {\bf B}}{\rho} + \nu\nabla^2 {\bf U}, 
\end{equation}
\begin{equation}
\frac{\partial {\bf B}}{\partial t}=\nabla\times \left( {\bf U}\times {\bf B}\right)+\eta\nabla^2{\bf B}, 
\end{equation}
\begin{equation}
\nabla\cdot {\bf U}=0,~~~\nabla\cdot {\bf B}=0,
\end{equation}
where $\rho$ is the constant density, ${\bf U}$ is the velocity, $P$ is the thermal pressure, ${\bf B}$ is the magnetic field, ${\bf J}=\mu_0^{-1}\nabla \times {\bf B}$ is the current density with $\mu_0$ being the magnetic permeability of vacuum. The fluid has constant kinematic viscosity $\nu$ and ohmic resistivity $\eta$. 

Consider a cylindrical Taylor-Couette (TC) setup -- a basis for the DRESDYN-MRI experiment, which contains liquid sodium as an incompressible conducting fluid in the cylindrical coordinate system $(r,\phi,z)$ (Fig \ref{Fig1: TC_Setup}a). In this TC setup, the inner and outer cylinders with radii $r_{in}$ and $r_{out}$ rotate, respectively, with angular velocities $\Omega_{in}$ and $\Omega_{out}$. In the DRESDYN-MRI machine, the ratio of the inner and outer cylinder radii is fixed to $r_{in}/r_{out}=0.5$ and the aspect ratio $L_z/r_{in}=10$ is large, where $L_z$ is the length (height) of the cylinders, so we can assume them to be approximately infinitely long, while the ratio $\mu=\Omega_{out}/\Omega_{in}$ can be varied (Table \ref{Table1: Physical_parameters}). A current-carrying solenoid surrounding the outer cylinder imposes a constant magnetic field in the axial direction, $B_{0z}$, while a central linear current along the cylinder axis creates azimuthal magnetic field $B_{0\phi}=\beta B_{0z}(r_{in}/r)$, where $\beta\geq 0$ is a non-dimensional parameter measuring the strength of $B_{0\phi}$ relative to $B_{0z}$; at $\beta=0$, the background field is purely axial, corresponding to the SMRI regime. The resultant helical magnetic field, ${\bf B}_0=B_{0\phi}{\bf e}_{\phi}+B_{0z}{\bf e}_z$, is current-free between the cylinders, ${\bf J_0}=\mu_0^{-1}\nabla \times {\bf B_0}=0$, and therefore does not exert any Lorentz force on the fluid. So, in the present case of infinite cylinders, the equilibrium azimuthal flow ${\bf U}_0=(0,r\Omega(r),0)$ with the classical hydrodynamical TC profile of angular velocity
\begin{equation*}
\Omega(r) = C_1+\frac{C_2}{r^2}
\end{equation*}
where the coefficients $C_1$ and $C_2$ are
\begin{equation*}
C_1=\frac{\Omega_{out}r_{out}^2-\Omega_{in}r_{in}^2}{r_{out}^2-r_{in}^2}, \hspace{0.3cm} C_2=\frac{(\Omega_{in}-\Omega_{out})r_{in}^2r_{out}^2}{r_{out}^2-r_{in}^2},
\end{equation*}
is an exact solution of Eqs. (1)-(3). In real experiments, of course, there is always some deviation of the equilibrium flow profile from the TC one mainly in the Ekman-Hartmann boundary layers near endcaps \cite{Hollerbach_Fournier2004,Szklarski2007,Stefani_etal2009, Gissinger_etal2012}.
 	
We study the linear stability of this base TC flow ${\bf U}_0$ with the imposed helical magnetic field ${\bf B}_0$, against small perturbations, ${\bf u}={\bf U}-{\bf U}_0$, $p=P-P_0$, ${\bf b}={\bf B}-{\bf B}_0$.  We assume the perturbations to have the modal form  $\propto \exp(\gamma t+ i m\phi+ik_z z)$, where $\gamma$ is the (complex) eigenvalue and $m$ and $k_z$ are, respectively, the azimuthal and axial wavenumbers. A positive real part of $\gamma$ implies the presence of instability in the flow. Normalizing time by $\Omega_{in}^{-1}$, length by $r_{in}$, $\Omega(r)$ by $\Omega_{in}$, ${\bf B}_0$ by $B_{0z}$, ${\bf b}$ by $Rm B_{0z}$, ${\bf u}$ by $r_{in}\Omega_{in}$, $p$ by $\rho r_{in}^2\Omega_{in}^2$ and linearizing main Eqs. (1)-(3), we obtain the system of non-dimensional perturbation equations \cite{Ruediger_etal_2018_PhysRepo}:
\begin{multline}\label{momen_Ur}
(\gamma+im\Omega) u_r= 2\Omega u_\phi-\frac{dp_t}{dr}+\frac{Ha^2}{Re}\left(ik_z+\frac{im\beta}{r^2}\right)b_r\\-\frac{Ha^2}{Re}\cdot\frac{2\beta}{r^2} b_\phi+\frac{1}{Re}
\left(\Delta u_r-\frac{u_r}{r^2}-\frac{2im}{r^2}u_{\phi}\right) ,
\end{multline} 
\begin{multline}\label{momen_Uphi}
(\gamma+im\Omega)u_\phi= -\left(2\Omega+r \frac{d\Omega}{dr}\right) u_r-\frac{im}{r}p_t+\\
\frac{Ha^2}{Re}\left(ik_z+\frac{im\beta}{r^2}\right)b_{\phi} +\frac{1}{Re}\left(\Delta u_{\phi}-\frac{u_{\phi}}{r^2}+\frac{2im}{r^2}u_r\right)
\end{multline}
	\begin{equation}\label{momen_Uz}
(\gamma+im\Omega)u_z=-ik_zp_t +\frac{Ha^2}{Re}\left(ik_z+\frac{im\beta}{r^2}\right)b_z+\frac{1}{Re}\Delta u_z,
\end{equation}
\begin{equation}\label{induc_Br}
Rm(\gamma+im\Omega)b_r=\left(ik_z+\frac{im\beta}{r^2}  \right)u_r+\Delta b_r  -\frac{b_r}{r^2}-\frac{2im}{r^2}b_{\phi},
\end{equation}
\begin{multline}\label{induc_Bphi}
Rm(\gamma+im\Omega)b_{\phi}=\left(ik_z+\frac{im\beta}{r^2}  \right)u_{\phi}+Rm\cdot r\frac{d\Omega}{dr}b_r\\+\frac{2\beta}{r^2}u_r+\Delta b_{\phi}-\frac{b_{\phi}}{r^2}+\frac{2im}{r^2}b_r
\end{multline}
\begin{equation}\label{induc_Bz}
Rm(\gamma+im\Omega)b_z=\left(ik_z+\frac{im\beta}{r^2}  \right)u_z+\Delta b_z,
\end{equation}
\begin{equation}\label{vel_incompress}
\frac{du_r}{dr}+\frac{u_r}{r}+\frac{im}{r}u_{\phi}+ik_z u_z=0,~~~
\frac{db_r}{dr}+\frac{b_r}{r}+\frac{im}{r}b_{\phi}+ik_z b_z=0,
\end{equation}
where we have introduced the total (thermal+magnetic) pressure perturbation 
\[
p_t=p+\frac{Ha^2}{Re}\left(b_z+\frac{\beta}{r}b_{\phi}\right)
\]
and $\Delta$ is the Laplace operator in cylindrical coordinates
\[
\Delta=\frac{1}{r}\frac{d}{dr}\left(r\frac{d}{dr}\right)-\frac{m^2}{r^2}-k_z^2.
\]
In these equations, the parameters $Ha$, $Re$, $Rm$ are the Hartmann, Reynolds and magnetic Reynolds numbers, respectively, whose definitions are given in Table \ref{Table2: non_dimentional}. Here the Hartmann number is defined for the vertical magnetic field $B_{0z}$, while below we also use the Hartmann number $Ha_{\phi}$ in terms of the azimuthal field at the inner cylinder, $B_{0\phi}(r_{in})$, which is also given in Table \ref{Table2: non_dimentional}. The magnetic Prandtl number $Pm=\nu/\eta=Rm/Re$ is very small and fixed to the value for liquid sodium  at  T=130$^{\circ}$C, $Pm=7.77\times 10^{-6}$. It is well known that SMRI is governed primarily by Lundquist number $Lu=Ha\cdot Pm^{1/2}$ and $Rm$ \cite{Kirillov_Stefani_2010ApJ, Kirillov_etal2014, Ruediger_etal_2018_PhysRepo}, which we also choose as a main parameter in our analysis. To find the optimal regimes for the detection of SMRI in the DRESDYN-MRI experiment, we adopt the ranges of the key parameters of the flow achievable in this experiment, which are given in Table \ref{Table1: Physical_parameters} and in nondimensional form in Table \ref{Table2: non_dimentional}

In this paper, we consider only axisymmetric perturbations with $m=0$ (i.e., $\partial/\partial \phi=0$), which, as we checked, are the dominant unstable ones for small to moderate values of $\beta \leq 4$ and other parameter regimes considered here that are pertinent to the experiment, whereas for $\beta \gg 1$, non-axisymmetric $m=1$ AMRI modes dominate instead \cite{Hollerbach_Rudiger_2010}. For axisymmetric perturbations and $\beta=0$, Eqs. (\ref{momen_Ur}) - (\ref{vel_incompress}) reduce to the main equations of \cite{Rudiger_Schultz_Shalybkov_2003PhRvE} for the analysis of SMRI, while for $\beta \ne 0$ they coincide with those used by \cite{Hollerbach_Rudiger_2005} to study HMRI. We focus on the Rayleigh-stable regime, i.e., $\mu=\Omega_{out}/\Omega_{in} > (r_{in}/r_{out})^2=0.25$, ensuring that instabilities in the flow are solely of magnetic nature, and consider values of $\mu$ up to quasi-Keplerian $\mu=0.35$, which is of immediate interest for astrophysical disks.

\begin{table}
	\begin{tabular}[b]{cc}
			\hline
			\textbf{Physical Parameter} & \textbf{Values} \\
			\hline
			$r_{in}$     & 0.2 m\\
			$r_{out}$ & 0.4 m\\
			$L_z$  & 2 m\\
			$\Omega_{\, in}$ & $\leq 2 \pi \cdot 20$ Hz\\
			$\Omega_{\, out}$ & $\leq 2 \pi \cdot 6$ Hz\\
			Axial magnetic field ($B_{0z}$) & $\leq 150$ mT \\
			Current through central rod (I) & $ \leq 50$ kA \\
			Conductivity $ (\sigma)$ & $9.5\times10^6 S/m$ \\
			Viscosity$ (\nu)$ & $6.512 \times 10^{-7} m^2/s$ \\
			Density $(\rho)$ & 920 $kg/m^3$\\
			\hline
		\end{tabular}
		\caption{Physical parameters of the new DRESDYN-MRI experiment with liquid sodium. The material parameters are for a sodium temperature of T=130$^{\circ}$C.}
		\label{Table1: Physical_parameters}
		\end{table}
	
	\begin{table}
	\resizebox{\columnwidth}{!}{%
		\begin{tabular}[b]{ccc}			\hline
			\textbf{Dimensionless Parameter} & \textbf{Definition} & \textbf{Values} \\
			\hline
			$\mu$  & $\Omega_{out}/ \Omega_{in} $ & (0.25,0.35]\\
			$\beta$ & $B_{0\phi}(r_{in})/B_{0z}$ & [0, 4]\\
			Normalised height of the TC device & $L_z/r_{in}$ & 10\\
			\textit{Renolds number} ($Re$) & $\Omega_{in} r_{in}^2/\nu$ & $\leq 7.72\times10^{6}$ \\
			\textit{Hartmann number} ($Ha$) & $B_{0z}r_{in}/\sqrt{\rho\mu_0\nu\eta}$ & $\leq 3778$\\
			\textit{Magnetic Prandtl Number} ($Pm$) & $\nu/\eta $ & $7.77 \times 10^{-6}$\\
			\textit{Magnetic Reynolds Number} ($Rm$) & $RePm$ & $\leq 40$\\
			\textit{Lundquist Number} ($Lu$) & $Ha \sqrt{Pm}$& $\leq 10$\\
			
			Azimuthal \textit{Hartmann number} ($Ha_{\phi}$) & $B_{0\phi}(r_{in}) r_{in}/\sqrt{\rho\mu_0\nu\eta}$ & $\leq 1259$\\
			Azimuthal \textit{Lundquist Number} ($Lu_\phi$) & $Ha_\phi \sqrt{Pm}$& $\leq 3.51$\\
			\hline
			
		\end{tabular}
		}
		\caption{Non-dimensional system parameters of the DRESDYN-MRI experiment with liquid sodium based on the values in Table I, assuming material parameters of liquid sodium at T=130$^{\circ}$C}\label{Table2: non_dimentional}

		\end{table}
	
\section{Results}
	
Equations (\ref{momen_Ur})-(\ref{vel_incompress}) together with appropriate boundary conditions for the velocity and magnetic field constitute the 1D (in the radial direction) eigenvalue problem for finding $\gamma$ and corresponding eigenfunctions as a function of other parameters. In the resulting eigenvalue problem, the radial structure is solved by spectral collocation method using Chebyshev polynomials typically up to $N=30-40$ \cite{Hollerbach_Rudiger_2005, Priede_Gunter_2009,Hollerbach_Rudiger_2010, Mamatsashvili_Stefani_Hollerbach_Rudiger_2019}. Due to the large difference (by a factor $\sim 8$) between the conductivity of liquid sodium and the cylinder walls made of stainless steel, we apply insulating boundary conditions for the magnetic field \cite{goodman_ji_2002, Rudiger_Schultz_Shalybkov_2003PhRvE}. For the velocity, standard no-slip boundary conditions are used. The above equations with these boundary conditions are reduced to a large ($4N\times4N$) matrix eigenvalue problem where the growth rate of the instability is determined by the (positive) real part of $\gamma$. We assume that at least one full wavelength of an unstable mode should fit into the axial extension, $L_z$, of the device. Thus, in most of the analysis below we set the minimum wavenumber for unstable modes to be equal to $k_{z, min}=2\pi/L_z$ and maximize their growth rate over larger wavenumbers $k_z\geq k_{z, min}$, hence excluding modes with larger axial scale, i.e., with wavenumbers smaller than $k_{z,min}$. The modifications of the resulting marginal stability curves when including those larger-scale modes are very small, as shown below.
	
\subsection{SMRI for $\beta=0$}
	
\begin{figure*}
\begin{minipage}{.32\textwidth}
\centerline{\includegraphics[width=\textwidth]{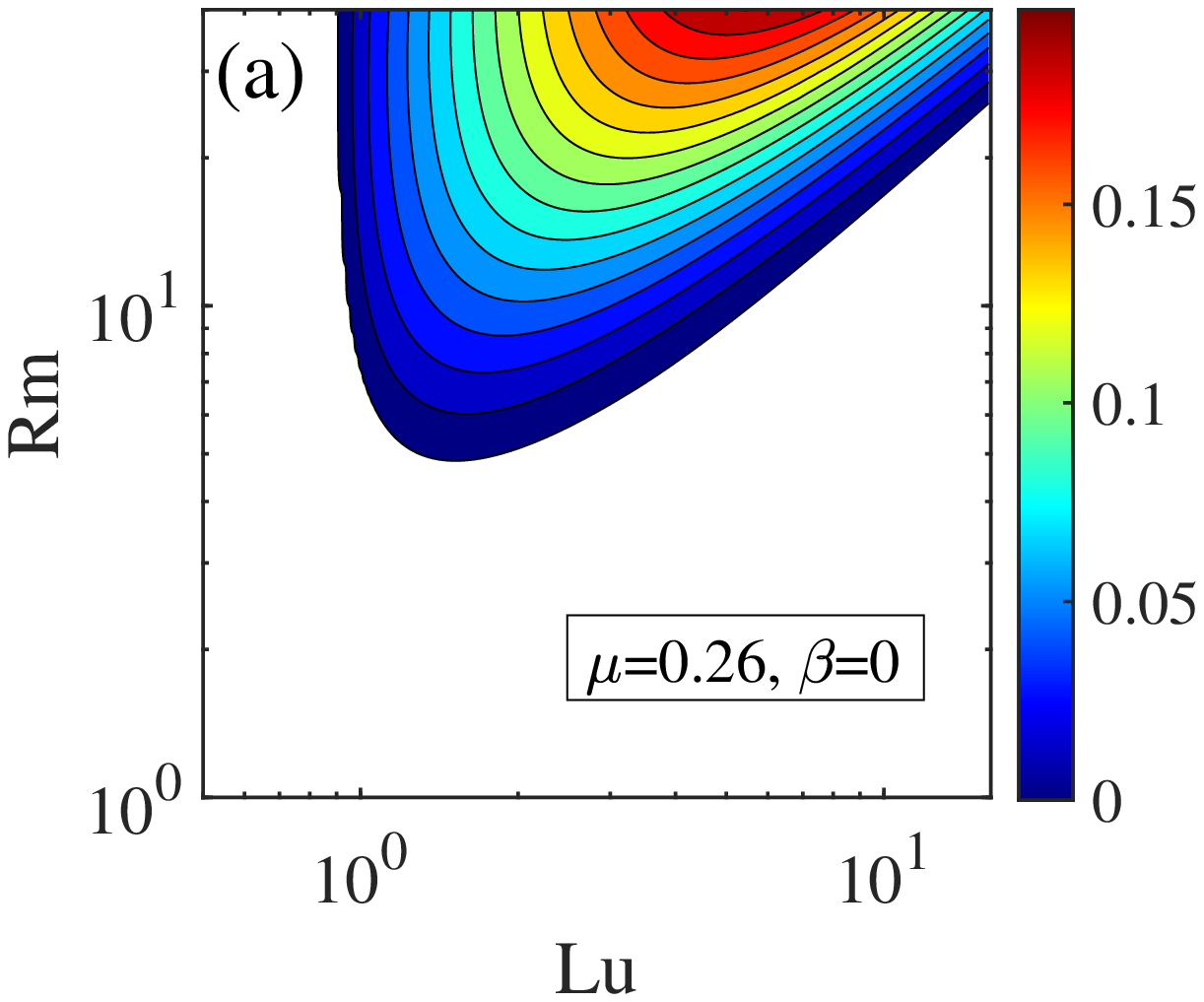}}
\end{minipage}
\vspace{0.7cm}
\begin{minipage}{.32\textwidth}
\centerline{\includegraphics[width=\textwidth]{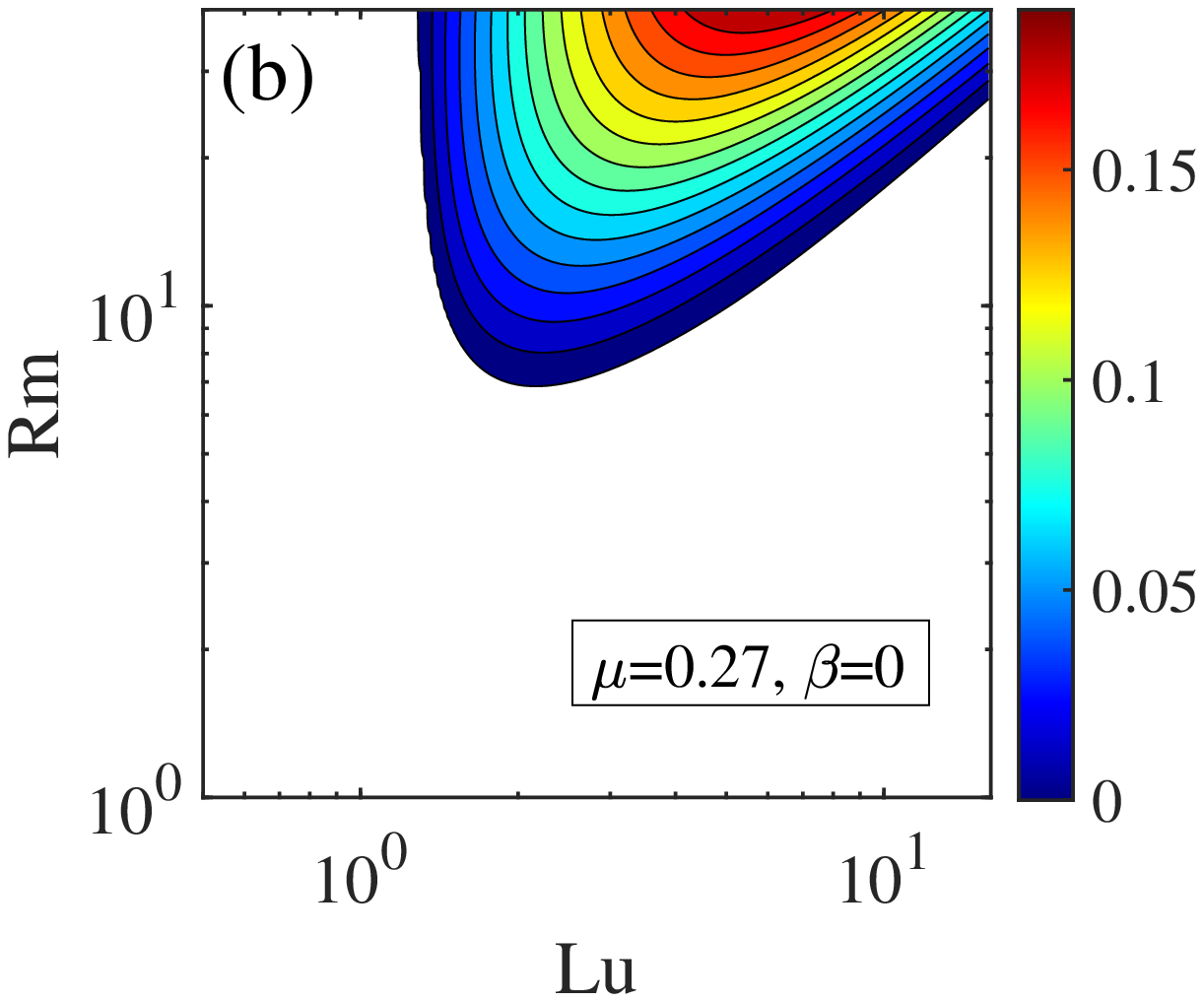}}
\end{minipage}
\begin{minipage}{.32\textwidth}
\centerline{\includegraphics[width=\textwidth]{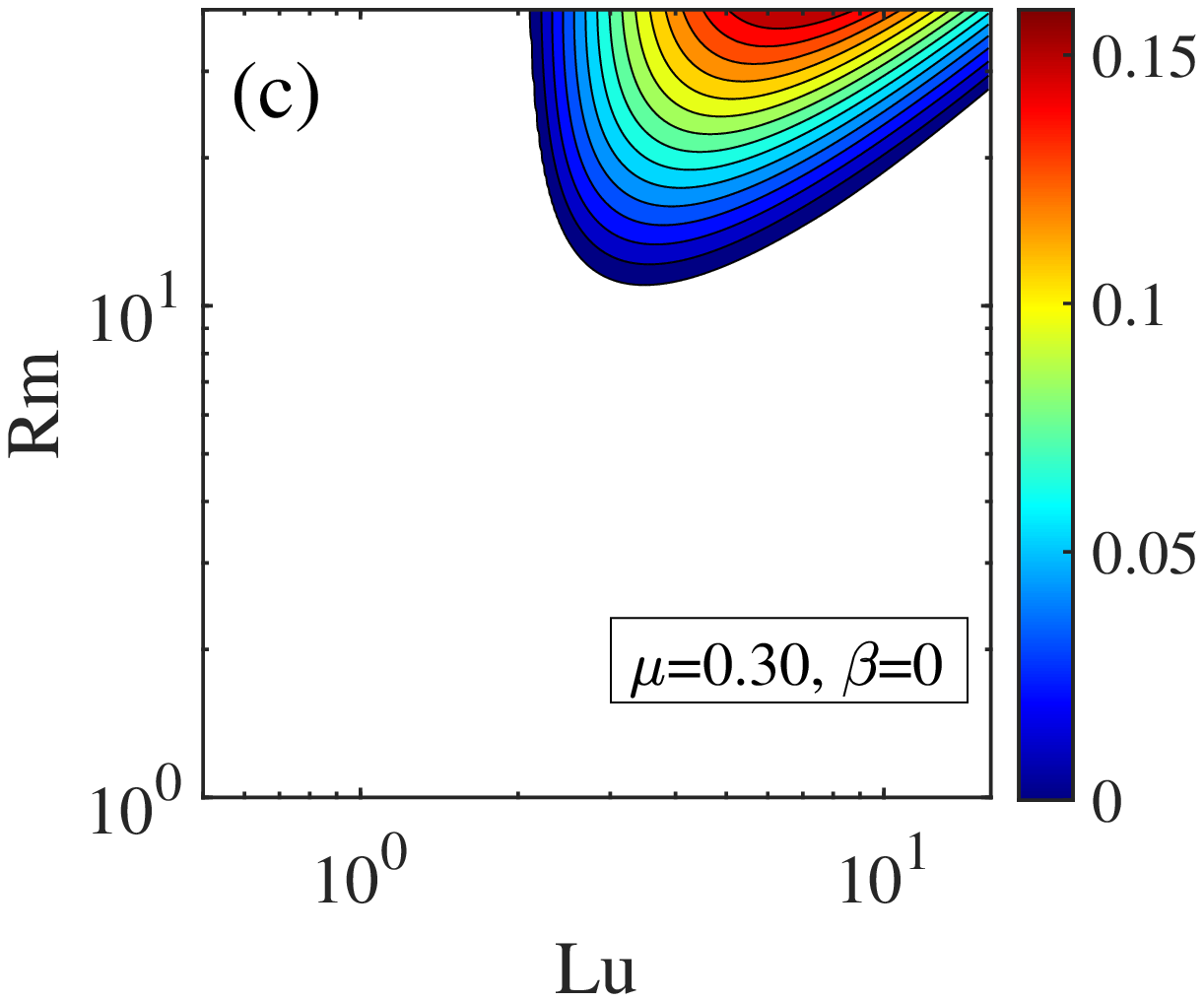}}
\end{minipage}\\
\begin{minipage}{.32\textwidth}
\centerline{\includegraphics[width=\textwidth]{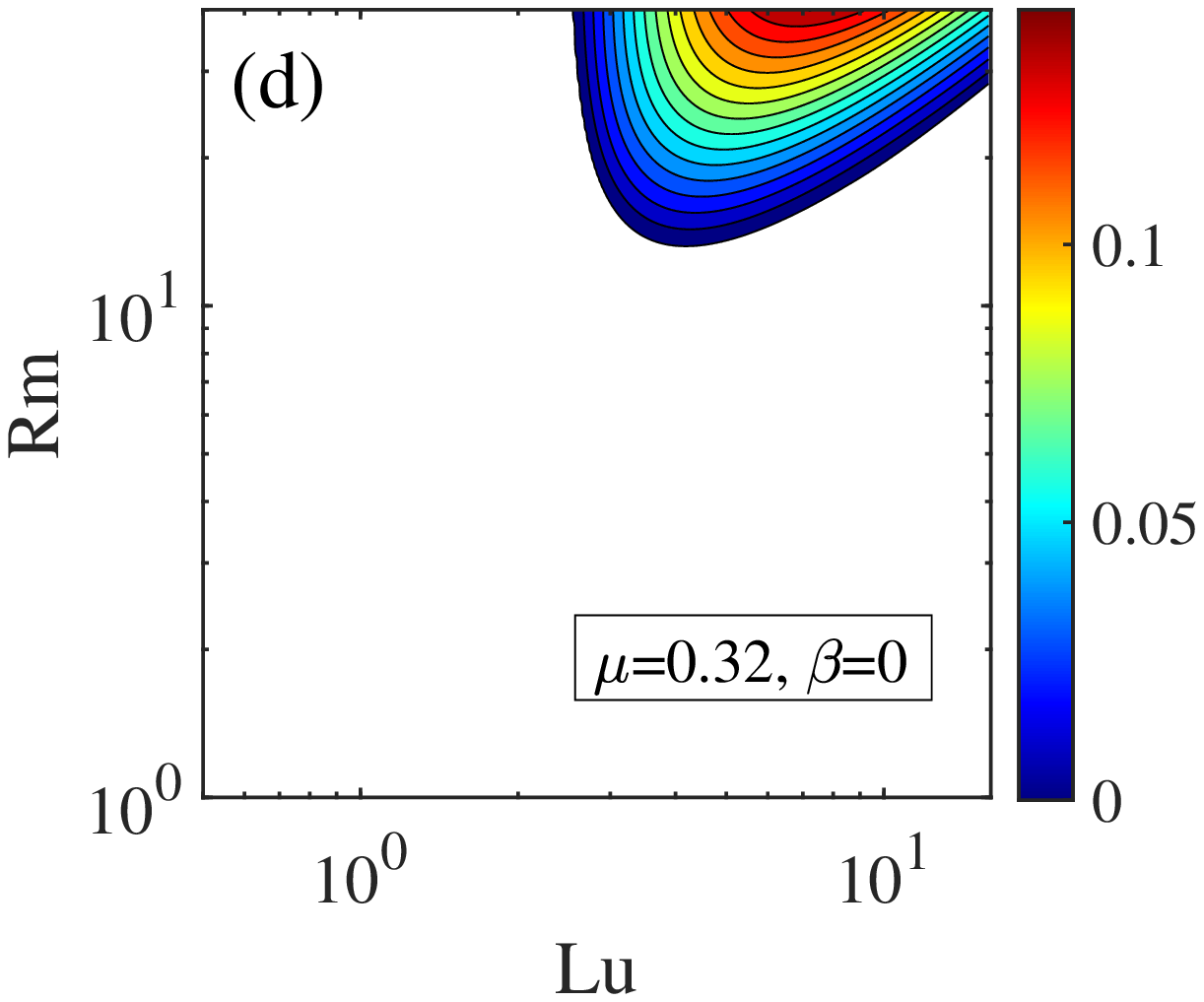}}
\end{minipage}
\begin{minipage}{.32\textwidth}
\centerline{\includegraphics[width=\textwidth]{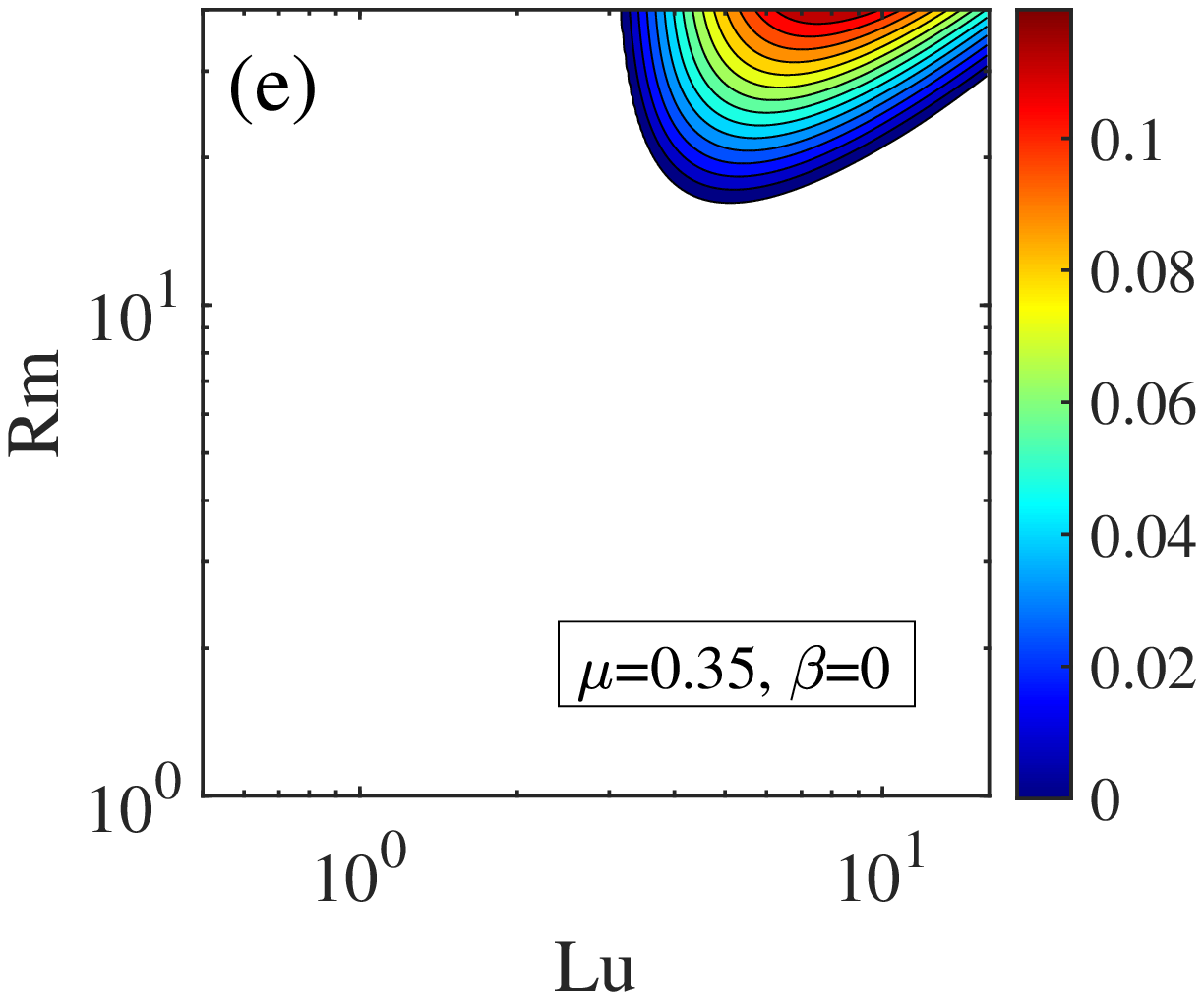}}
\end{minipage}
\begin{minipage}{.31\textwidth}
\centerline{\includegraphics[width=0.9\textwidth]{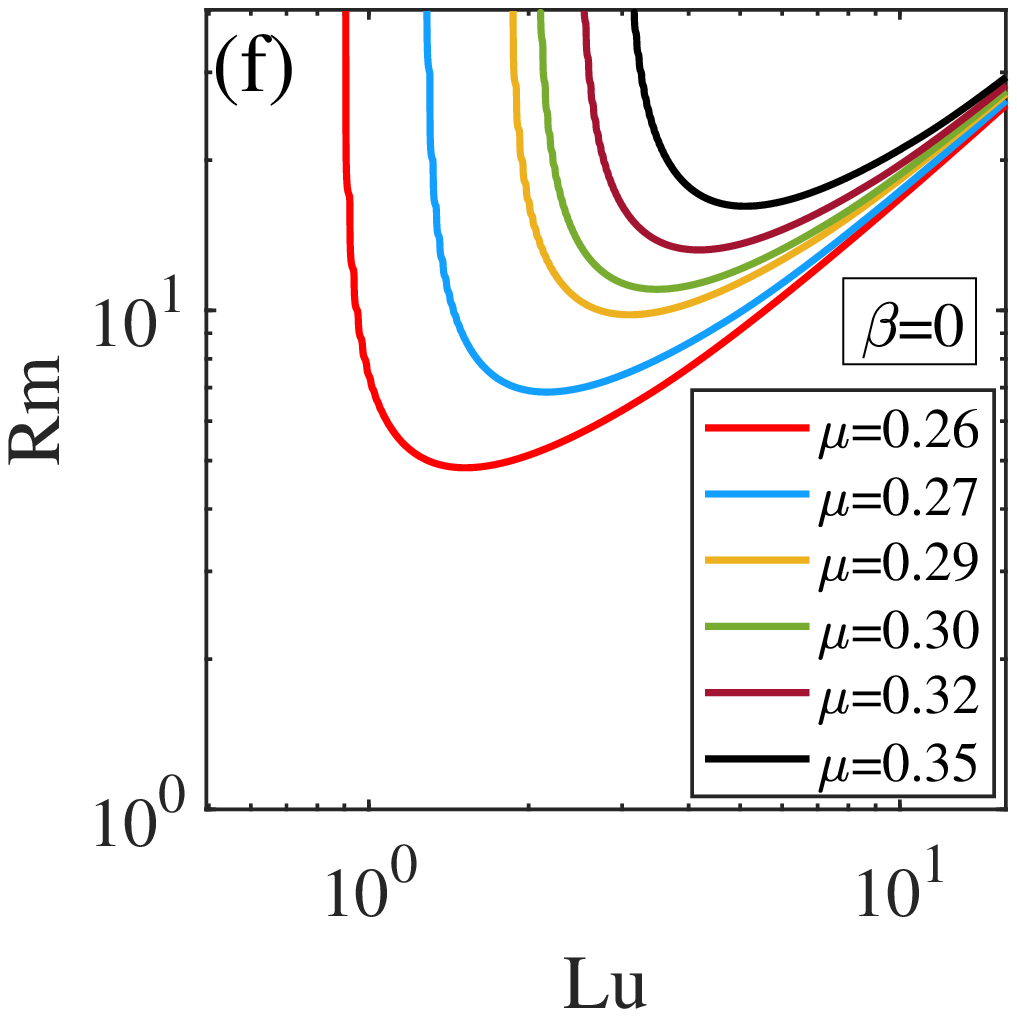}}
\end{minipage}			
\caption{ (a) - (e) SMRI growth rate ${\rm Re}(\gamma)\geq0$ in the $(Lu, Rm)$-plane, maximised over $k_z\geq k_{z,min}$, for fixed $\beta=0, Pm = 7.77 \times 10^{-6}$ and varying $\mu$. (f) The corresponding marginal stability curves for these values of $\mu$, which better indicate a shift of the unstable area to higher $Lu$ and $Rm$, implying a greater stability for increasing $\mu$.}\label{Fig2: Beta0_LuRm}
\end{figure*}

To obtain the regions of SMRI in the parameter space, in Fig. \ref{Fig2: Beta0_LuRm} we plot the growth rate, ${\rm Re}(\gamma) \geq 0$, of the most unstable axisymmetric mode, maximized over all axial wavenumbers $k_z\geq k_{z, min}$, in the ($Lu, Rm$)-plane for varying $\mu$ and a purely axial magnetic field with $\beta=0$. As mentioned above, we focus on the Rayleigh-stable regime, $\mu>0.25$. It is seen that the instability region in the ($Lu, Rm$)-plane decreases and moves towards higher $Lu$ and $Rm$ with increasing $\mu$. Consequently, the critical $Lu_c$ and $Rm_c$ for the onset of SMRI also increase monotonically with $\mu$, as evident in Fig. \ref{Fig2: Beta0_LuRm} (f), showing the corresponding marginal stability (${\rm Re}(\gamma)=0$) curves in the ($Lu, Rm$)-plane. For example, the critical $Lu_c=1.522$ and $Rm_c=4.838$ for $\mu=0.26$, while $Lu_c=5.094$ and $Rm_c=16.17$ for the quasi-Keplerian rotation with $\mu=0.35$. Since the maximum achievable Lundquist and magnetic Reynolds numbers in DRESDYN-MRI experiments are, respectively, 10 and 40 (Table I), it follows from these results that SMRI can be, in principle, well detectable in these experiments for $\mu$ in the range $(0.25, 0.35]$, which includes quasi-Keplerian rotation too, in contrast to previous studies in the smaller PROMISE experiment.
	
\subsection{HMRI and helically modified SMRI for $\beta \ne 0$}

\begin{figure*}
\centering
\begin{minipage}{.31\textwidth}
\centerline{\includegraphics[width=\textwidth]{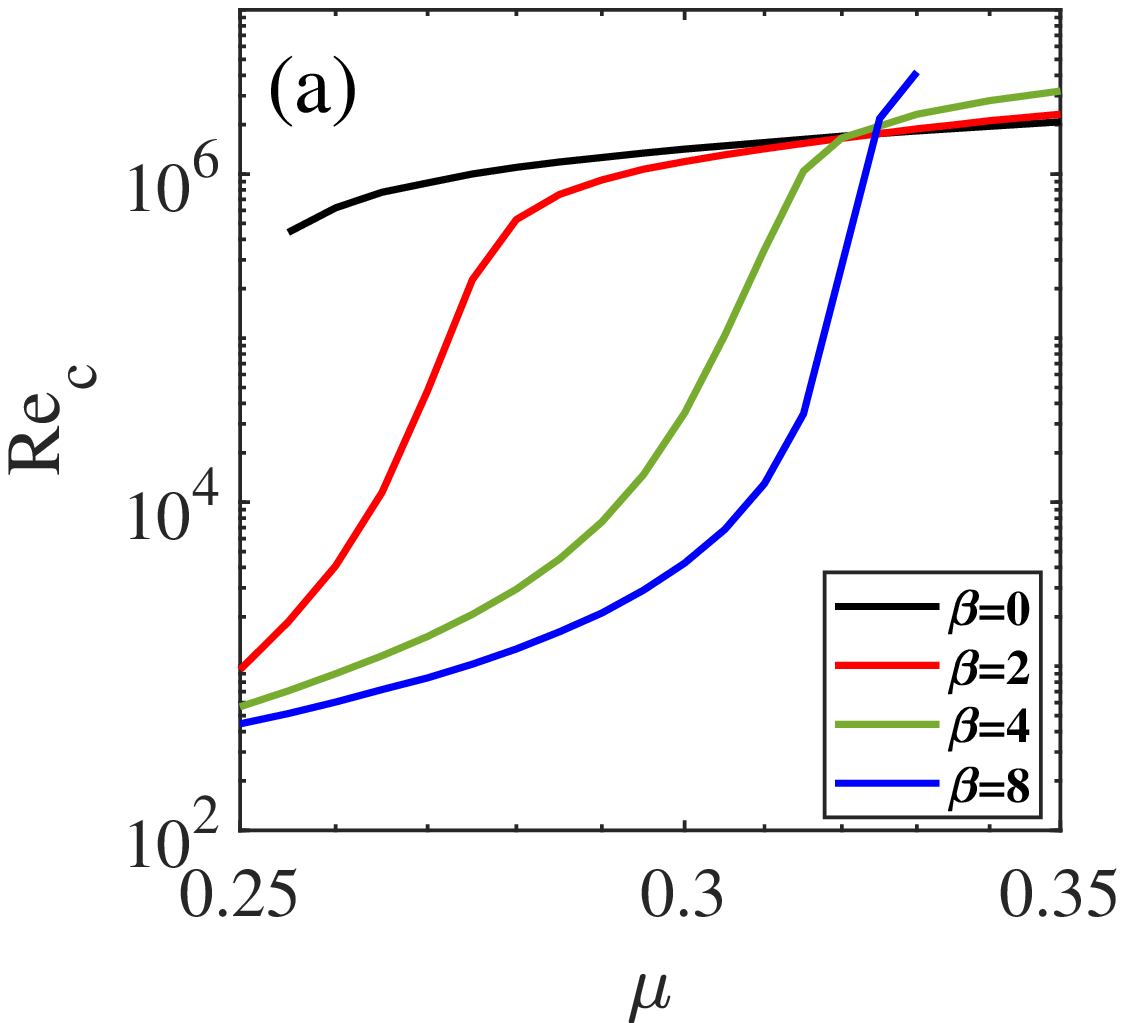}}
\end{minipage}
\hspace{2em}
\begin{minipage}{.32\textwidth}
\centerline{\includegraphics[width=\textwidth]{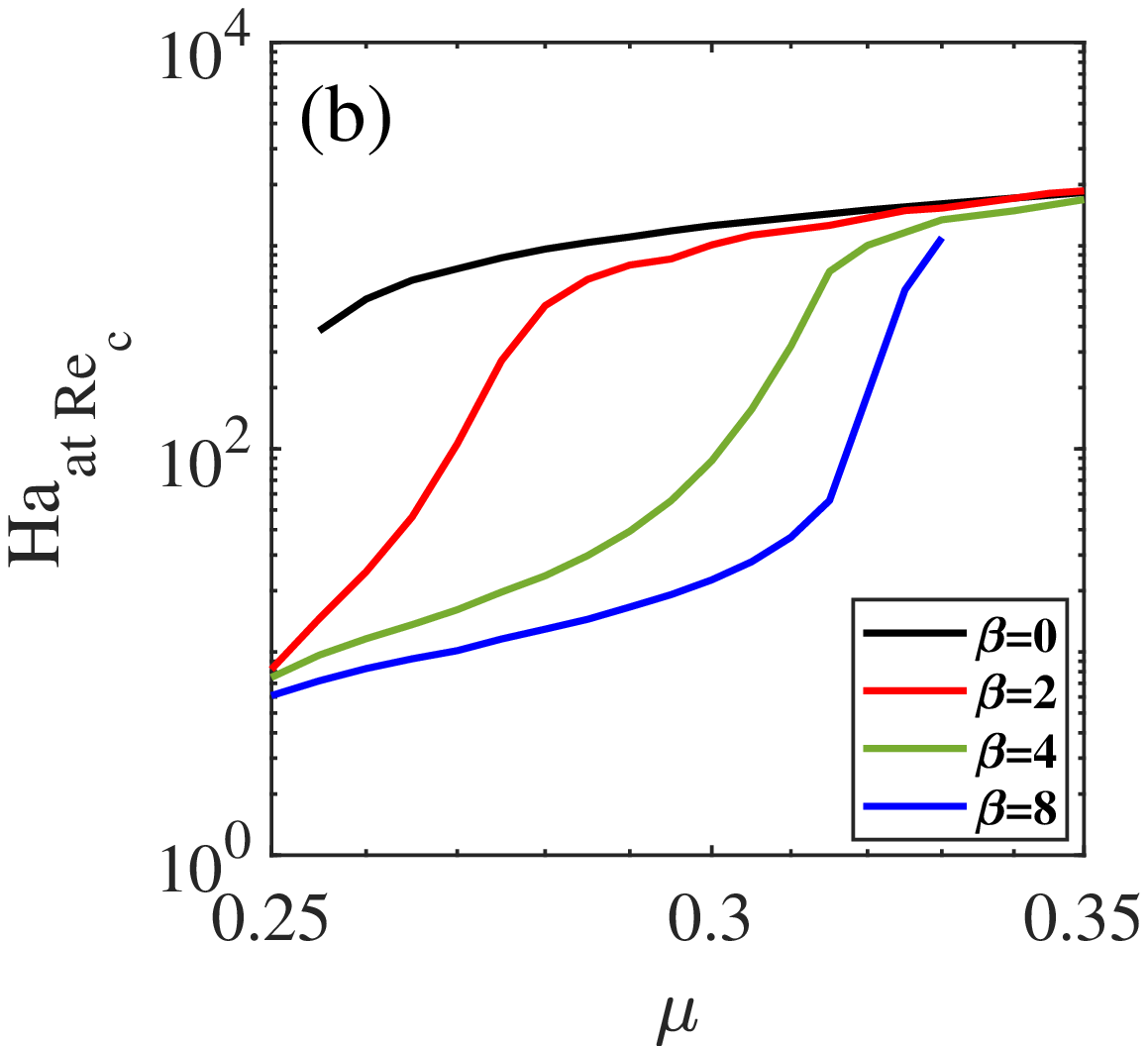}}
\end{minipage}
\caption{(a) Critical Reynolds numbers $Re_c$ of the instability, minimized over $k_z\geq k_{z,min}$, and (b) the corresponding Hartmann $Ha_c$ numbers at those $Re_c$, as functions of $\mu$ for different $\beta= 0, 2, 4, 8$. The black curve at $\beta=0$ corresponds to SMRI, while the other curves at $\beta=2,4,8$ correspond to HMRI at lower $\mu$, which eventually go smoothly into a plateau with much higher $Ha_c\sim 10^3$ and $Re_c\sim 10^6$, pertaining to the H-SMRI regime.}\label{Fig3: mu_vs_Rec}
\end{figure*}

In the DRESDYN-MRI experiments, we plan to approach SMRI regime from the well-known HMRI regime by gradually decreasing the azimuthal magnetic field, that is, for a fixed $\mu$ we start with a certain non-zero $\beta$ and gradually decrease it to 0. To examine the transition from HMRI to SMRI, in Fig. \ref{Fig3: mu_vs_Rec} we plot the critical $Re_c$ and $Ha_c$ as a function of $\mu$ at several $\beta=0, 2, 4, 8$. For SMRI at $\beta = 0$, the critical $Re_c\sim 10^6$ and $Ha_c\sim 10^3$ for $\mu > 0.25$ while for $\beta=2$ both these numbers fall to much lower values at $\mu < 0.275$, which lie in the well-known HMRI regime \cite{Hollerbach_Rudiger_2005}. As $\mu$ increases further, $Ha_c$ and $Re_c$ also steeply increase and at about $\mu=0.275$ level off near $Ha_c\sim 10^3$ and $Re\sim 10^6$, increasing beyond that only slowly with $\mu$. The latter range with much higher critical Reynolds and Hartmann (Lundquist) numbers clearly corresponds to the SMRI, or more precisely in \emph{the helically modified SMRI (H-SMRI) regime}, because of the presence of azimuthal field together with the axial one. Thus, we observe a continuous and monotonous transition from HMRI to H-SMRI as $\mu$ increases along the marginal stability curve at a given $\beta=2$, with the transition point being around $\mu=0.275$. It is seen in Fig. \ref{Fig3: mu_vs_Rec} that the marginal curves for larger $\beta = 4$ and $8$ exhibit a similar behavior with respect to $\mu$, except that $Re_c$ and $Ha_c$ fall to even lower values in the HMRI regime at smaller $\mu$ as compared to $\beta=2$ case and level off at higher $\mu=0.32$ and 0.33, respectively, in the H-SMRI regime. Thus, there is a gradual decrease of $Re_c$ and $Ha_c$ for the onset of HMRI with increasing $\beta$. By contrast, in the H-SMRI regime at larger $\mu$, $Re_c$ increases, though weakly, while $Ha_c$ remains nearly the same with increasing $\beta$. This implies that the background azimuthal field acts to enhance HMRI, but to stabilize H-SMRI. Note that for the quasi-Keplerian value $\mu=0.35$ there is only H-SMRI branch (see Sec. C). The dramatic reduction in the critical values of the system parameters for the onset of HMRI compared to the experimentally challenging ones for (H-)SMRI, as first demonstrated in Ref. \cite[][see their Fig. 1, which is similar to our Fig. 3]{Hollerbach_Rudiger_2005}, had enabled successful detection of HMRI in the PROMISE experiments \cite{Stefani_Gundrum_Gerbeth_etal_2006PhRvL, Stefani_Gerbeth_Gundrum_Etal_2009PhysRevE}. However, the (H-)SMRI had remained still elusive in those experiments.

\begin{figure*}
\begin{minipage}{.32\textwidth} 
\centerline{\includegraphics[width=\textwidth]{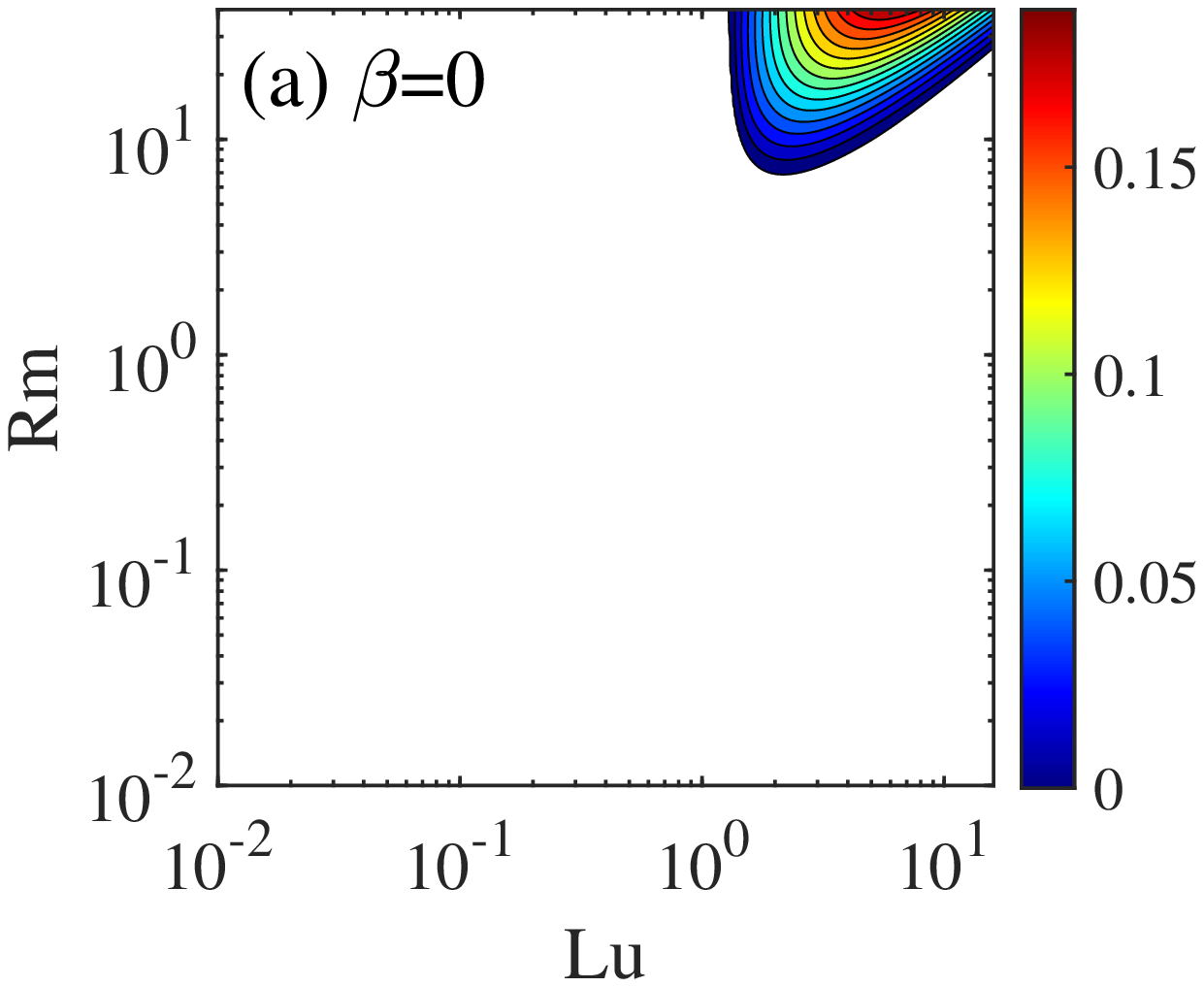}}
\end{minipage}
\vspace{0.7cm}
\begin{minipage}{.32\textwidth}
\centerline{\includegraphics[width=\textwidth]{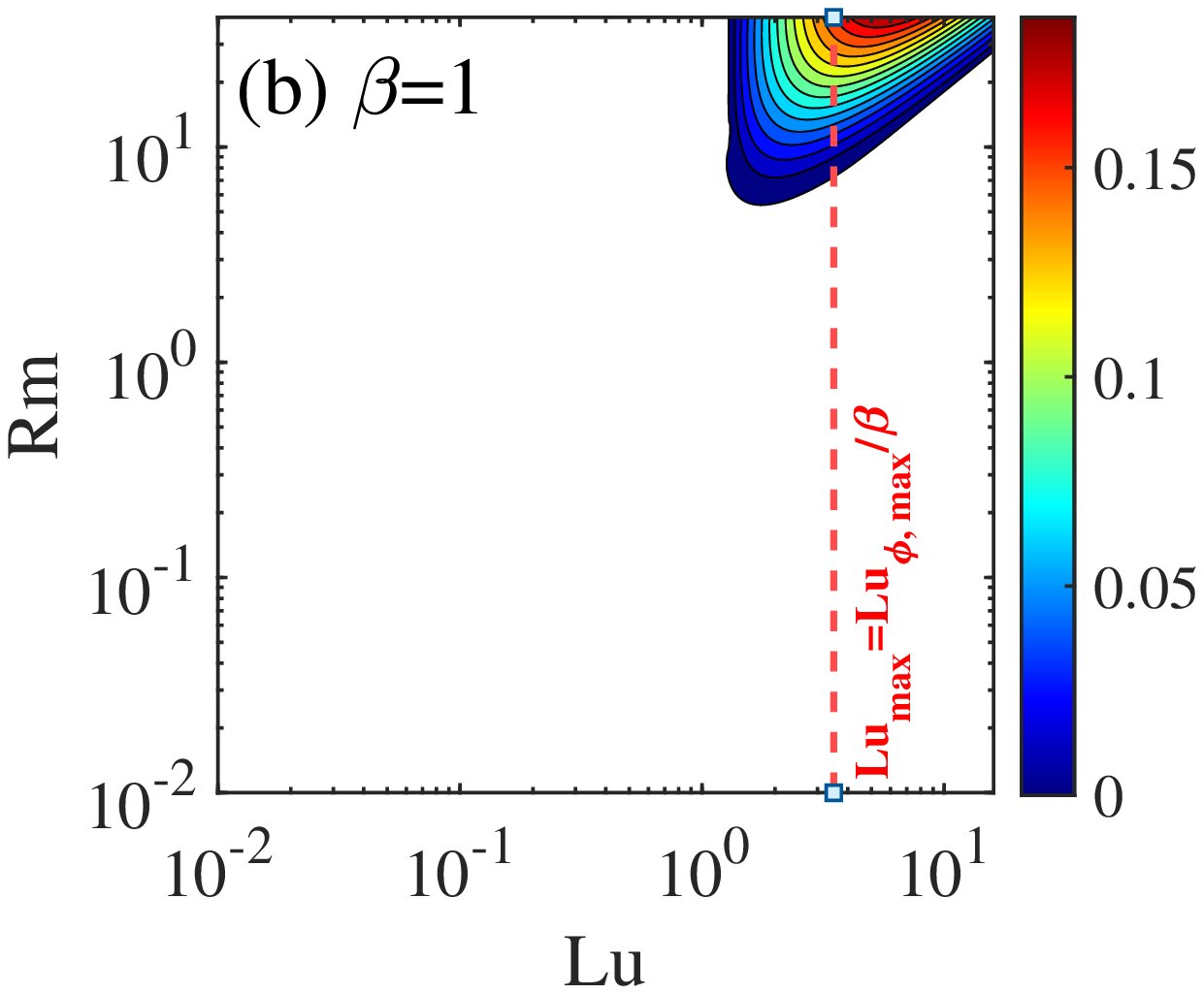}}
\end{minipage}
\begin{minipage}{.32\textwidth}
\centerline{\includegraphics[width=\textwidth]{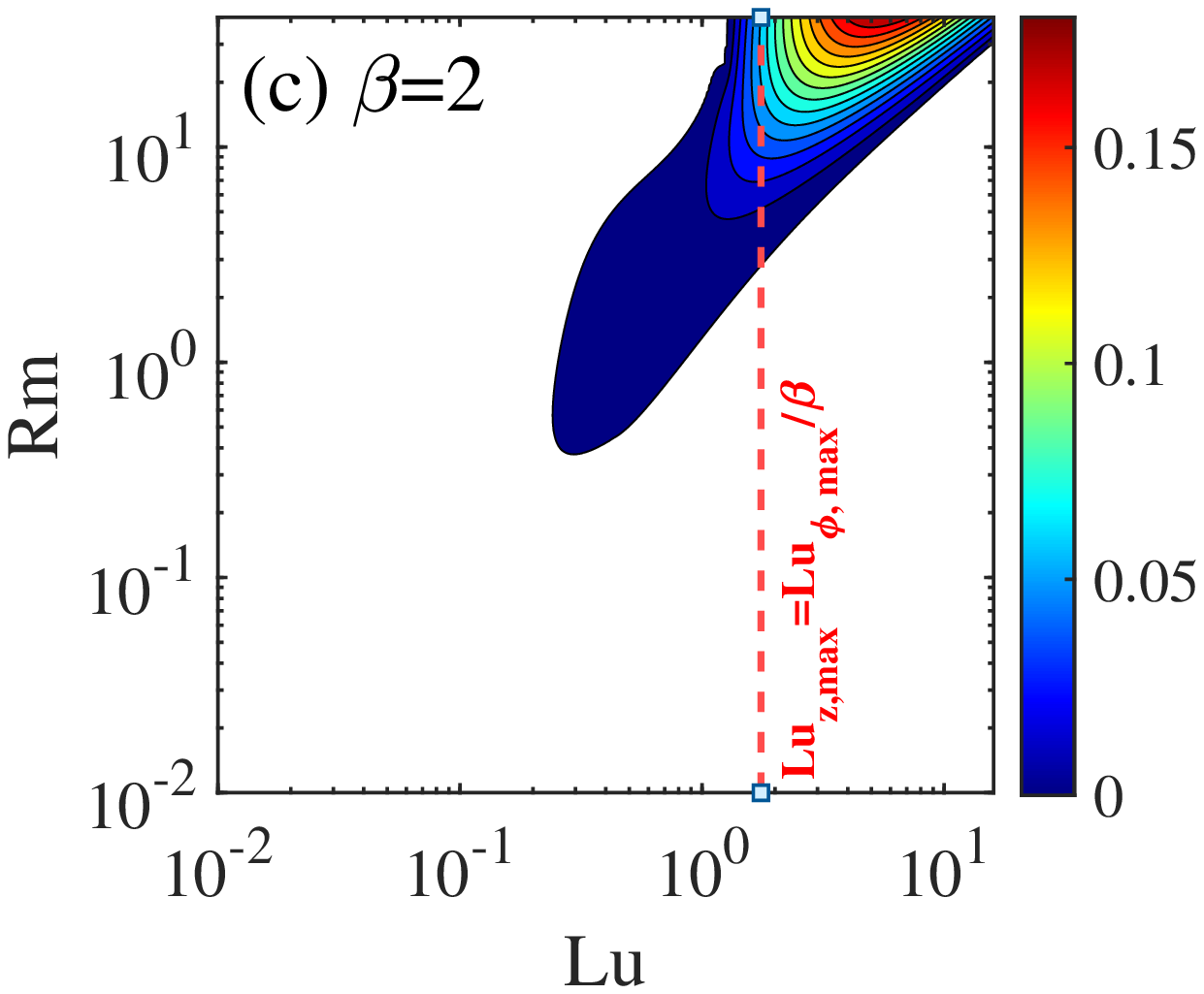}}
\end{minipage}
\begin{minipage}{.32\textwidth}
\centerline{\includegraphics[width=\textwidth]{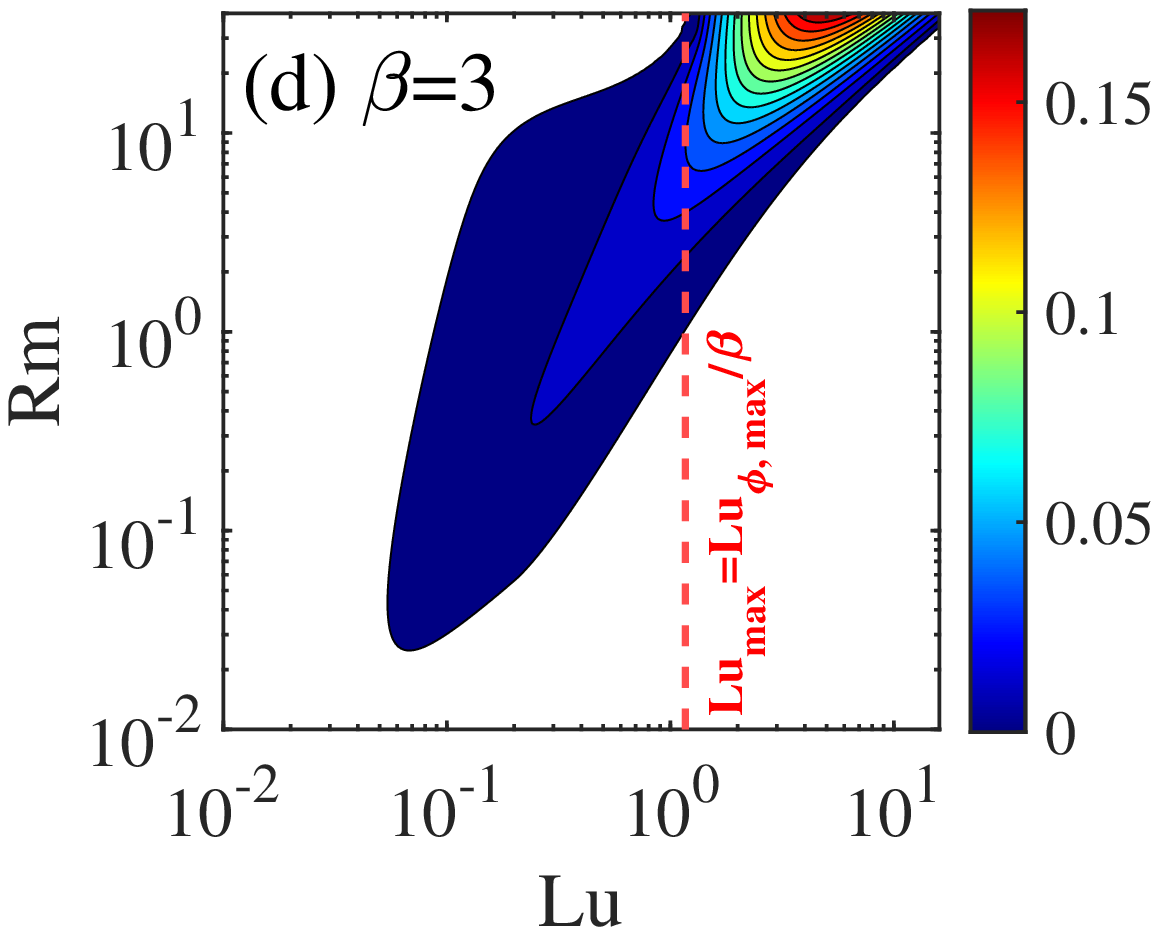}}
\end{minipage}
\begin{minipage}{.32\textwidth}
\centerline{\includegraphics[width=\textwidth]{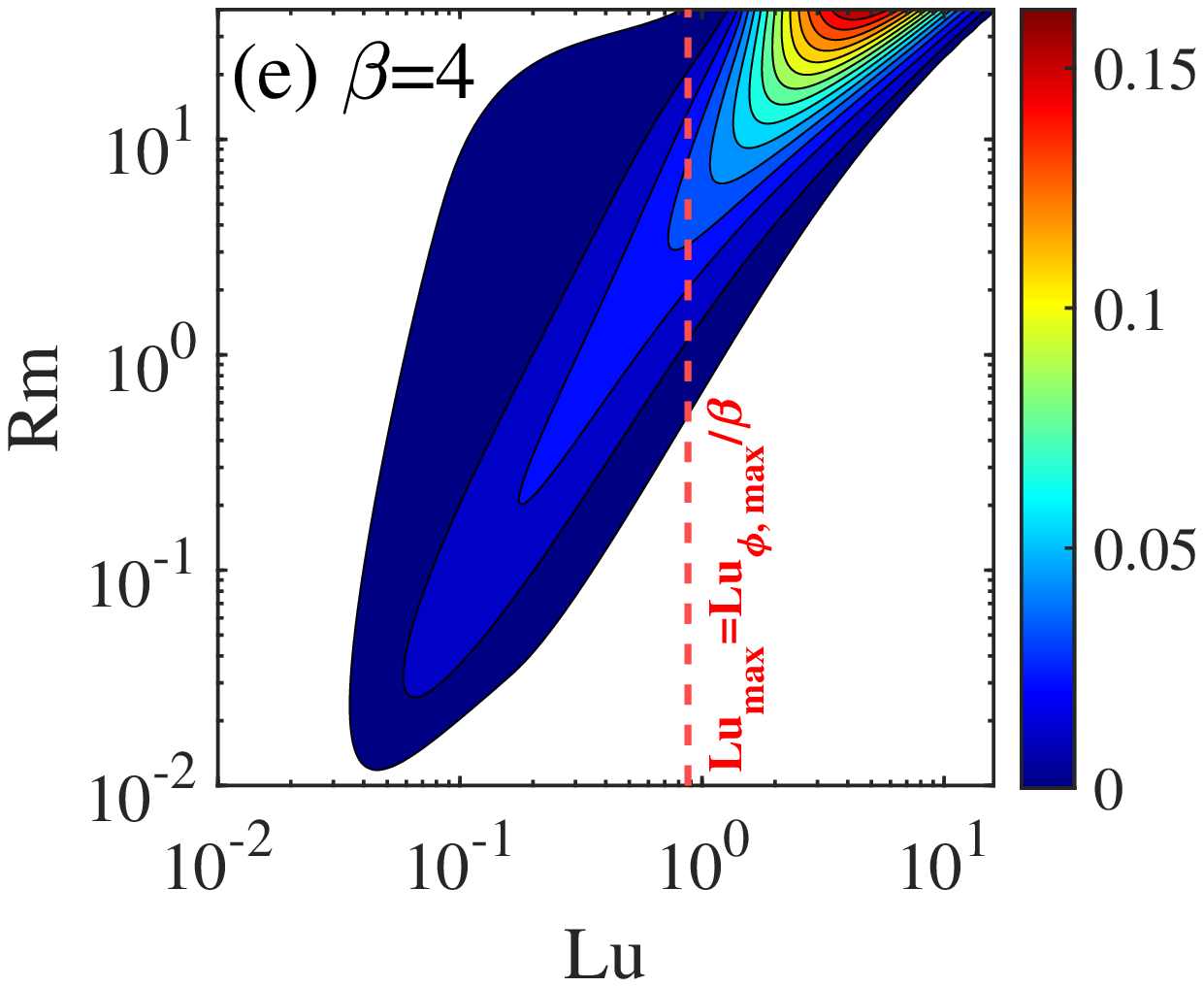}}
\end{minipage} 
\begin{minipage}{.31\textwidth}
\centerline{\includegraphics[width=0.85\textwidth]{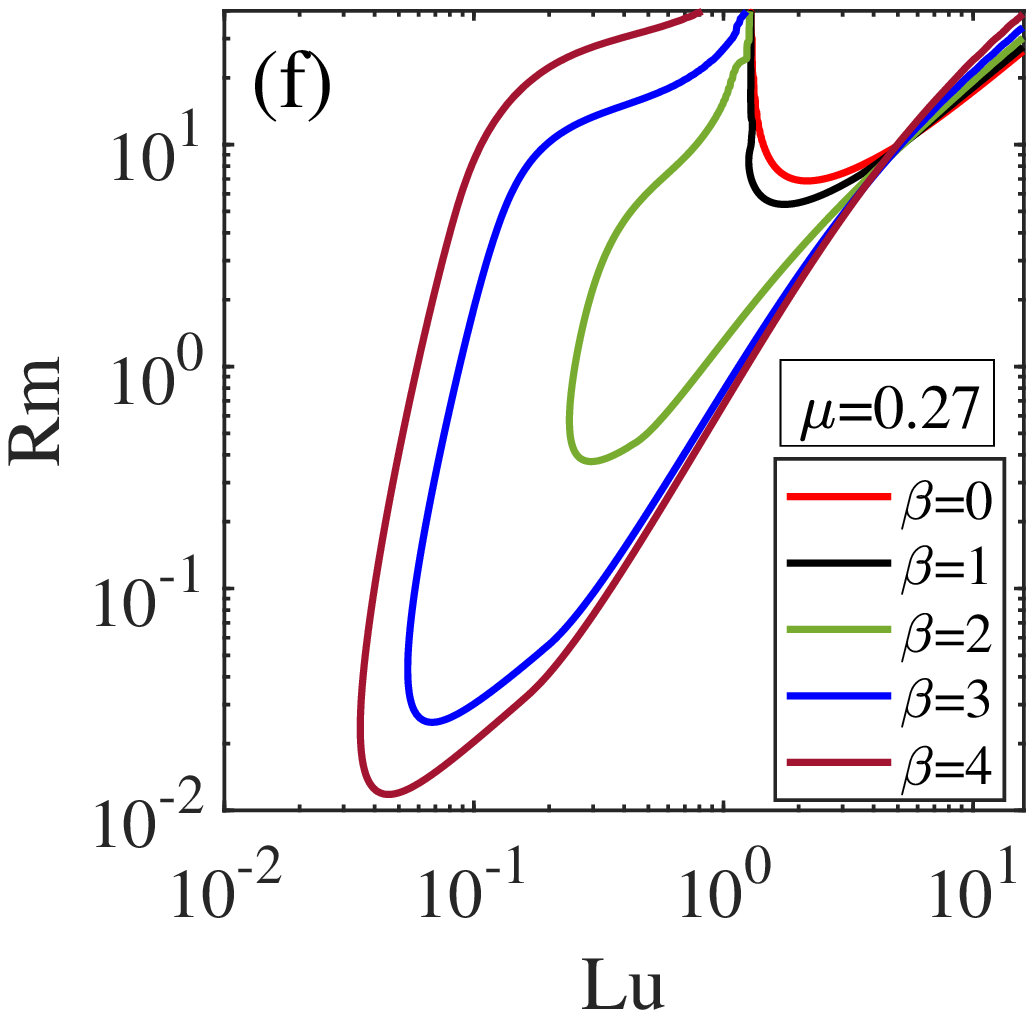}}
\end{minipage}
\caption{(a) - (e) Growth rate $\rm Re(\gamma)\geq0$ in the $(Lu,Rm)$-plane, maximized over $k_z\geq k_{z,min}$, for fixed $\mu=0.27, Pm = 7.77 \times 10^{-6}$ and different $\beta$. The vertical red-dashed line in the panels marks the limit on maximum $Lu$ for each $\beta$ and fixed maximum azimuthal $Lu_{\phi}$ that can be attained in the experiment (see Table \ref{Table2: non_dimentional}). This line shifts to the left (to smaller $Lu$) as $\beta$ increases. In panel (a), for $\beta=0$ only SMRI exists in the flow. In panels (b) - (e), the more extended instability branch for $\beta\ne 0$ at lower $Lu$ and $Rm$ represents HMRI (dark blue), while the instability region at higher $Lu$ and $Rm$ is H-SMRI (blue/yellow/red). (f) The corresponding marginal stability curves for the same values of $\beta$ and fixed $\mu=0.27$ in the $(Lu, Rm)$-plane. As $\beta$ increases, the HMRI stability curve branches off the SMRI curve (red at $\beta=0$) and further extends to lower and lower $Lu$ and $Rm$.} \label{Fig4: mu27_beta}
\end{figure*}

To further explore the transition from HMRI to SMRI, in Fig. \ref{Fig4: mu27_beta} we show the growth rate, ${\rm Re}(\gamma) \geq 0$, maximized over axial wavenumbers $k_z\geq k_{z,min}$, in the ($Lu, Rm$)-plane as $\beta$ is increased from 0 to 4 for a fixed $\mu=0.27$. For $\beta=0$, the imposed magnetic field is purely axial, so only SMRI takes place with a critical $Lu_c=2$ and $Rm_c=6.861$ (Fig. \ref{Fig4: mu27_beta}a). On reaching $\beta=1$, by applying an azimuthal magnetic field externally, the critical $Lu_c$ and $Rm_c$ decrease slightly and the instability region starts  changing its shape predominantly near these critical values, as seen in Fig. \ref{Fig4: mu27_beta}(b), indicating the onset (branching out) of HMRI mode and  the modification of SMRI mode due to helical magnetic field. In the experiments, we have an upper limit on the axial current and therefore on the azimuthal magnetic field measured by azimuthal Lundquist number $Lu_{\phi}=\beta Lu \leq Lu_{\phi,max}$ (see Table \ref{Table2: non_dimentional} for definition and value), which in turn imposes a stricter constraint on $Lu$ for each $\beta$, that is, $Lu < Lu_{max}=Lu_{\phi,max}/\beta$. To take this into account, we have added vertical red-dashed lines in Figs. \ref{Fig4: mu27_beta}(b)-\ref{Fig4: mu27_beta}(e) that mark the corresponding upper limit, $Lu_{max}$, on the axial Lundquist number that can be achieved in the experiments for a given $\beta$ and the maximum azimuthal $Lu_{\phi, max}=3.51$ (Table II). It is seen in this figure that as $\beta$ is increased, $Lu_{max}$ decreases (the red dashed line shifts left) due to the constraint on the maximum $Lu_{\phi}$. As a result, larger and larger unstable area of the $(Lu, Rm)$-plane beyond this threshold, $Lu>Lu_{max}$, which primarily belongs to H-SMRI, becomes experimentally inaccessible. For this reason, in the experiments, we plan to start from a reasonable value of $\beta\sim 4$ and decrease it, thus allowing for the gradual increase of $Lu$ until reaching the H-SMRI regime. 
	
On reaching $\beta=2$, the unstable region spreads out further down in the $(Lu, Rm)$-plane (Fig. \ref{Fig4: mu27_beta}c). Since the applied azimuthal magnetic field is stronger now, the new extended instability region (dark blue) at smaller $Lu$ and $Rm$ can be interpreted as a HMRI mode, whereas the instability region at larger $Lu$ and $Rm$, which originally (at $\beta=0$) corresponded to SMRI and out of which HMRI has branched, represents now the H-SMRI branch (light blue/yellow/red). However, it is worth noting that the continuity of the overall instability region is maintained, in the sense that there is no sharp boundary separating these two instability types. This classification of the instability into essential HMRI and H-SMRI in the present global 1D analysis is consistent with a similar classification made in the local WKB study  \cite{Kirillov_Stefani_2010ApJ}. As $\beta$ increases to 3 and 4 in Figs. \ref{Fig4: mu27_beta}(d) and \ref{Fig4: mu27_beta}(e), respectively, the HMRI branch broadens and extends further down in the lower-left segment of the ($Lu,Rm$)-plane, leading to significantly reduced critical $Lu_c$ and $Rm_c$ in contrast to H-SMRI, as we have already observed in Fig. \ref{Fig3: mu_vs_Rec}. Note also in this figure that in the presence of azimuthal magnetic field, i.e., $\beta \ne 0$, the maximum growth rate of H-SMRI is lower than that of pure SMRI at $\beta=0$ in the given domain of $Lu$ and $Rm$, implying that the azimuthal field, while giving rise to HMRI, also modifies SMRI (referred to as H-SMRI) and reduces its growth rate. In Fig. \ref{Fig4: mu27_beta}(f), we plot the corresponding marginal stability (${\rm Re}(\gamma)=0$) curves for $\mu=0.27$ and different $\beta$. It is seen how the stability curve of HMRI gradually branches off the SMRI curve (red at $\beta=0$) and extends to the smaller values of $Lu$ and $Rm$ as the applied azimuthal magnetic field ($\beta$) increases with respect to the axial one.

\begin{figure}
\centering
\includegraphics[width=0.6\columnwidth]{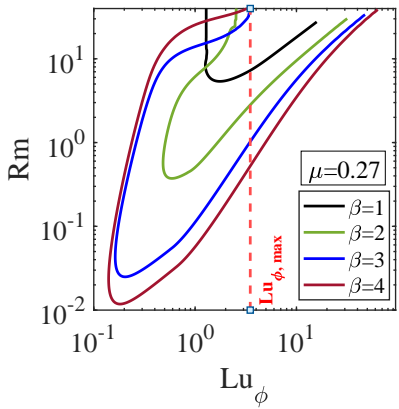}
\caption{Marginal stability curves, $\rm Re(\gamma)=0$, optimized over $k_z\geq k_{z,min}$, for fixed $\mu=0.27, Pm = 7.77 \times 10^{-6}$ and different $\beta$, as shown in Fig. \ref{Fig4: mu27_beta}(f) but now in the $(Lu_\phi, Rm)$-plane. The vertical red-dashed line marks the experimental limit on the maximum $Lu_{\phi}$ (see Table \ref{Table2: non_dimentional}).}
\label{fig:mu27_Lu_phi}
\end{figure}

Figure \ref{fig:mu27_Lu_phi} shows the same marginal stability curves for $\mu=0.27$ and different $\beta$ as in Fig. \ref{Fig4: mu27_beta} (f), but as a function of the azimuthal Lundquist number $Lu_{\phi}$ and $Rm$. The vertical red-dashed line indicates the maximum value, $Lu_{\phi, max}$, achievable in the DRESDYN experiments. As seen in Fig. \ref{fig:mu27_Lu_phi}, for $\beta \leq 1$, the most of the H-SMRI region is well attainable in the experiment. However, for larger $\beta \geq 1$, the H-SMRI region appears to be mostly located at $Lu_{\phi}>Lu_{\phi,max}$, or equivalently $Lu>Lu_{\phi,max}/\beta$, outside the range of experimentally achievable Lundquist numbers, as is also seen in Figs. \ref{Fig4: mu27_beta}(c)-\ref{Fig4: mu27_beta}(e).  

\begin{figure*}
\begin{minipage}{.45\textwidth}
\centerline{\includegraphics[width=0.7\textwidth]{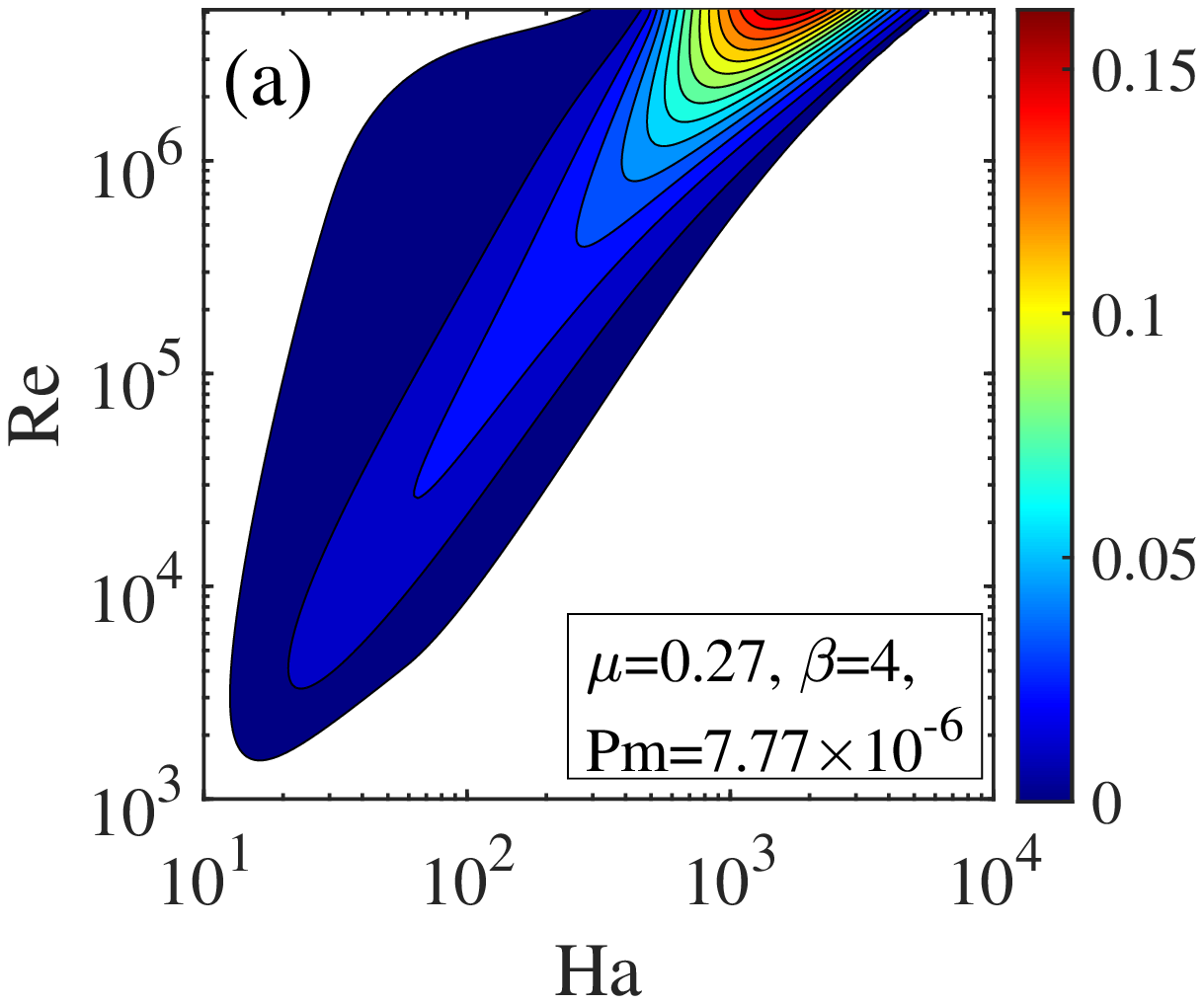}}
\end{minipage}
\vspace{0.7cm}
\begin{minipage}{.45\textwidth}
\centerline{\includegraphics[width=0.7\textwidth]{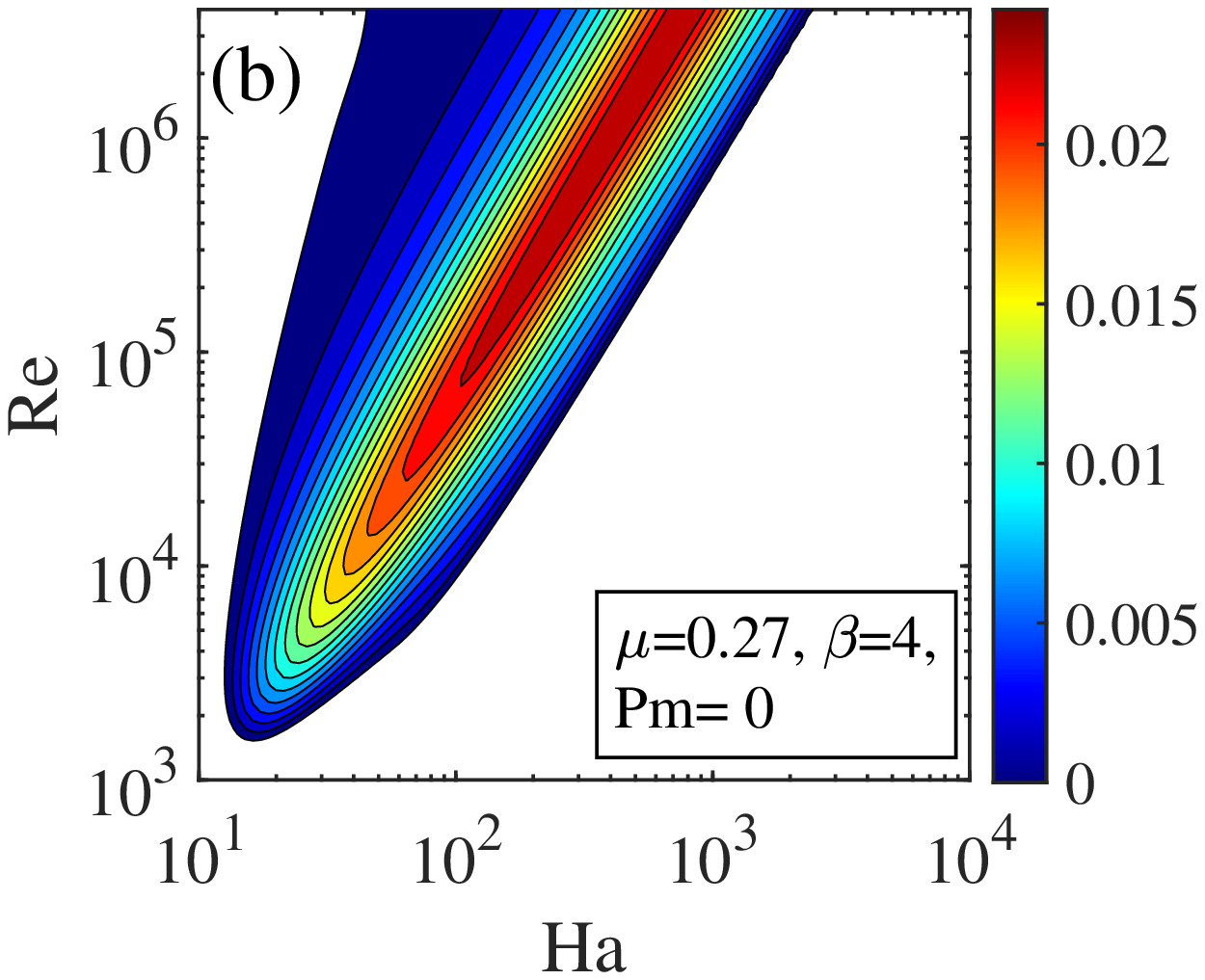}}
\end{minipage}
\begin{minipage}{.45\textwidth}
\centerline{\includegraphics[width=0.7\textwidth]{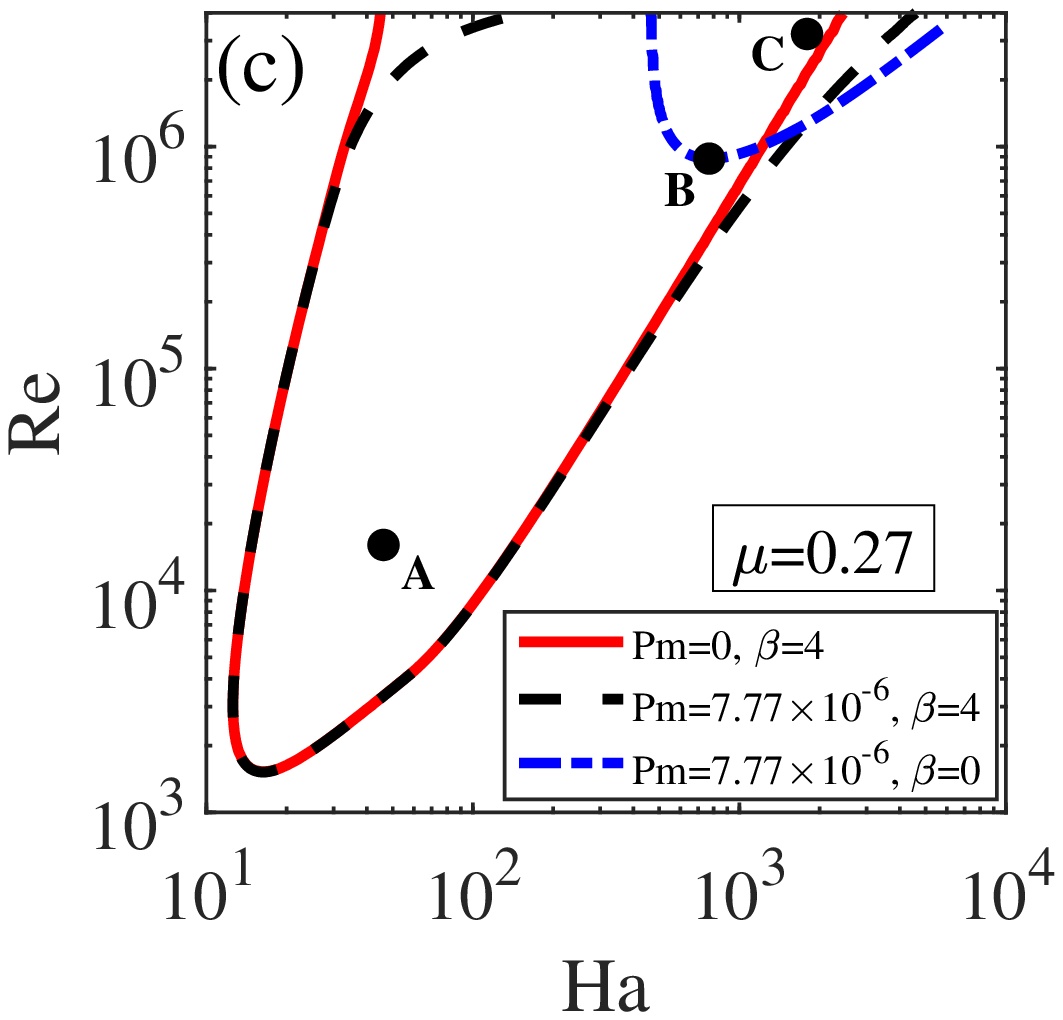}}
\end{minipage}
\begin{minipage}{.45\textwidth}
\centerline{\includegraphics[width=0.7\textwidth]{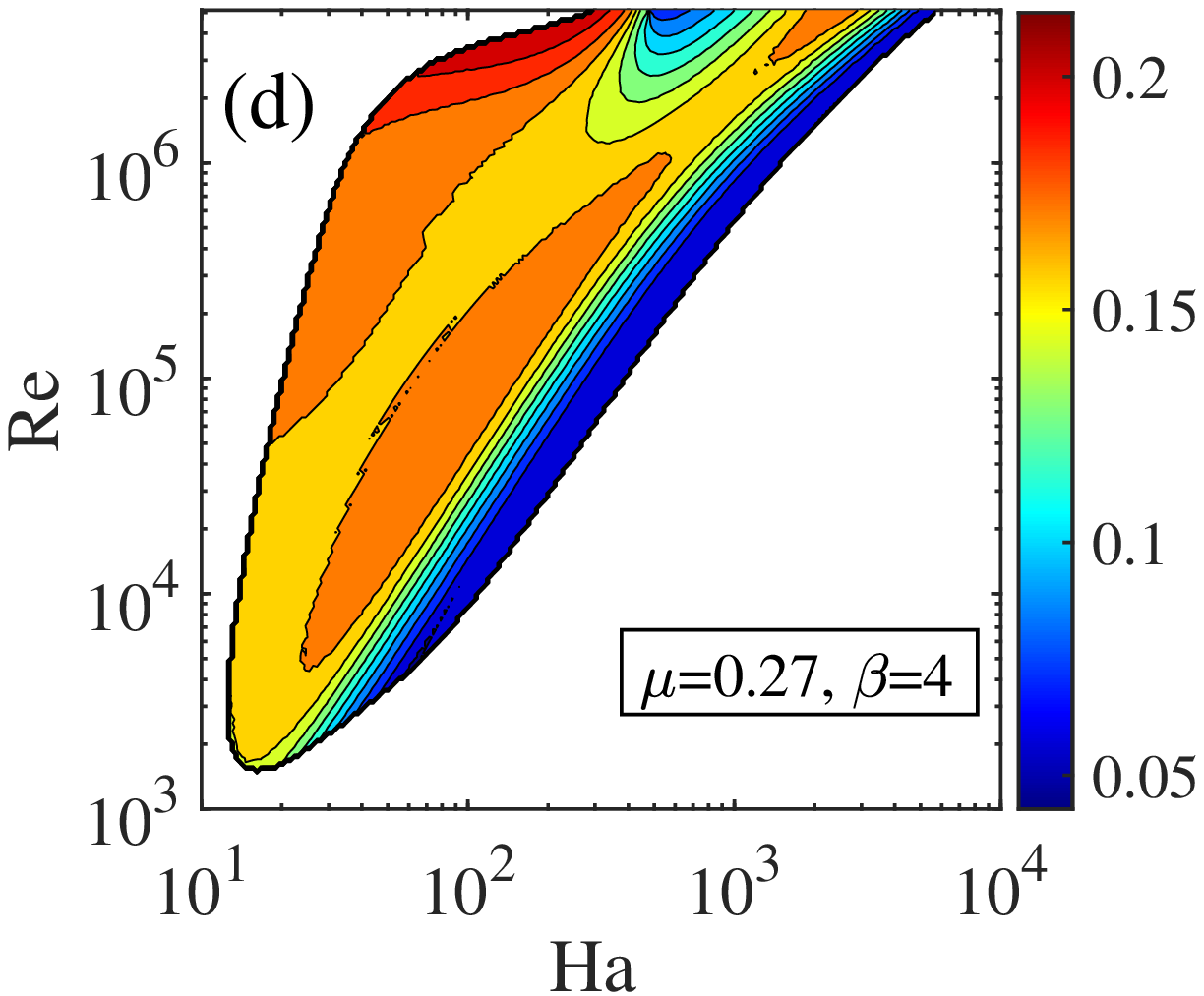}}
\end{minipage}
\caption{(a) Growth rate $\rm Re(\gamma)\geq0$ at fixed $\mu=0.27, Pm = 7.77 \times 10^{-6}$ and $\beta=4$ in the $(Ha, Re)$-plane. (b) shows the same as (a) but in the inductionless limit $Pm =0$, which represents the essential HMRI. (c) The associated marginal stability curves in the $(Ha, Re)$-plane at $Pm = 0$, $\beta=4$ for HMRI (red), at $Pm = 7.77 \times 10^{-6}, \beta=4$ (dashed black) comprising HMRI in the lower extended part and H-SMRI upper part and, given for reference, SMRI curve for $Pm = 7.77 \times 10^{-6}, \beta=0$ (blue dashed). Points A, B and C are analysed in more detail in Figs. \ref{Fig7: Gr_k} - \ref{transition_region} and \ref{Fig: mu27_Abs_instability}. (d) Eigenfrequency distribution in the $(Ha,Re)$-plane corresponding to the growth rate map in (a).} \label{Fig6: mu27_Pm_comparison}.
\end{figure*}
	
To confirm that the broadening instability regions with increasing $\beta$ in Fig. \ref{Fig4: mu27_beta} indeed correspond to the essential HMRI, in Fig. \ref{Fig6: mu27_Pm_comparison} we show the growth rate in the $(Ha,Re)$-plane (a) at small but finite $Pm=7.77\times10^{-6}$ normally used in this paper and (b) in the inductionless approximation, $Pm=0$, with the same $\mu=0.27$ and $\beta=4$. In the inductionless limit, only essential HMRI survives, whereas H-SMRI disappears  \cite{Priede_Grants_Gerbeth_2007PhRvE}, so Fig. \ref{Fig6: mu27_Pm_comparison}(b) in fact depicts the most unstable HMRI mode, whose shape in the $(Ha,Re)$-plane is almost identical to that of the extended instability branch in Fig. \ref{Fig6: mu27_Pm_comparison}(a) for the smaller $Ha$ and $Re$ in this plane. Note also that the maximum growth rate in the case of finite $Pm$, associated with H-SMRI, is larger than that of the inductionless HMRI. To further demonstrate this similarity, in Fig. \ref{Fig6: mu27_Pm_comparison}(c) we show the associated marginal stability curves for the inductionless HMRI (red line), for the finite $Pm$ case (black dashed line) and, as a reference, for SMRI (blue dashed line). It is evident from this figure that the lower extended part of the stability curve at finite $Pm$ and smaller $Ha$ and $Re$ identically matches that of the inductionless essential HMRI, hence scaling with these numbers, and gradually deviates from the latter with increasing $Ha$ and $Re$ as HMRI smoothly transitions into H-SMRI and scales instead with $Lu$ and $Rm$. Although the domains of SMRI and H-SMRI overlap in this plot, the domain of SMRI does not fully lie within the domain of H-SMRI as a result of modification by the azimuthal field.

Figure \ref{Fig6: mu27_Pm_comparison}(d) shows the distribution of the eigenfrequency in the $(Ha,Re)$-plane at finite $Pm$ and $\beta=4$. Two distinct smoothly interconnecting, broad high-frequency region (yellow/orange/red) and one small low-frequency region (blue) at higher $Re$, are evident in this plot, which correspond, respectively, to the unstable regions of low and high growth rate in Fig. \ref{Fig6: mu27_Pm_comparison}(a). In other words, the less unstable HMRI in the lower extended part of the $(Ha,Re)$-plane has noticeably larger frequency than that of the more unstable H-SMRI in the upper part of this plane. It is known that HMRI represents the instability of inertial waves and hence has the frequency of these waves \cite{Liu_Goodman_Herron_Ji_2006PhRvE, Kirillov_etal2014}, which is mostly independent of the magnetic field. On the other hand, (H-)SMRI is the instability of magneto-coriolis waves \cite{Nornberg_Ji_etal_2010_PhysRevLett, Kirillov_Stefani_2010ApJ}, whose frequency is zero for SMRI at $\beta=0$ and linearly increases with $\beta$ for H-SMRI (see also Fig. 13 in Sec. IV). 
	
\begin{figure*}
\includegraphics[width=0.30\textwidth]{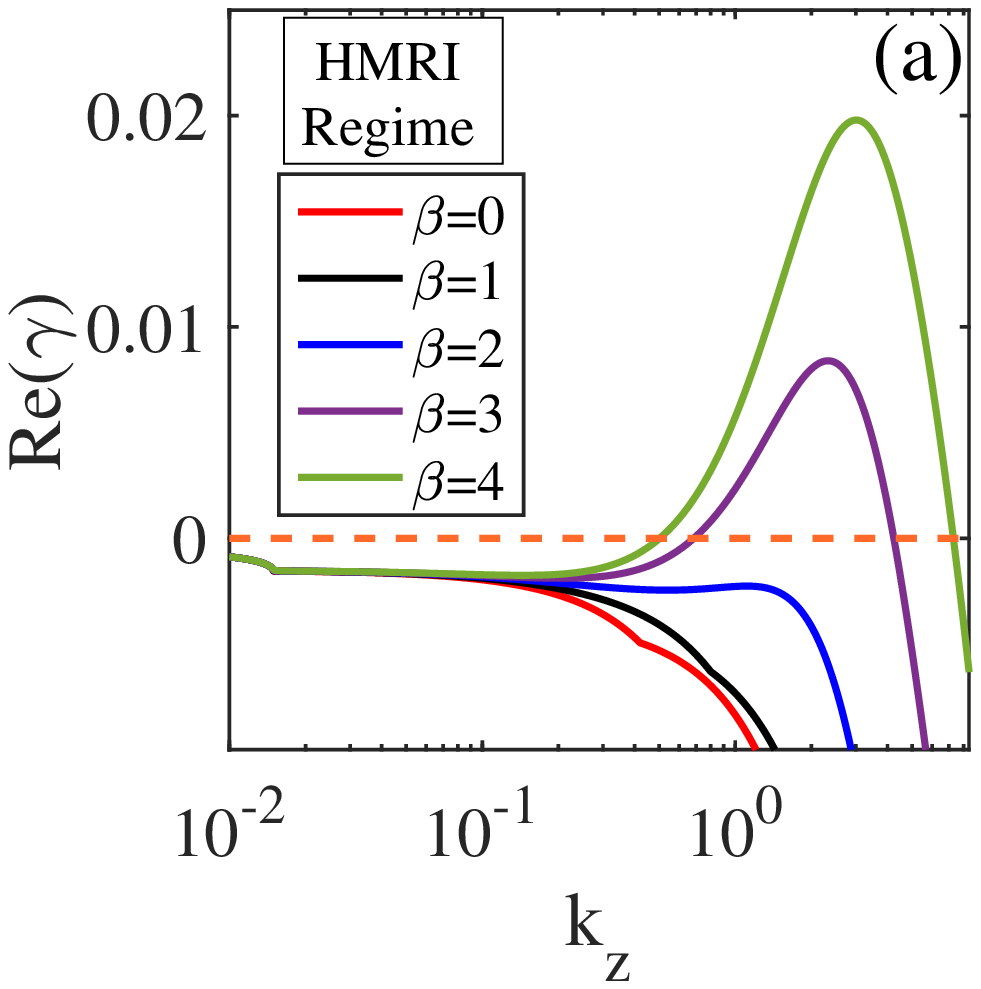}
\hspace{2.em}
\vspace{0.6cm}
\includegraphics[width=0.30\textwidth]{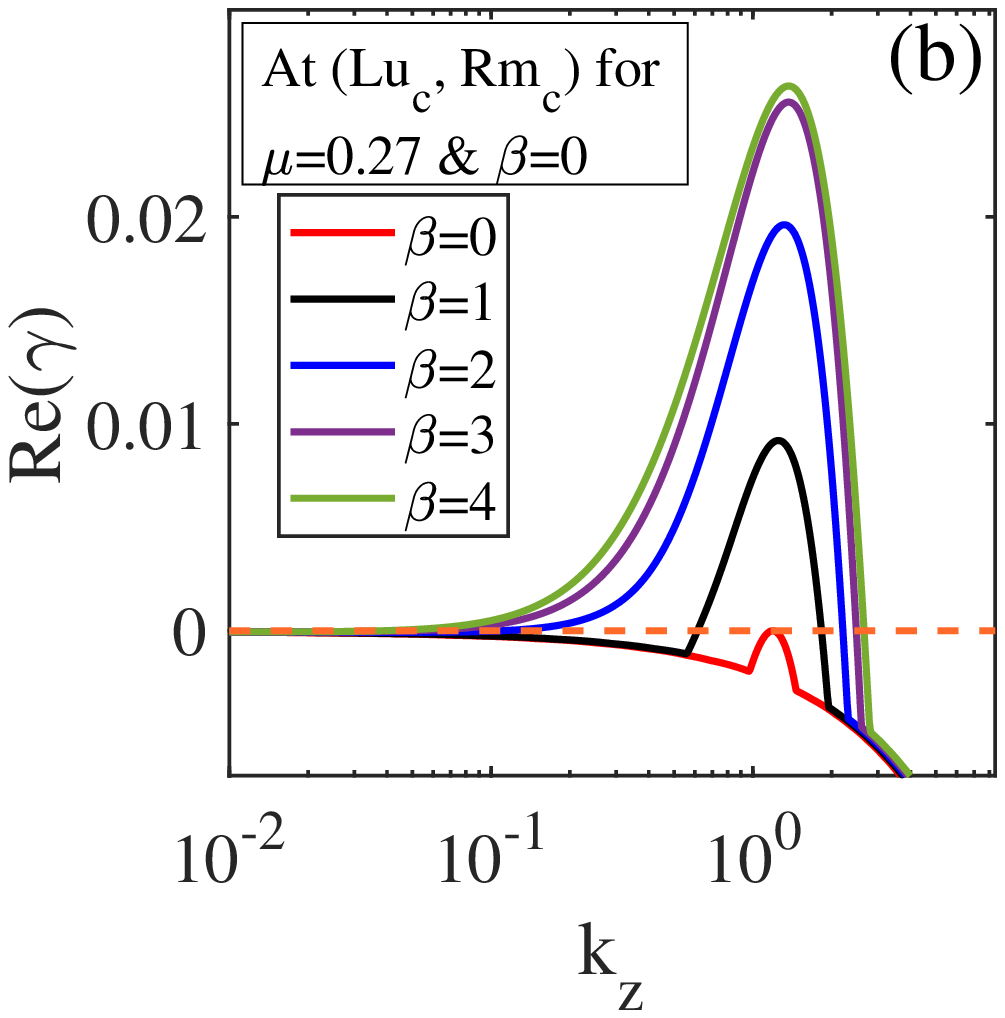}
\hspace{2.em}
\includegraphics[width=0.30\textwidth]{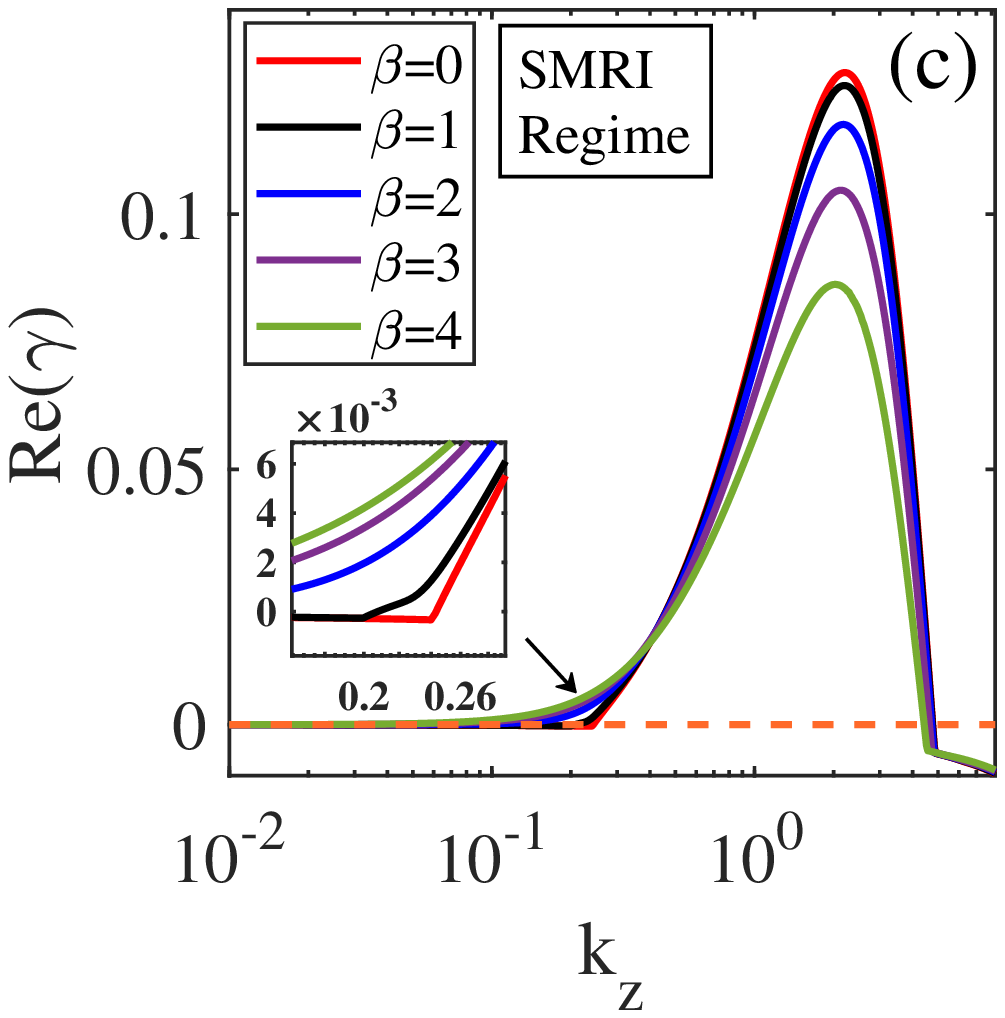}
\includegraphics[width=0.30\textwidth]{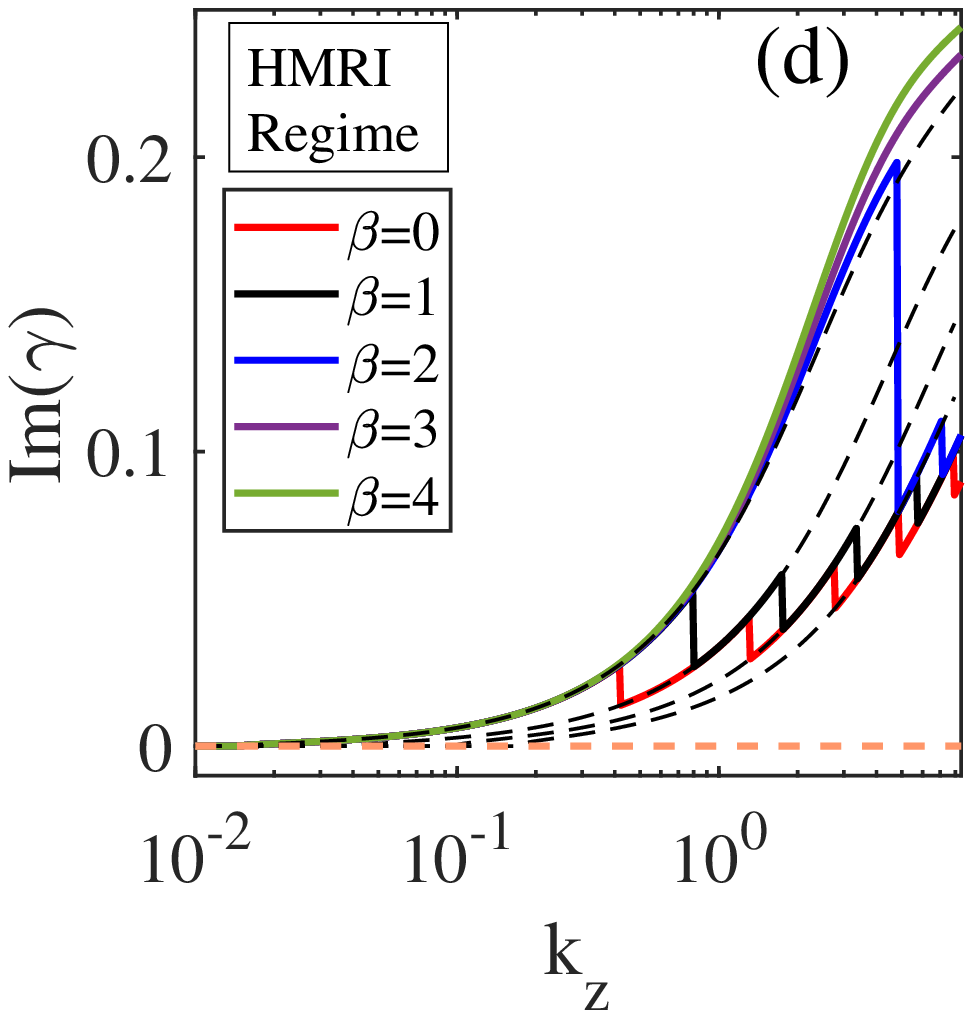}
\hspace{2.em}
\includegraphics[width=0.30\textwidth]{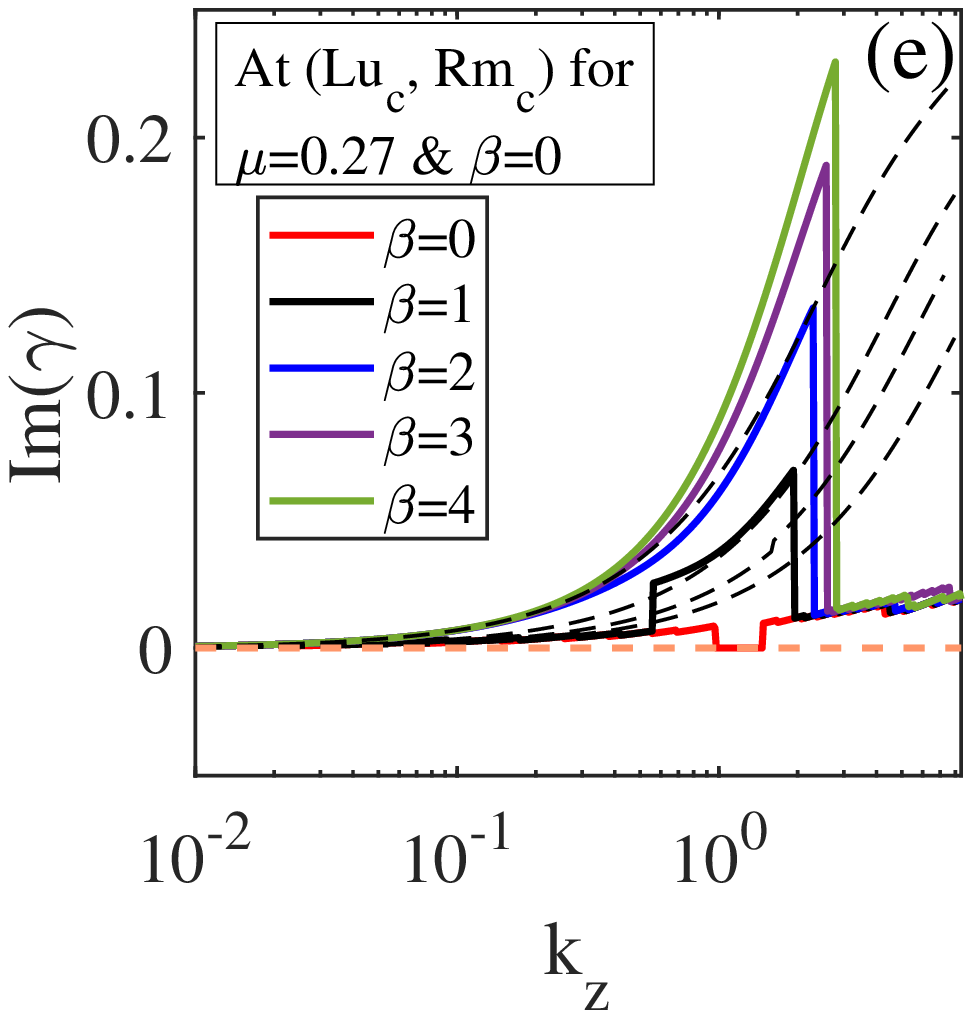}
\hspace{2.em}
\includegraphics[width=0.30\textwidth]{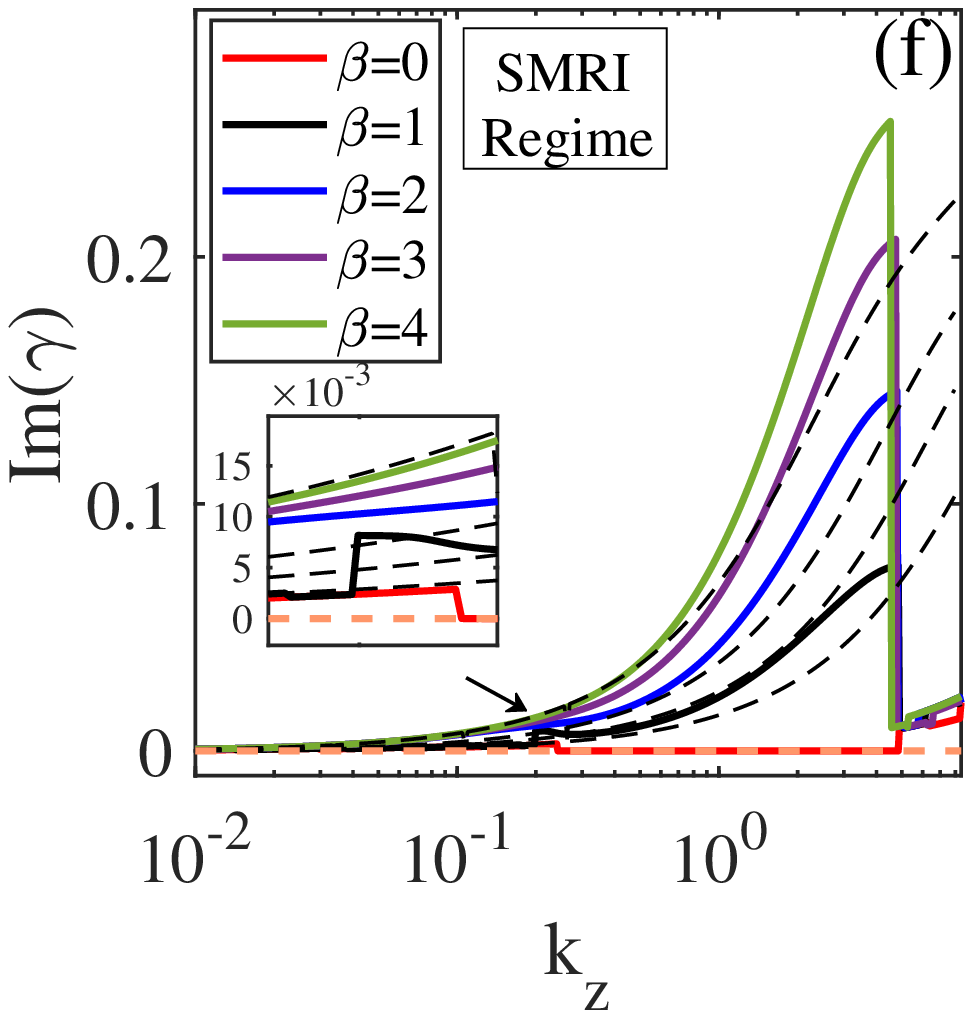}
\caption{(a)-(c) Growth rate, ${\rm Re}(\gamma)$, and (d)-(f) frequency, ${\rm Im}(\gamma)$, as a function of $k_z$ for $\mu=0.27, Pm = 7.77 \times 10^{-6}$ and different $\beta$. The three characteristic cases at $Lu=0.1287, Rm=0.1247$ (a, d), at the critical $Lu_c=2.1461, Rm_c=6.8610$ (b, e) and at $Lu=5, Rm=25$ (c, f) correspond, respectively, to the points A, B and C in Fig. \ref{Fig6: mu27_Pm_comparison}(c). The black dashed lines in the frequency plots in the bottom row show the frequency of various, purely hydrodynamical, inertial wave modes.} \label{Fig7: Gr_k}
\end{figure*}
	
To better understand the transition from HMRI to H-SMRI, in Fig. \ref{Fig7: Gr_k}, we show the growth rate, ${\rm Re}(\gamma)$, and frequency, ${\rm Im}(\gamma)$, as a function of axial wavenumber $k_z$ for fixed $\mu=0.27$ and varying  $\beta$ in three characteristic cases denoted by points A, B and C in Fig. \ref{Fig6: mu27_Pm_comparison}(c). These are: 1. the HMRI regime at lower $Lu=0.1287, Rm=0.1247$ (point A), 2. the transition region at intermediate $Lu=2.1461, Rm=6.8610$ (point B, chosen such that SMRI would be marginally stable) and 3. the H-SMRI region at higher $Lu=5, Rm=25$ (point C). 

In the first case, there is only HMRI at $\beta=3$ and $4$, whereas the values $\beta=1$ and $2$ as well as SMRI at $\beta=0$ are stable (Fig. \ref{Fig7: Gr_k}a). It is seen that with increasing $\beta$, the growth rate increases and the interval of unstable $k_z$ broadens. As mentioned above, the HMRI results from the destabilization of inertial oscillations. This is demonstrated in Fig. \ref{Fig7: Gr_k}(d) showing the corresponding frequency of the HMRI mode, which well coincides with the frequency of purely hydrodynamical inertial waves (dashed black lines) and hence does not much depend on $\beta$, as typical of HMRI. 

In the transition case, we see that the growth rate and the instability interval along $k_z$ also increase monotonically with $\beta$ (Fig. \ref{Fig7: Gr_k}b), although more weakly than in the first case. This is natural as the point B in Fig. \ref{Fig6: mu27_Pm_comparison}(c) is marginally SMRI-unstable initially at $\beta=0$, while becoming increasingly more HMRI-unstable as $\beta$ increases. In this regime, there is some competition between HMRI and H-SMRI modes. Figure \ref{Fig7: Gr_k}(e) shows the behavior of the corresponding frequency with $\beta$. For $\beta=0$, emerging marginally SMRI-unstable slow magneto-coriolis waves are stationary with zero frequency. With increasing $\beta$, the frequency undergoes a ``jump'', after which it follows the characteristics of the HMRI mode (the inertial oscillations shown by black dashed lines) with non-zero frequency in the unstable region, converging to a certain value as $\beta$ increases. We return to this transition process and discuss this in more detail in Section \ref{subsec_transition_region}.
	
In contrast to the behavior in the above two cases depicted in Figs. \ref{Fig7: Gr_k}(a) and \ref{Fig7: Gr_k}(b), in the third H-SMRI case shown in Fig. \ref{Fig7: Gr_k}(c), the growth rate decreases with increasing $\beta$, while the unstable range along $k_z$ does not change with $\beta$ and remains similar to that of SMRI. Figure \ref{Fig7: Gr_k}(f), shows the behavior of the corresponding frequency, which starting from zero for the stationary SMRI mode, increases with $\beta$ for H-SMRI and deviates from that of inertial waves in contrast to that of HMRI that is independent of $\beta$.

\begin{figure*}
\begin{minipage}{\textwidth}
\centerline{\includegraphics[width=\textwidth]{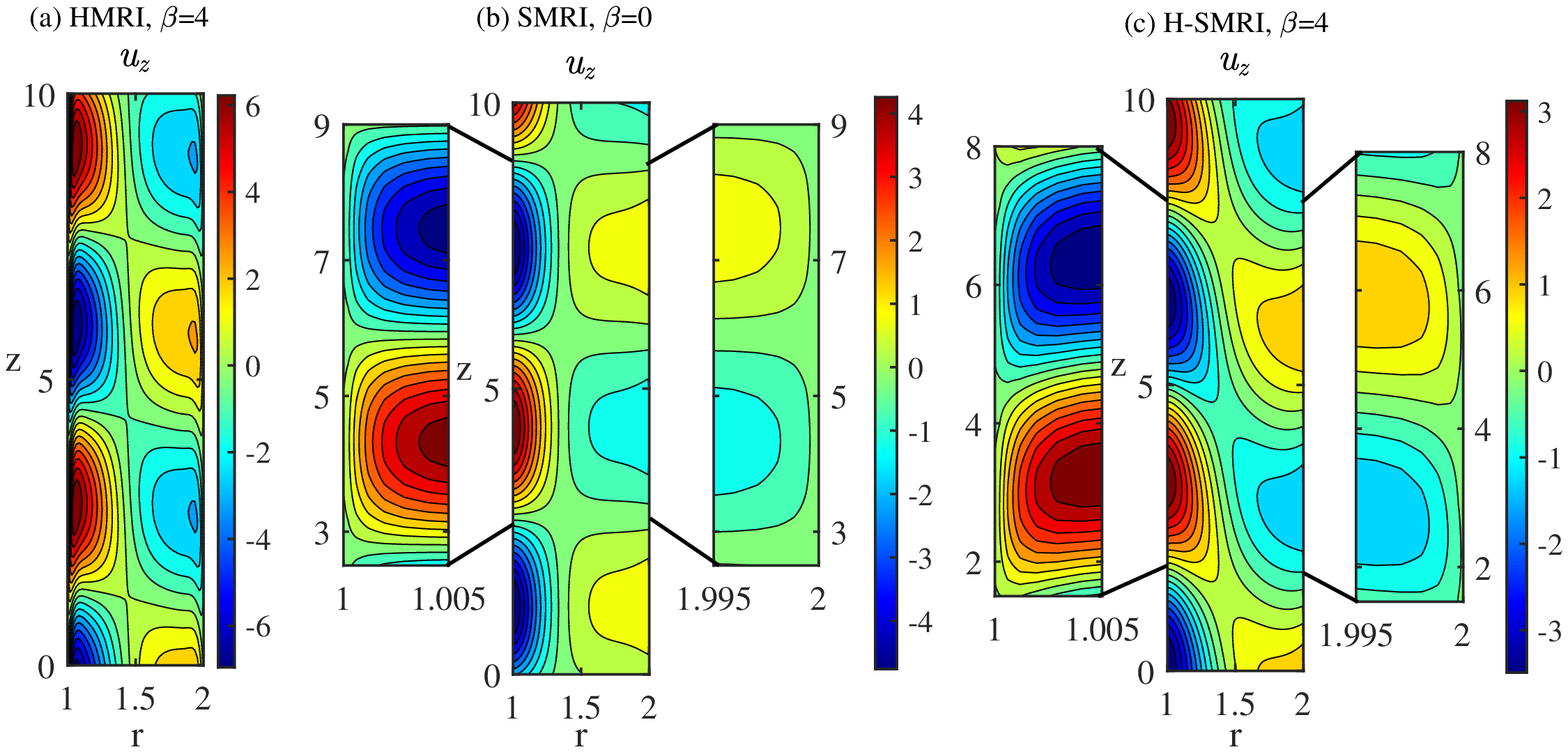}}
\end{minipage}
\begin{minipage}{0.55\textwidth}
\centerline{\includegraphics[width=\textwidth]{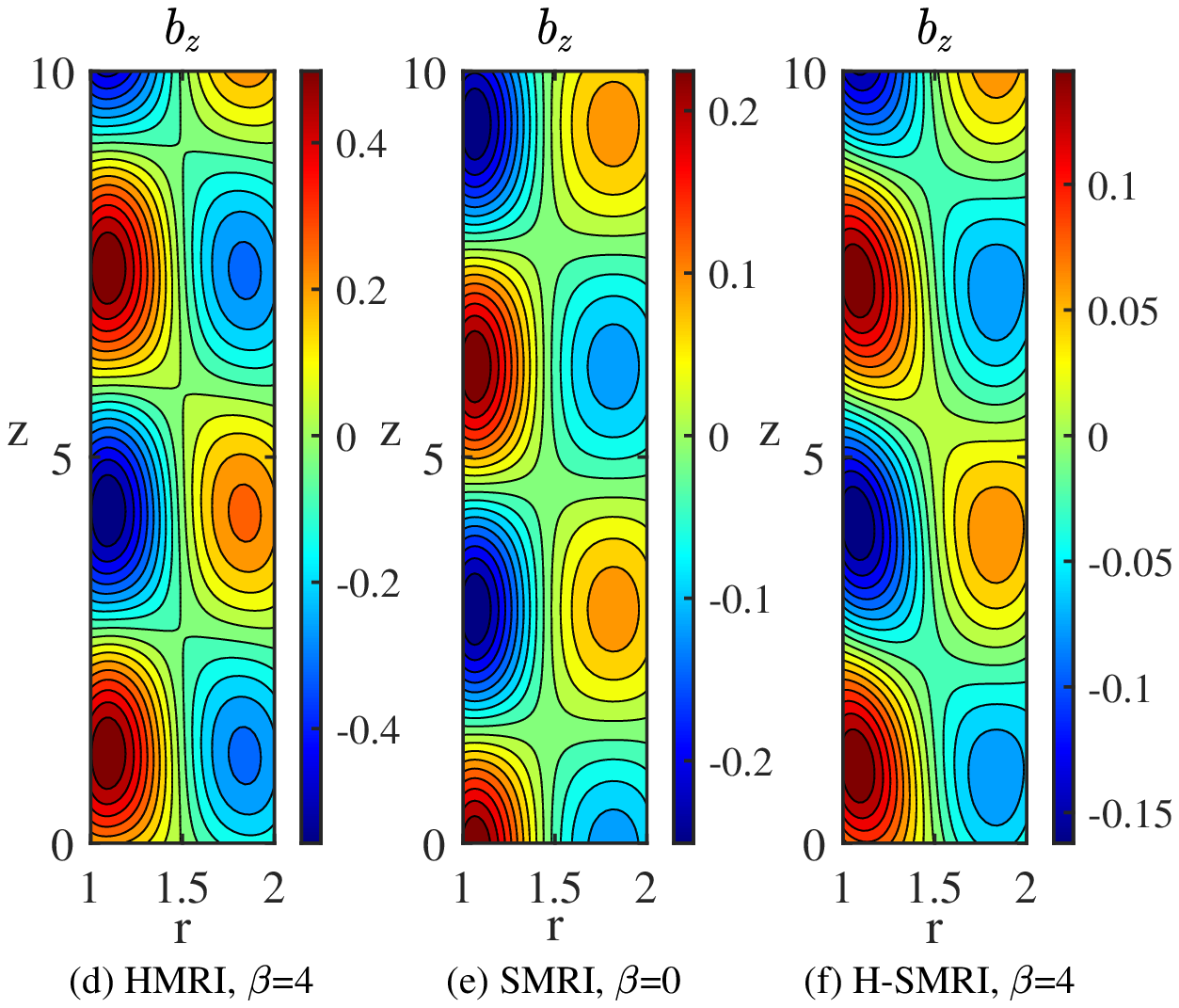}}
\end{minipage}
\caption{(a-c) Axial velocity $u_z$ and (d-f) magnetic field $b_z$ eigenfunctions for fixed $\mu=0.27, Pm=7.77\times10^{-6}$. (a), (d) show HMRI eigenfunctions at $Lu=0.1287, Rm=0.1247, k_z=3.05$ and $\beta=4$ (point A in Fig. \ref{Fig6: mu27_Pm_comparison}c). (b),(e) and (c),(f) have $Lu=5, Rm=25, k_z=2.2$ (point C in Fig. \ref{Fig6: mu27_Pm_comparison}c), showing SMRI and H-SMRI eigenfunctions for $\beta=0$ and $\beta=4$, respectively. In (b) and (c), insets zoom into the boundary layers, which are rather thin because of very high $Re=Rm/Pm=3.22\times 10^6$, and clearly illustrate that $u_z$ indeed vanishes at the cylinder walls in these layers in accordance with no-slip boundary conditions.} \label{Eigenfunctions}
\end{figure*}

\subsubsection{Eigenfunctions}

Here we explore the structure of the eigenfunctions for the instability modes. Figure \ref{Eigenfunctions} shows the eigenfunctions for the axial velocity $u_z$ and axial magnetic field $b_z$ in the $(r,z)$-plane for fixed $\mu=0.27$ at $Lu=0.1287$, $Rm=0.1247, k_z=3.05, \beta=4$ for HMRI (point A), at the same $Lu=5$, $Rm=25, k_z=2.2$ for H-SMRI with $\beta=4$ (point C) and for SMRI with $\beta=0$. The axial velocity eigenfunction of HMRI mode shows small shearing in the middle of the flow domain and quite significant shearing along the cylinder walls (Fig. \ref{Eigenfunctions}a) while the eigenfunction for the SMRI mode does not show such a shearing in the middle of the flow or along the boundary but is more concentrated towards the inner cylinder wall (Fig. \ref{Eigenfunctions}b). The axial velocity eigenfunction for the H-SMRI mode (Fig. \ref{Eigenfunctions}c) is, however, characteristically different from those of HMRI or SMRI in that it has much stronger axial shear in the middle of the flow than the HMRI mode while having no shear along the boundary similar to that of SMRI mode. This, indeed, shows that the helical magnetic field modifies SMRI by increasing the axial shearing in the middle of the flow in the resulting H-SMRI mode. Note also that because of very high $Re$ associated with SMRI and H-SMRI, both their velocity eigenfunctions have rather thin boundary layers near the cylinder walls (insets in Figs. \ref{Eigenfunctions}b and \ref{Eigenfunctions}c), where the velocity perturbations drop steeply to zero at the walls according to the no-slip boundary conditions. Similarly, the axial magnetic field eigenfunction for the HMRI mode has a slight shearing in the middle of the flow (Fig. \ref{Eigenfunctions}d), while the eigenfunction of the SMRI mode does not have such shearing but instead is more concentrated near the inner cylinder wall (Fig. \ref{Eigenfunctions}e). However, as evident from Fig. \ref{Eigenfunctions}(f), the H-SMRI mode has significant stronger axial shearing in the middle of the domain than HMRI mode.
    
\begin{figure*}
\includegraphics[width=0.3\textwidth]{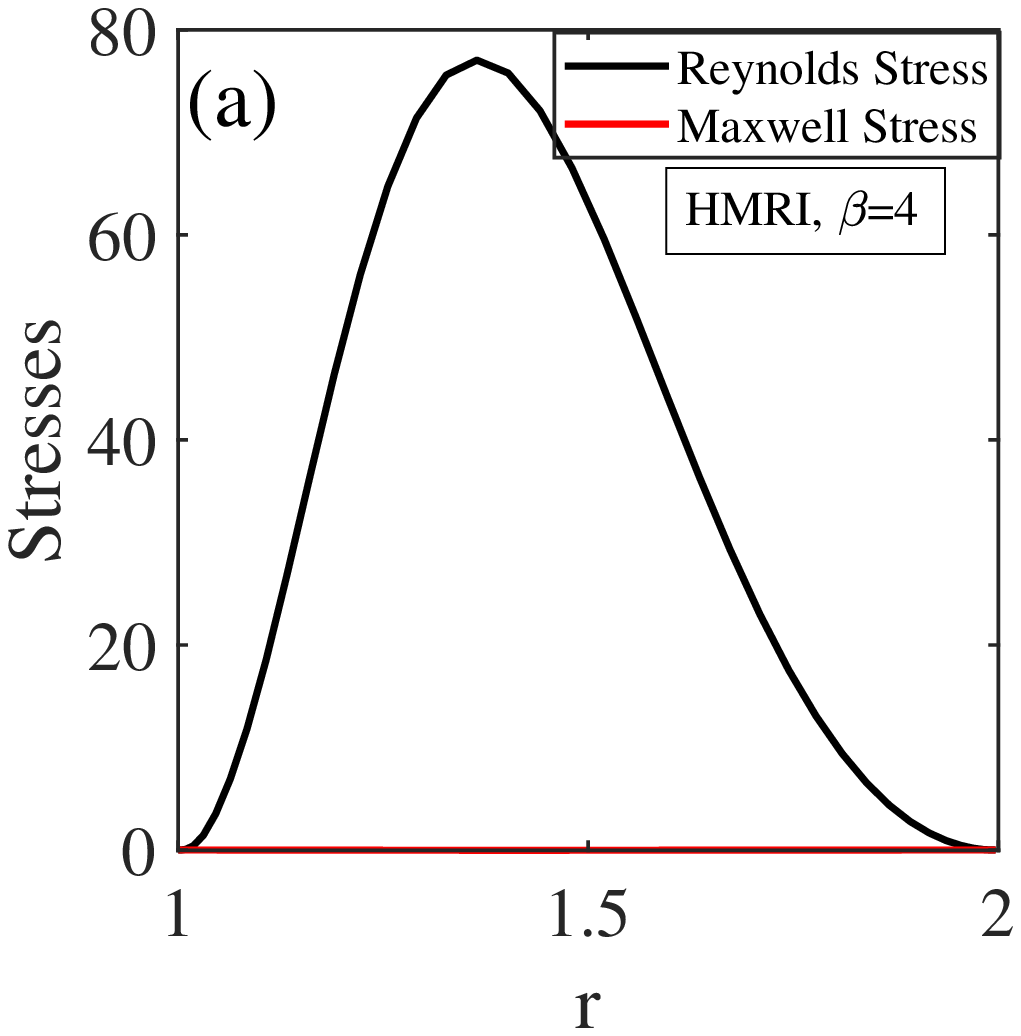}
\hspace{2em}
\includegraphics[width=0.31\textwidth]{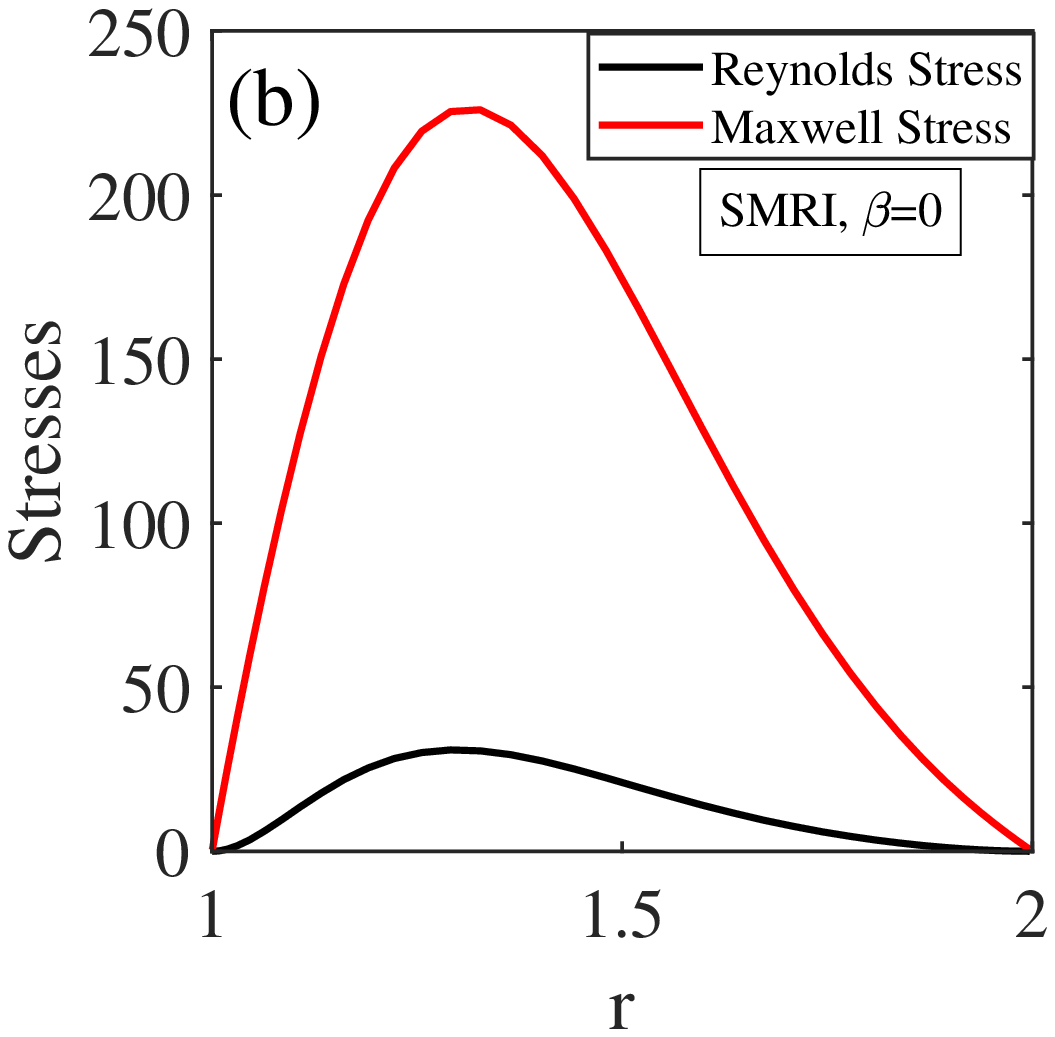}
\hspace{2em}
\includegraphics[width=0.296\textwidth]{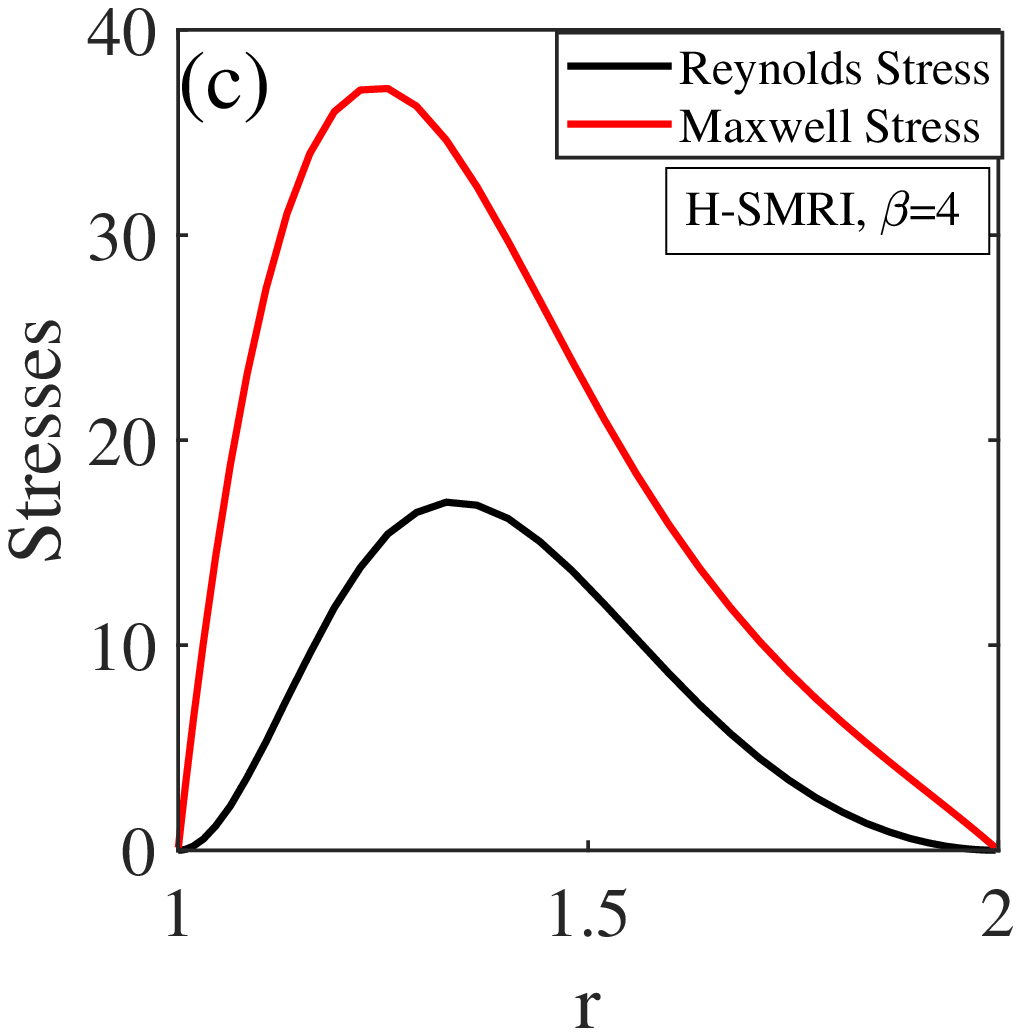}
\caption{(a)-(c) Maxwell and Reynolds stresses for $\mu=0.27, Pm=7.77\times10^{-6}$ in three instability regimes. (a) shows the HMRI regime at $Lu=0.1287, Rm=0.1247, k_z=3.05$ and $\beta=4$ (point A in Fig. \ref{Fig6: mu27_Pm_comparison} c),  (b) and (c) plots are done at $Lu=5, Rm=25, k_z=2.2$ (point C) and show, respectively, SMRI with $\beta=0$ and H-SMRI with $\beta=4$ regimes.} \label{contri_stresses}
	\end{figure*}
	
\subsubsection{Stresses}	
Both HMRI and (H-)SMRI are shear-driven instabilities, that is, gain free energy for growth from the background differentially rotating TC flow. This energy exchange between unstable perturbations and the flow is mediated by the radial components of Maxwell, $-2 {\rm Re}(b_rb_\phi^*)$, and Reynolds, $2 {\rm Re}(u_r u_\phi^*)$, stresses. However, either one of these two types of stresses is dominant for HMRI or (H-)SMRI. The Reynolds stress plays a main role in the energy supply for HMRI, in which velocity perturbations are much larger than the magnetic field ones, because the latter, due to high resistivity $Rm \ll 1$ in this regime, are proportional to $Rm$ and are therefore quite small \cite{Liu_Goodman_Herron_Ji_2006PhRvE, Mamatsashvili_etal2018}. On the other hand, for SMRI that operates at much higher $Rm \gtrsim 1$, magnetic field perturbations are more important and hence Maxwell stress plays a major role \cite{goodman_ji_2002}. Using these properties, below we compute the Reynolds and Maxwell stresses associated with the above eigenfunctions in order to better classify these two instabilities in the present case.  

Figure \ref{contri_stresses} shows the radial distribution of Maxwell and Reynolds stresses for the eigenfunctions at fixed $\mu=0.27$ and $Pm=7.77\times10^{-6}$ in the HMRI and H-SMRI cases. In the essential HMRI regime, as expected, the Maxwell stress is negligible compared to Reynolds one (Fig. \ref{contri_stresses}a). By contrast, in the SMRI regime at $\beta=0$, Maxwell stress is much higher than Reynolds stress (Fig. \ref{contri_stresses}b). We integrate Maxwell and Reynolds stress over the radius to determine the ratio of net Maxwell to Reynolds stresses. This ratio is very small, equal to $0.00045$, for HMRI and much larger, 8.5, for SMRI. Moving from pure SMRI to H-SMRI by increasing $\beta$, the contribution of Reynolds stress increases with respect to Maxwell stress, with the above ratio decreasing to 2.3 at $\beta=4$, but still Maxwell stress dominates (Fig. \ref{contri_stresses}c).    
	
\begin{figure*}
\includegraphics[width=0.3\textwidth]{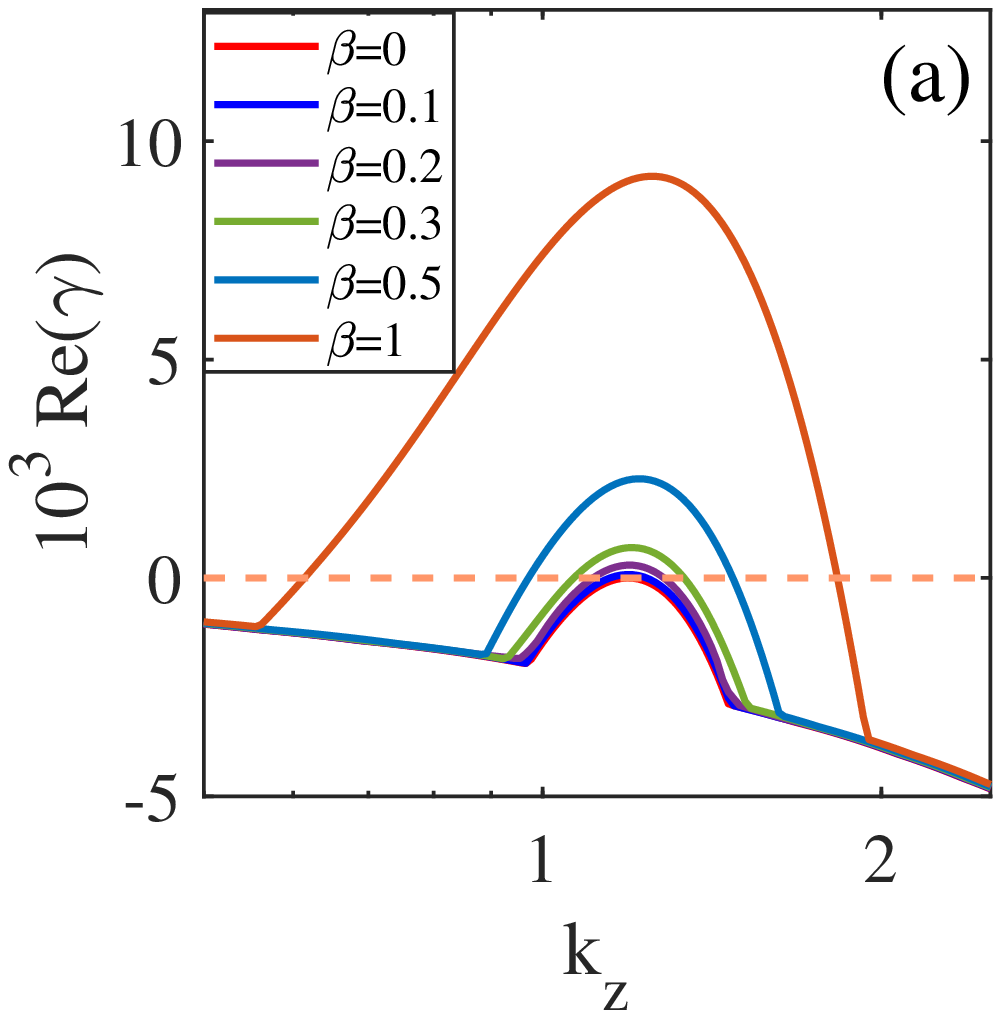}
\hspace{2em}
\includegraphics[width=0.3\textwidth]{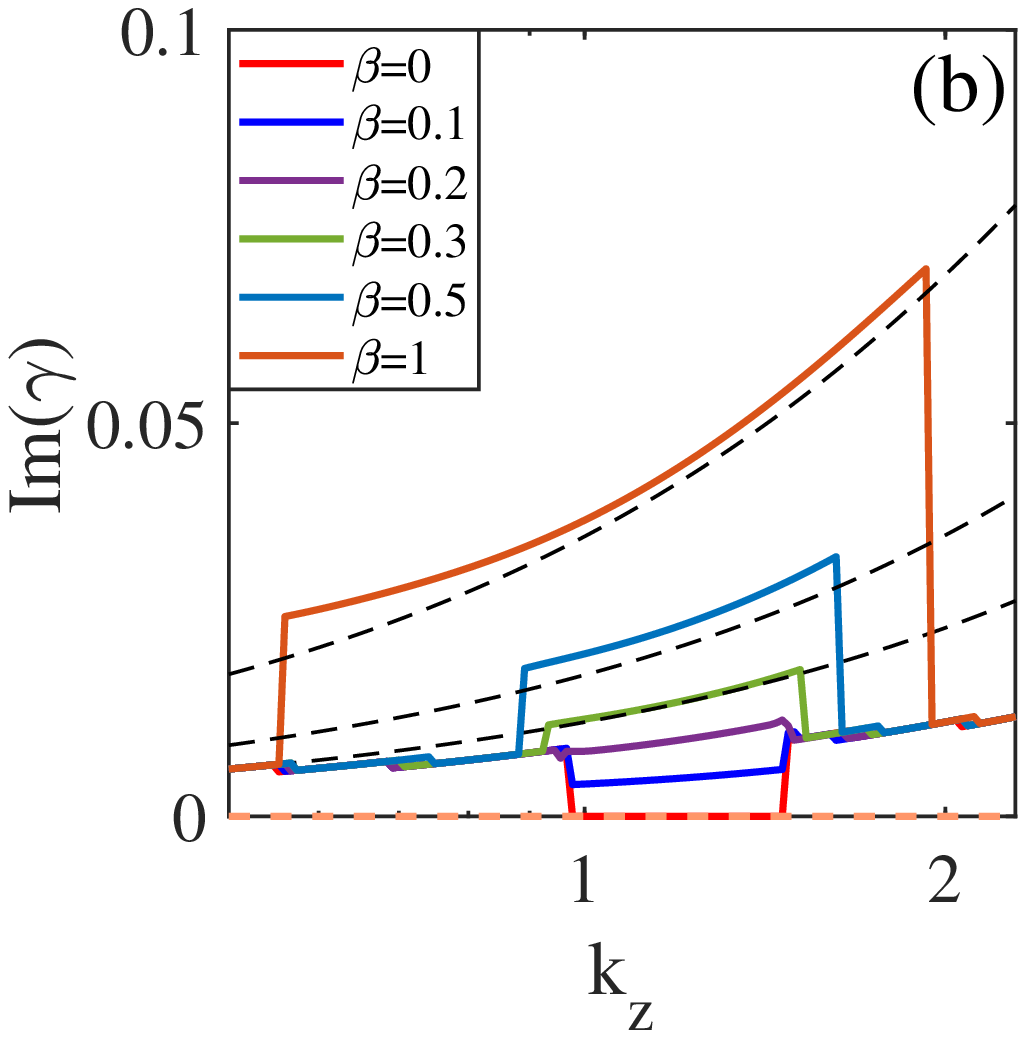}
\hspace{2em}
\includegraphics[width=0.3\textwidth]{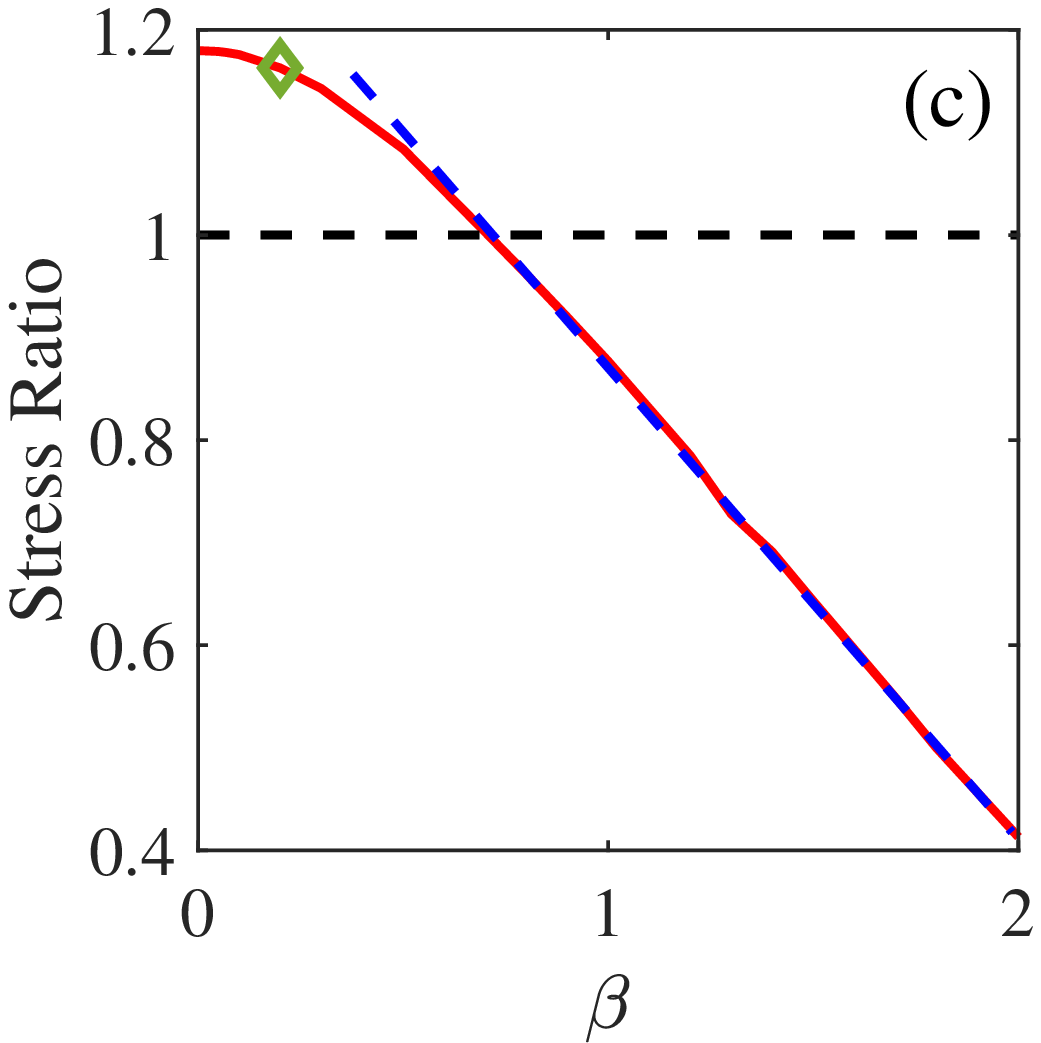}
\caption{(a) Growth rate and (b) eigenfrequency as a function of $k_z$ for $\mu=0.27, Pm=7.77\times10^{-6}\, \text{and}\, \beta \in [0,1]$ in the transition regime at $Lu_c=2.1461, Rm_c=6.8610$ (point B in Fig. \ref{Fig6: mu27_Pm_comparison}c). Black dashed lines in (b) show inertial modes. (c) Variation of the ratio of the Maxwell to Reynolds stresses with $\beta$. The green diamond at $\beta=0.2$ indicates the transition point from which the almost linear drop in stress ratio is observed. The blue dashed line indicates the linear drop in stress ratio with increasing $\beta$. The black dashed line at the unity level marks the equivalence of Maxwell and Reynolds stresses.} \label{transition_region}
\end{figure*}

\subsubsection{Transition between HMRI and H-SMRI -- a closer look} \label{subsec_transition_region}

Once we have analysed the behavior of Reynolds and Maxwell stresses for HMRI, SMRI and H-SMRI, we now return to the transition between HMRI and H-SMRI, which we presented above in Figs. \ref{Fig7: Gr_k}(b) and \ref{Fig7: Gr_k}(e), and analyse it in more detail, using additionally the ratio of the stresses as one of the indicators of this transition. In Figs. \ref{transition_region}(a) and \ref{transition_region}(b), we plot the growth rate and frequency, respectively, as a function of $k_z$ in a narrower range of $\beta \in [0,1]$, where this transition takes place. For small $\beta \leq 0.2$, the shape of the dependence of H-SMRI growth rate on $k_z$ and therefore the range of unstable wavenumbers are almost identical to that of SMRI at $\beta=0$. In other words, the instability region at small $\beta$ is still primarily determined by SMRI. This is evident also in the corresponding frequency curves (Fig. \ref{transition_region}b), where the frequency monotonically increases with $\beta$ in the same interval of unstable $k_z$ whose extent does not change with $\beta$. Thus, for small $0<\beta \leq 0.2$, H-SMRI mode with non-zero frequency branches off the marginally unstable SMRI (stationary magneto-coriolis) branch. From about $\beta=0.2$, when the frequency of H-SMRI approaches the frequency of inertial waves (violet curve), the instability changes its character -- its growth rate and the range of unstable $k_z$ start to increase noticeably with $\beta$, while the frequency now well matches different branches of inertial waves (black dashed curves). These features of the instability at $\beta > 0.2 $ are similar to that of essential HMRI, as shown in Figs. \ref{Fig7: Gr_k}(a) and \ref{Fig7: Gr_k}(d), so at $\beta=0.2$ we observe a continuous transition from the H-SMRI to HMRI regimes, when the HMRI branch smoothly emanates from the S-HMRI branch. In the present 1D global analysis, this transition point $\beta=0.2$, where the frequencies and growth rates of H-SMRI and HMRI become equal, is analogous to the spectral exceptional point in the local WKB analysis of Kirillov \& Stefani \cite{Kirillov_Stefani_2010ApJ}, where these two branches ``reconnect'' and exchange instabilities. 

Figure \ref{transition_region}(c) shows the variation of the ratio of the integrated over radius Maxwell to Reynolds stresses as a function of $\beta$, starting from SMRI, going to H-SMRI and finally to HMRI regimes. It is seen in this figure that this ratio is larger than unity and does not change much with $\beta$ for its small values, $\beta \leq 0.2$ in the (H-)SMRI regime, showing the dominance of Maxwell stresses for these $\beta$. After the transition point $\beta=0.2$ (green diamond), it falls almost linearly in the HMRI regime at $\beta > 0.2$, implying the increasing dominance of Reynolds stress over Maxwell stress.

\begin{figure*}
\begin{minipage}{.45\textwidth}
\centerline{\includegraphics[width=0.7\textwidth]{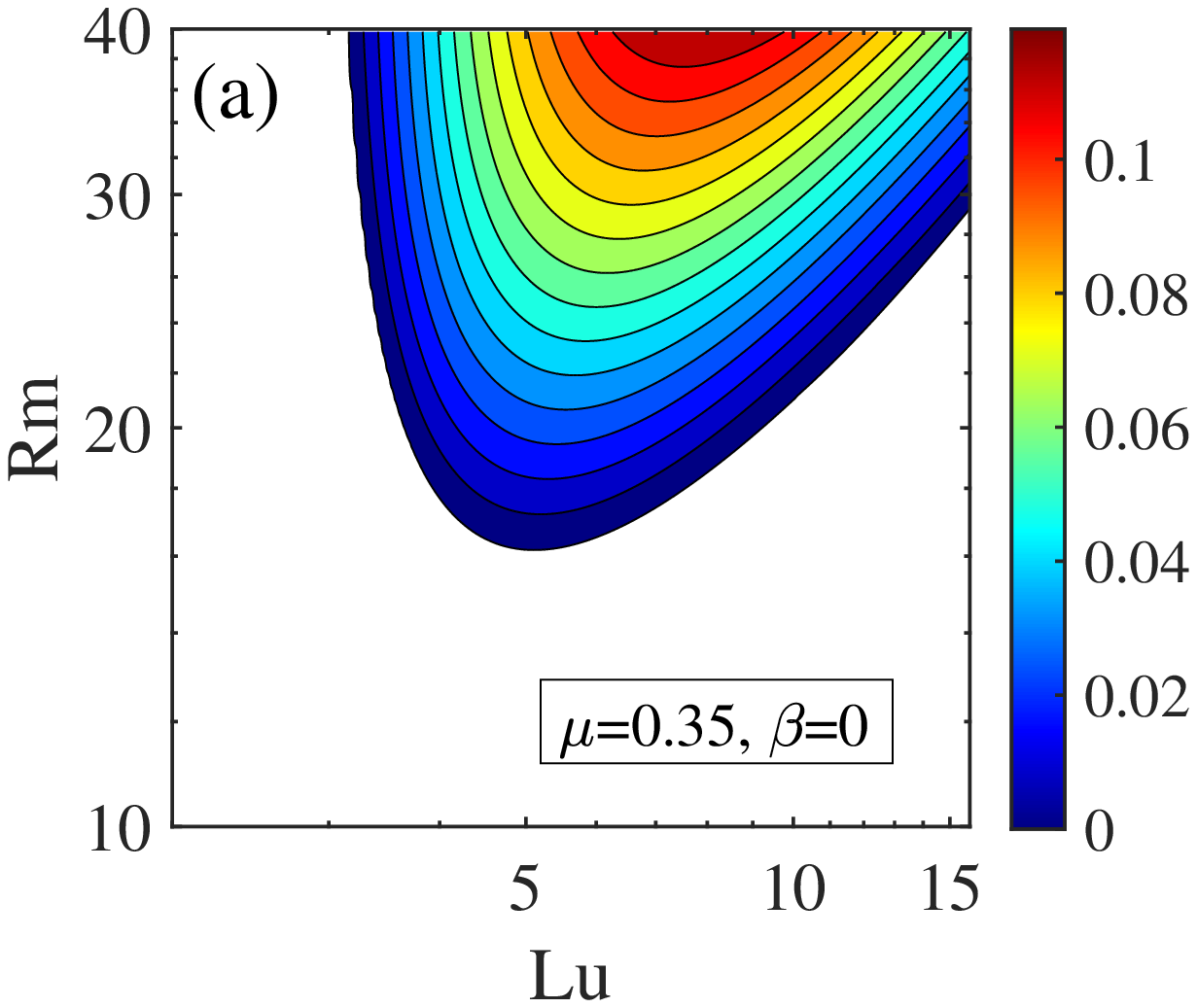}}
\end{minipage}
\vspace{0.6cm}
\begin{minipage}{.45\textwidth}
\centerline{\includegraphics[width=0.7\textwidth]{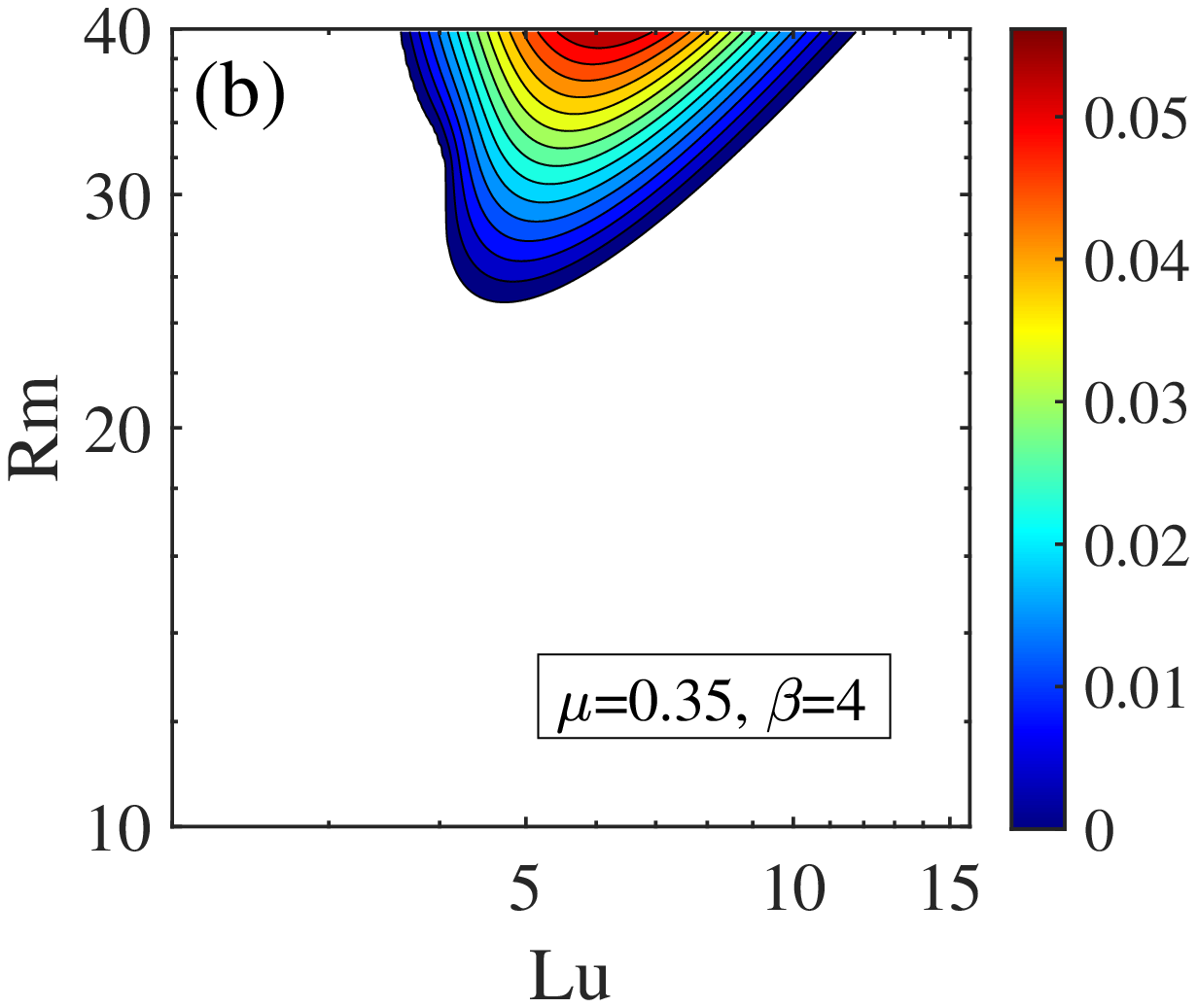}}
\end{minipage}
\begin{minipage}{.45\textwidth}
\centerline{\includegraphics[width=0.6\textwidth]{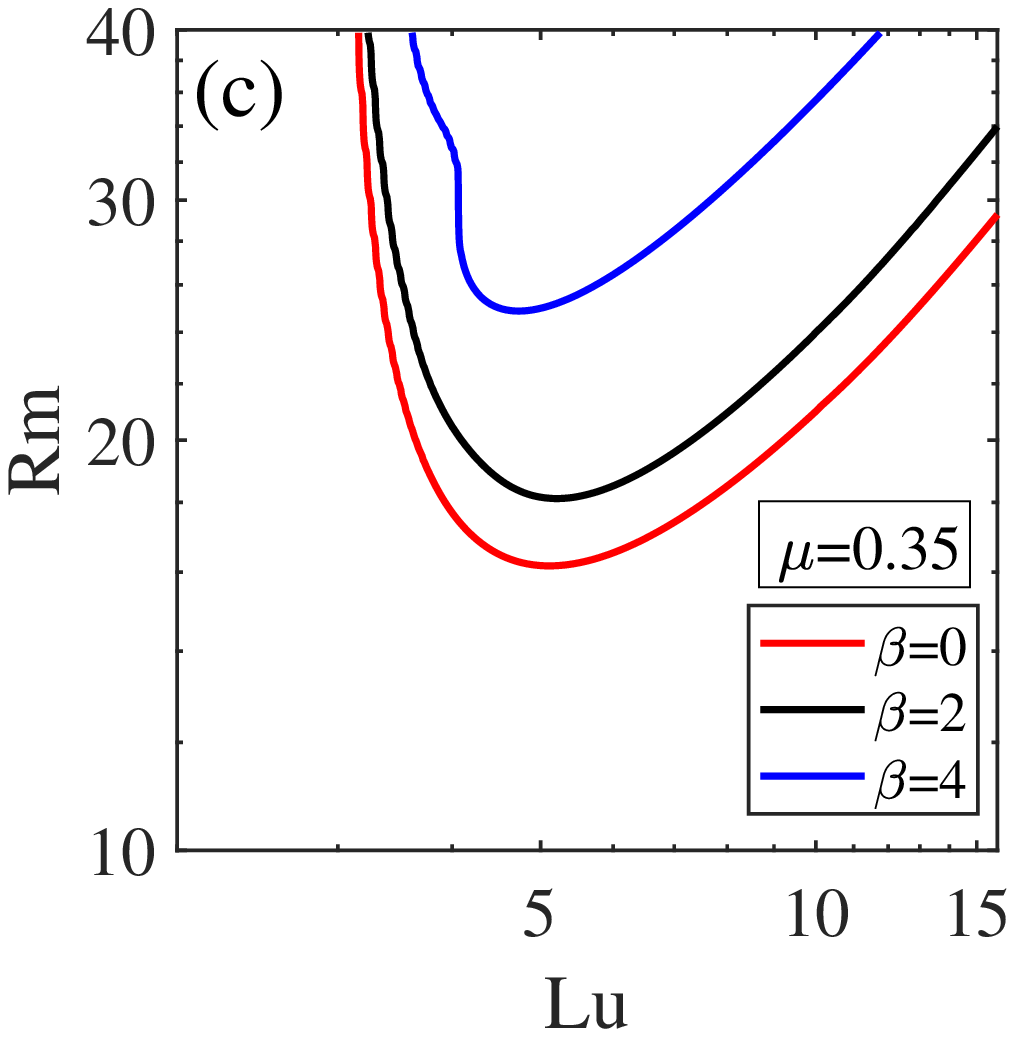}}
\end{minipage}
\begin{minipage}{.45\textwidth}
\centerline{\includegraphics[width=0.7\textwidth]{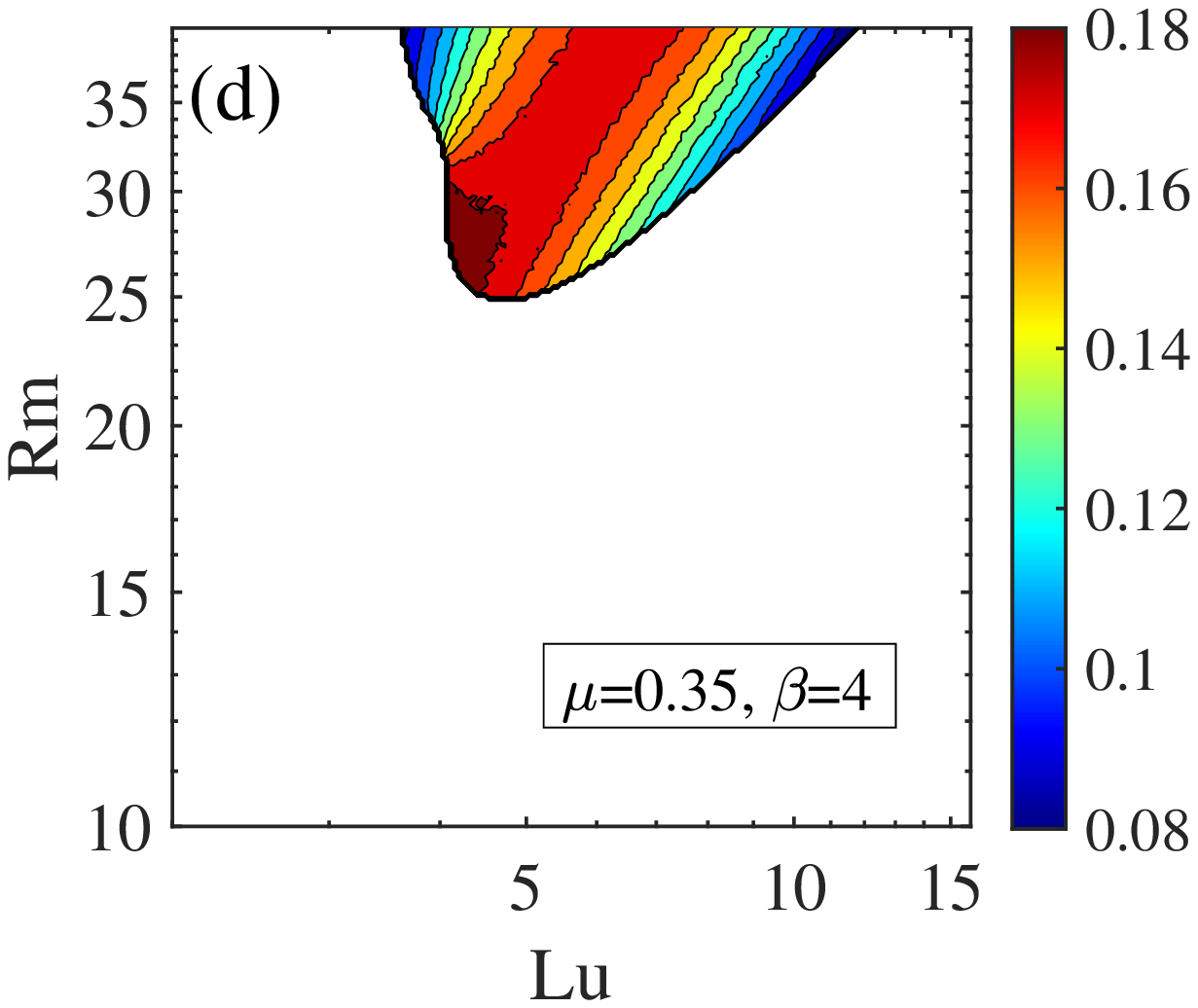}}
\end{minipage}
\caption{(a) Growth rate $\rm Re(\gamma)\geq0$ of SMRI in the $(Lu,Rm)$-plane for quasi-Keplerian rotation $\mu=0.35$ at $Pm = 7.77 \times 10^{-6}$ and $\beta=0$. (b) Same as (a) but for $\beta=4$. Note the reduced growth rate in the presence of azimuthal magnetic field. (c) Marginal stability curves for different $\beta$ and fixed $\mu=0.35$ in the $(Lu,Rm)$-plane. They clearly show better stability of the Keplerian flow with increasing $\beta$. (d) Frequency in the $(Lu,Rm)$-plane corresponding to the growth rate in (b).} \label{Fig: mu35_beta}
\end{figure*}

\subsection{Regime of quasi-Keplerian rotation ($\mu=0.35$)}
	
It is now interesting to consider HMRI and H-SMRI in the presence of quasi-Keplerian rotation profile corresponding to $\mu=0.35$ in our TC flow with $r_{in}/r_{out}=0.5$. Many studies have been conducted to study the effect of helical magnetic field on the stability of Keplerian flows \cite{Ruediger_etal_2018_PhysRepo}. Soon after the discovery of HMRI \cite{Hollerbach_Rudiger_2005}, it was conjectured that the Keplerian flow profile is stable against HMRI \cite{Liu_Goodman_Herron_Ji_2006PhRvE}. However, Hollerbach and R\"udiger \cite{Rudiger_Hollerbach_2007} suggested that Keplerian flow profile can be nevertheless unstable given at least one of the radial boundaries is conducting. Interestingly, the distinction between the convective and absolute variants of HMRI settled the discrepancy regarding the stability of Keplerian flow against HMRI. It was shown that the Keplerian flow is only convectively unstable due to HMRI for  at least one conducting boundary, while being absolutely stable against HMRI \cite{Priede_Gunter_2009, Priede_2011PhRvE}. It is well known that the absolute instability is more relevant than the convective one, because the former, having zero group velocity, tends to stay in the flow for a longer time, while the latter, travelling through the flow, decays at larger times at a given point \cite{Huerre_Monkewitz_1990, Chomaz_2005, Mishra_etal2021} (see also the next Sec. IV). Hence, it has been well established that the Keplerian flow profile is (absolutely) HMRI-stable regardless of the strength of azimuthal magnetic field ($\beta$ parameter) \cite{Priede_2011PhRvE}. For this reason, here we focus only on H-SMRI and analyze the effect of azimuthal magnetic field on it in relation to DRESDYN experiments. 
	
In Fig. \ref{Fig: mu35_beta}, we show the growth rate for the quasi-Keplerian value $\mu=0.35$ and varying $\beta$ in the $(Lu, Rm)$-plane. In contrast to the cases at smaller $\mu$ in Fig. \ref{Fig4: mu27_beta}, in this case, there is a single unstable region, which shrinks upward, mostly towards higher $Rm$, with increasing $\beta$, without developing, as expected, a broadening HMRI branch at lower $Lu$ and $Rm$. The largest growth rate and the broadest unstable region exist for SMRI at $\beta=0$ (Fig. \ref{Fig: mu35_beta}a), with the critical $Lu_c=5.094$ and $Rm_c=16.1714$ which, as mentioned above, fit within the parameter ranges of DRESDYN-MRI machine (Table II). \emph{This implies that SMRI in the astrophysically most relevant and important Keplerian regime can be captured in upcoming DRESDYN experiments.} The growth rate decreases about twice and the unstable area becomes smaller for H-SMRI at $\beta=4$ (Fig. \ref{Fig: mu35_beta}b). For even higher $\beta=6$ (not shown here), the instability region eventually disappears in the given $(Lu, Rm)$-domain in this figure. Thus, the azimuthal magnetic field has a stabilizing effect on SMRI, which is primarily driven by bending axial field lines. In Fig. \ref{Fig: mu35_beta}(c), we plot the corresponding marginal stability curves, which show how the shape of the unstable regions and their area change in the $(Lu,Rm)$-plane as $\beta$ is increased. In Fig. \ref{Fig: mu35_beta}(d), we plot the frequency distribution corresponding to the growth rate shown in Fig. \ref{Fig: mu35_beta}(b), which exhibits a stronger dependence on $Ha$ and $Rm$, unlike the frequency of inertial waves in the case of HMRI (Fig. \ref{Fig6: mu27_Pm_comparison}d).

\begin{figure*}
\begin{minipage}{.32\textwidth}
{\includegraphics[width=\textwidth]{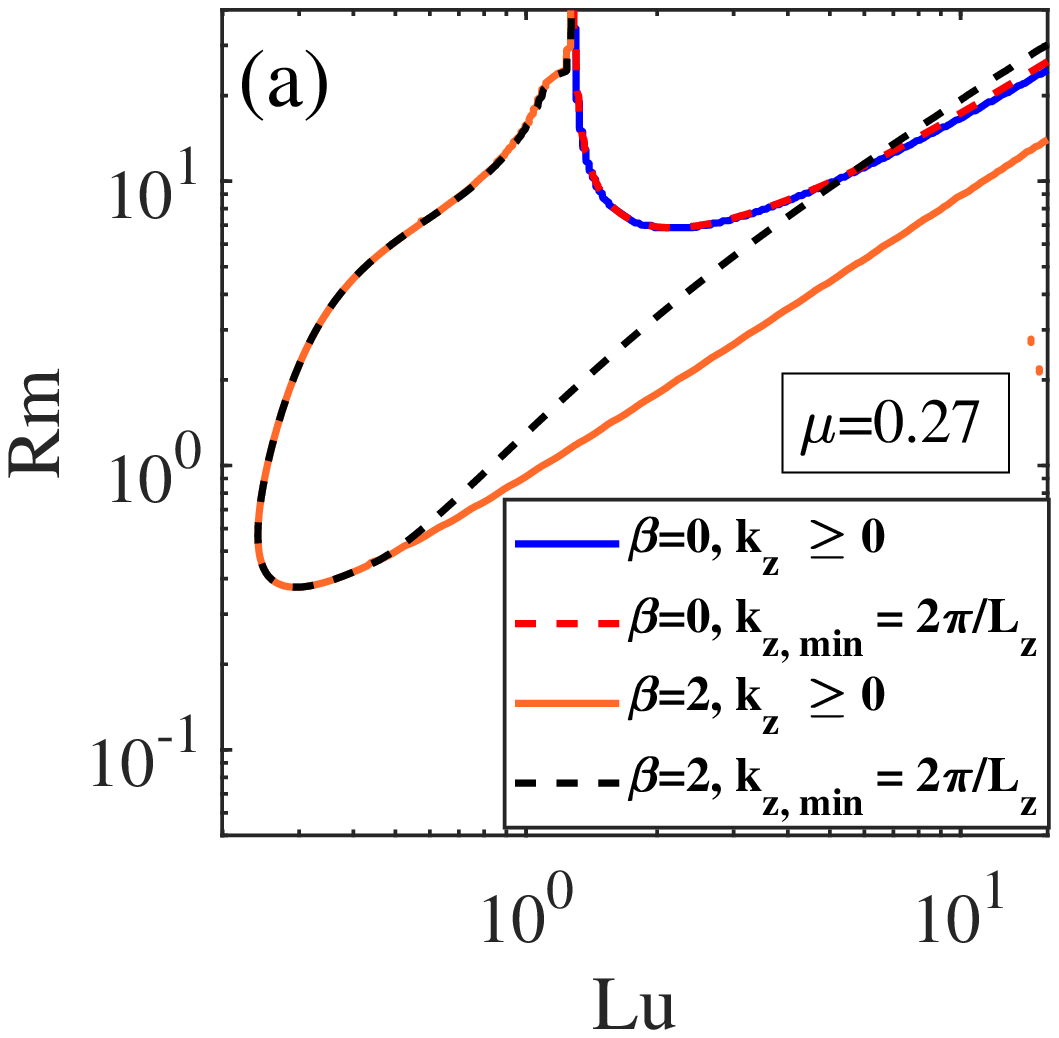}}
\end{minipage}
\begin{minipage}{.32\textwidth}
{\includegraphics[width=\textwidth]{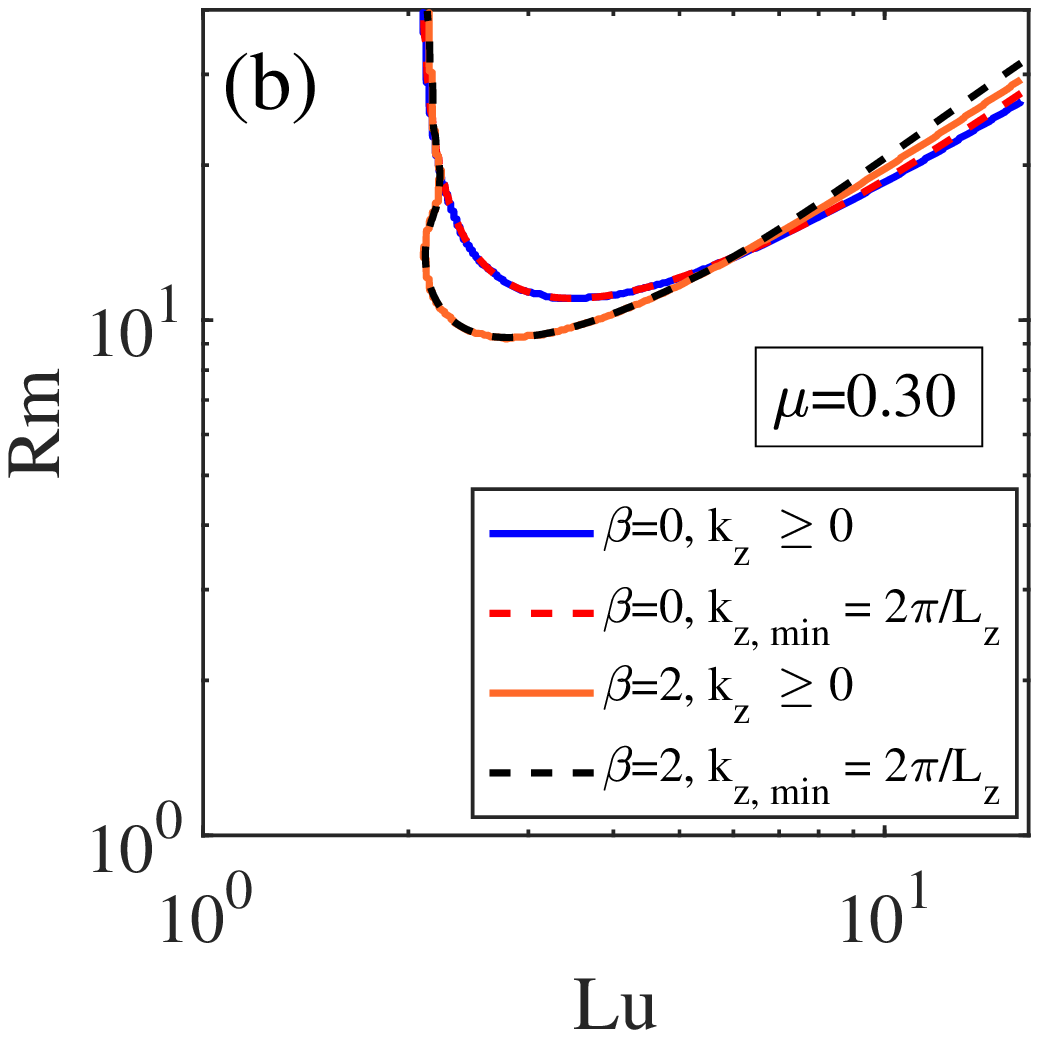}}
\end{minipage}
\begin{minipage}{.32\textwidth}
{\includegraphics[width=\textwidth]{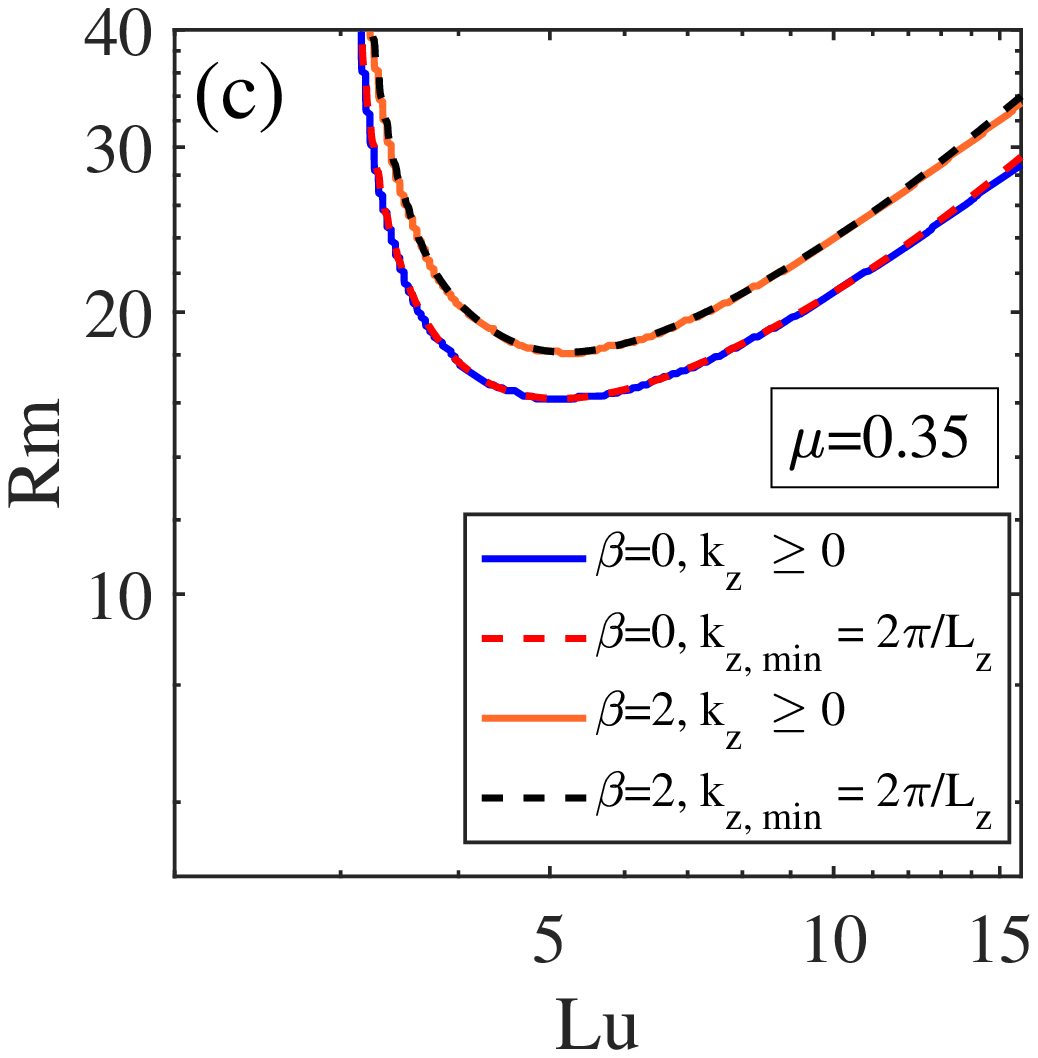}}
\end{minipage}
\caption{Marginal stability curves $\rm Re(\gamma)=0$ optimized over all positive wavenumbers, $k_z\geq 0$ (solid) and $k_z\geq k_{z,min}$ (dashed) for fixed $ Pm = 7.77 \times 10^{-6}$, varying $\beta$ and (a) $\mu=0.27$, (b) $\mu=0.30$, (c) $\mu=0.35$ in the $(Lu,Rm)$-plane. 
Note that the critical $Lu_c$ and $Rm_c$ do not change much whether the optimization wavenumber range is $k_z\geq 0$ or $k_z\geq k_{z,min}$.} \label{Fig:all_wavelength}
\end{figure*}
	
\begin{figure*}
\begin{minipage}{.32\textwidth}
\centerline{\includegraphics[width=\textwidth]{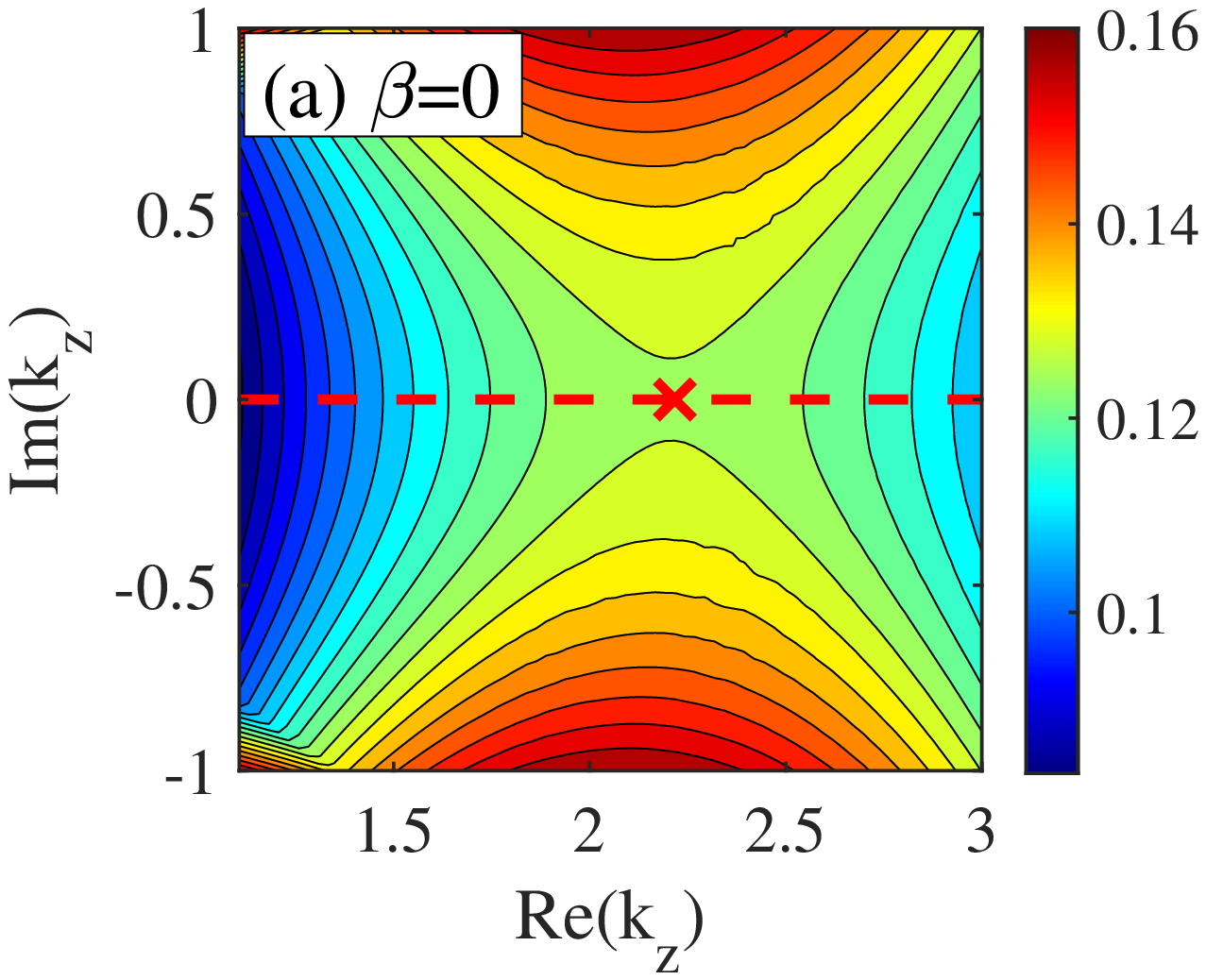}}
\end{minipage}
\vspace{0.6cm}
\begin{minipage}{.32\textwidth}
\centerline{\includegraphics[width=\textwidth]{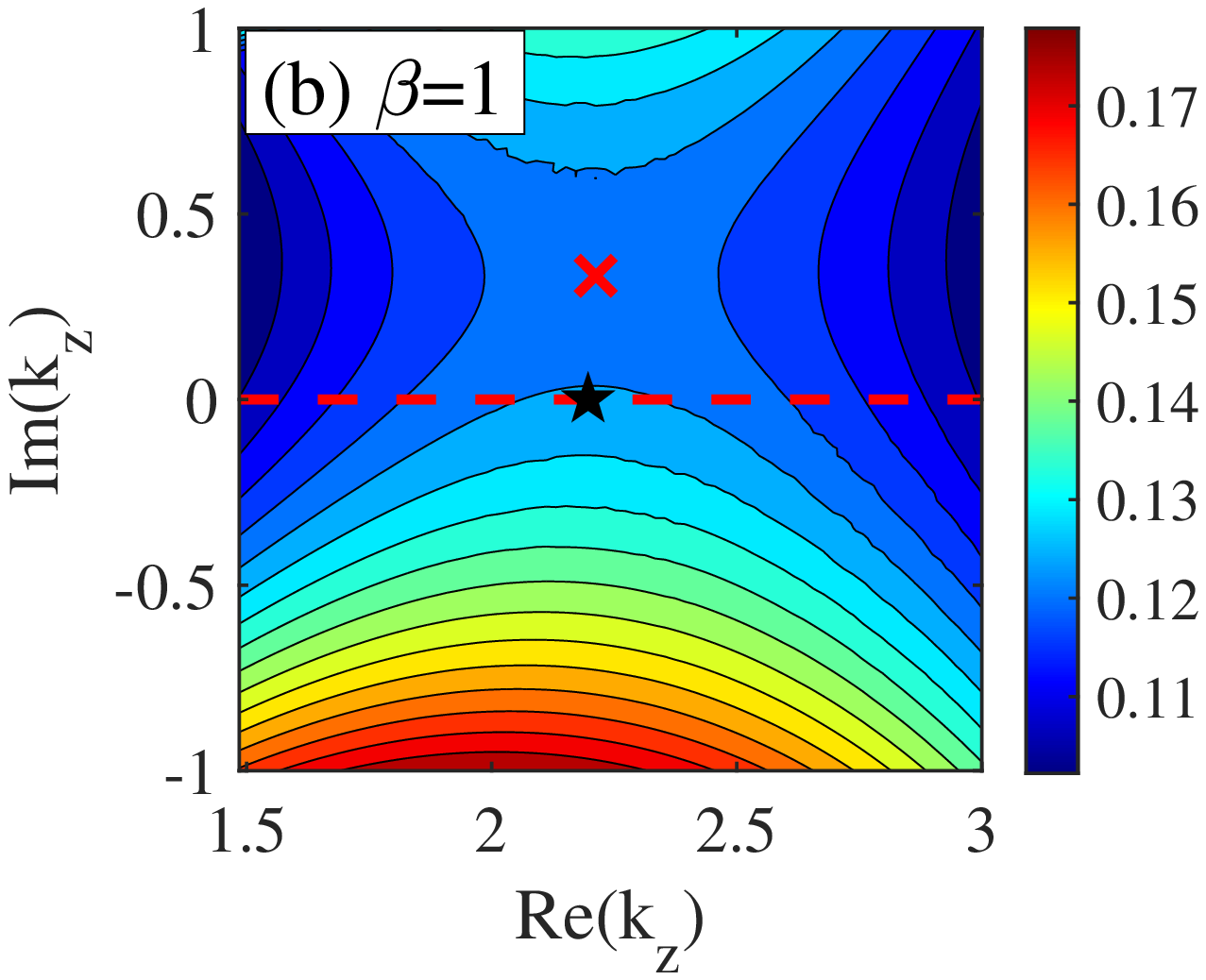}}
\end{minipage}
\begin{minipage}{.32\textwidth}
\centerline{\includegraphics[width=\textwidth]{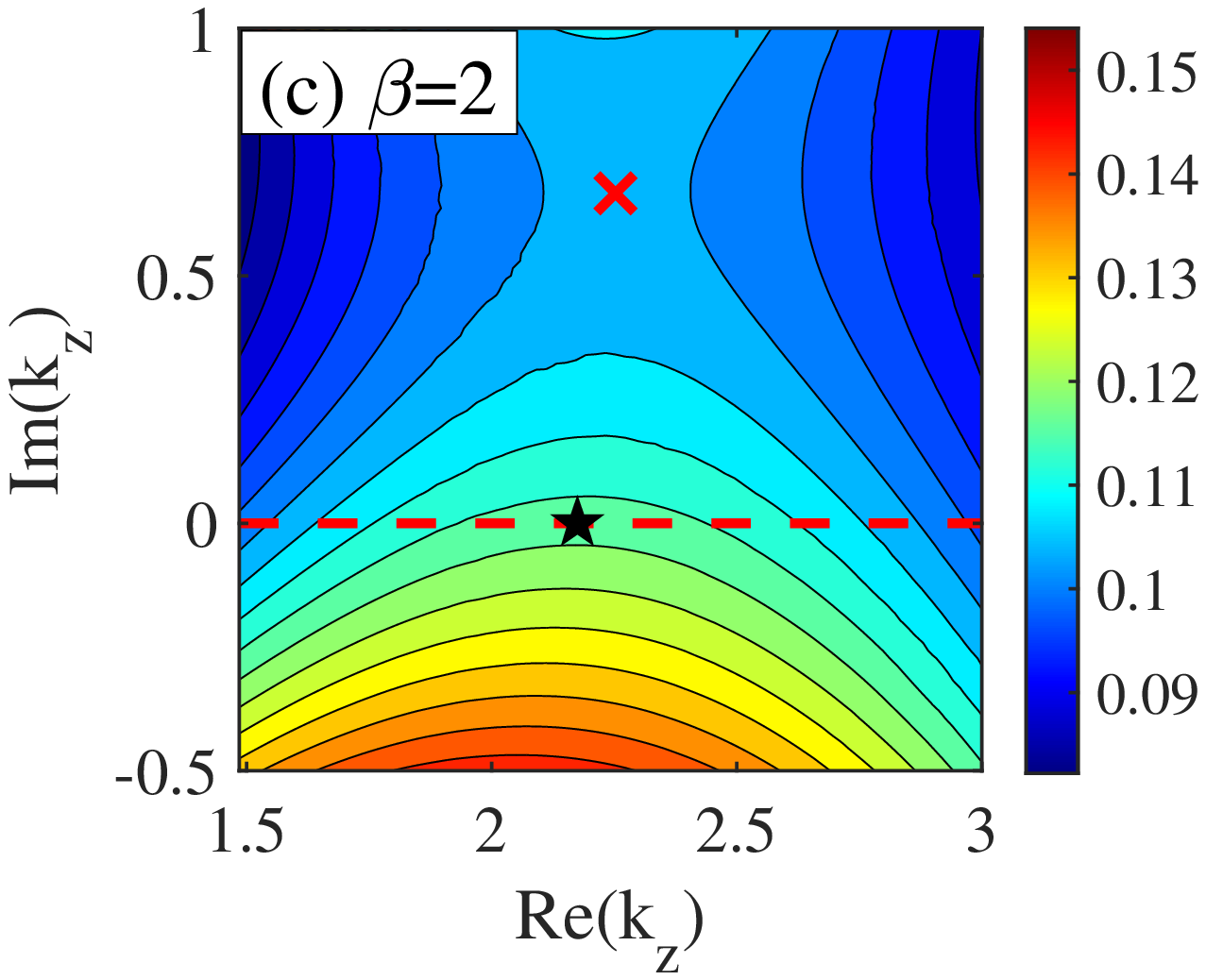}}
\end{minipage}
\begin{minipage}{.33\textwidth}
\centerline{\includegraphics[width=0.85\textwidth]{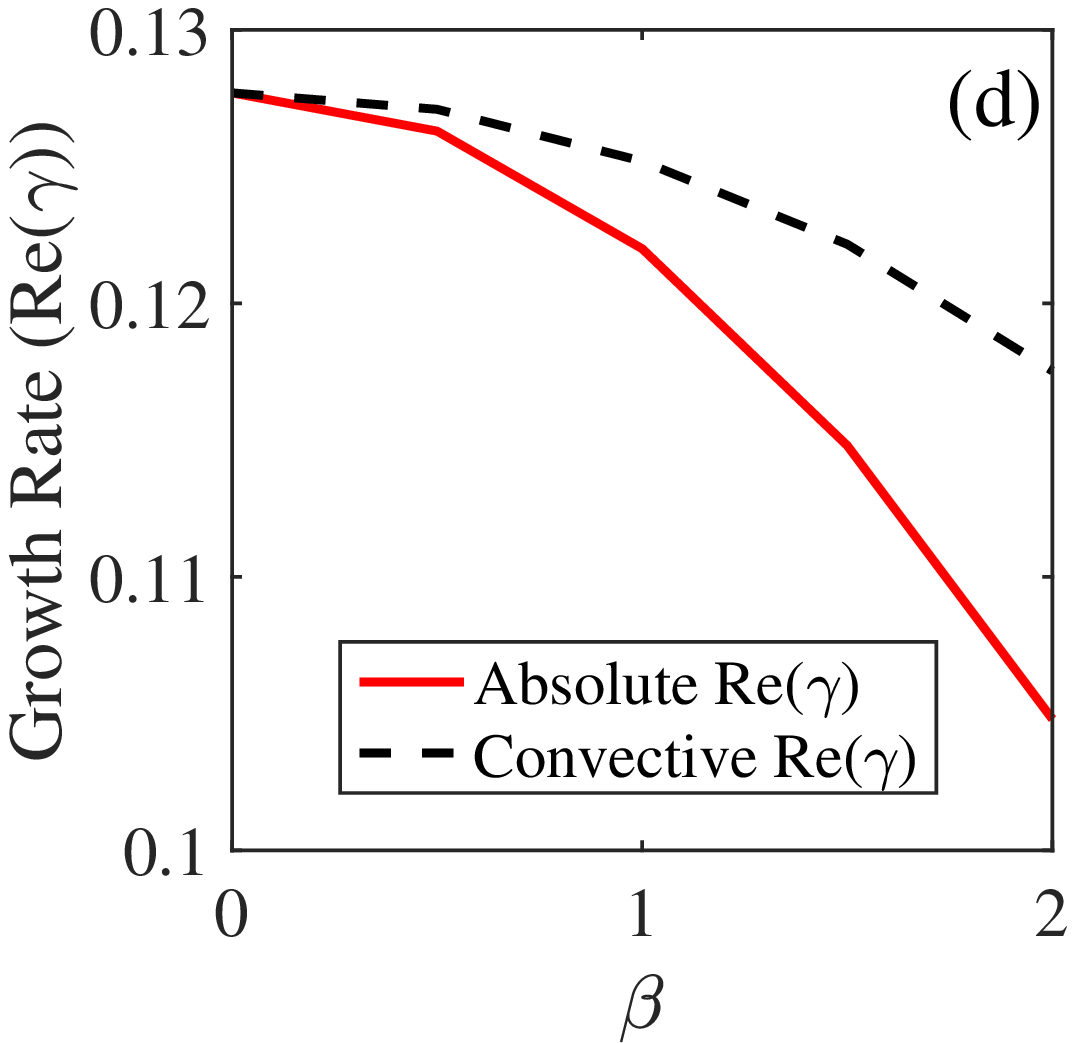}}
\end{minipage}
\begin{minipage}{.33\textwidth}
\centerline{\includegraphics[width=0.85\textwidth]{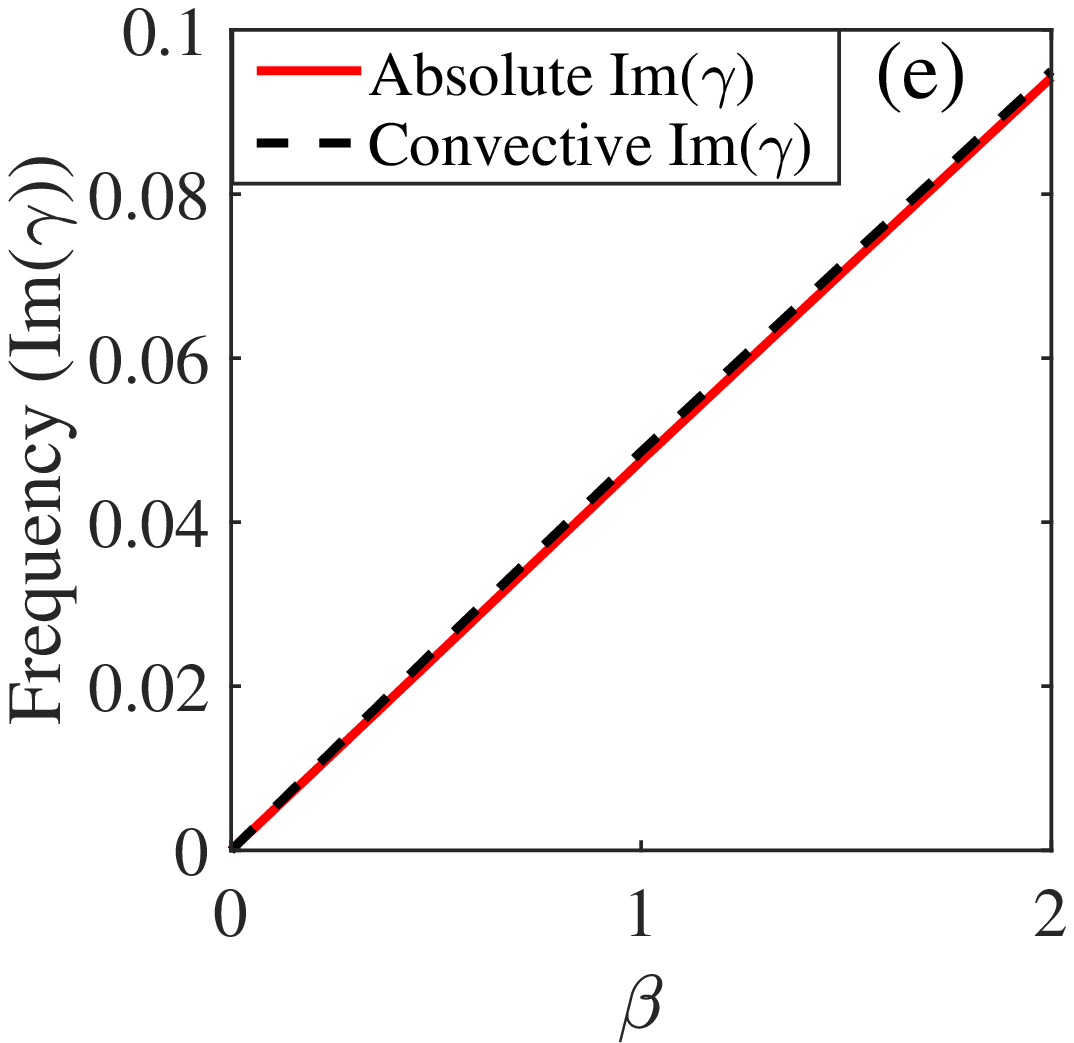}}
\end{minipage}
\caption{(a)-(c) Growth rate ${\rm Re}(\gamma)>0$ at $Lu=5$ and $Rm=25$ for fixed $\mu=0.27, Pm = 7.77 \times 10^{-6}$ and varied $\beta$ in the complex $k_z$-plane. The red cross indicates the saddle point marking the absolute instability, which for SMRI is $k_{z,s}=(2.21, 0)$ and for H-SMRI  are $k_{z,s}=(2.21,0.33)$ at $\beta=1$ and $k_{z,s}=(2.25,0.67)$ at $\beta=2$. The growth rate of the absolute H-SMRI given by ${\rm Re}(\gamma)$ at these saddle points are: ${\rm Re}(\gamma(k_{z,s}))=0.127$ for SMRI and for H-SMRI ${\rm Re}(\gamma(k_{z,s}))=0.12$  and  ${\rm Re}(\gamma_{kz,s})  =  0.1$, while the black stars mark the most unstable convective H-SMRI growth rate. Comparison of absolute (red) and convective (dashed black) growth rates ${\rm Re}(\gamma)$ (d) and corresponding frequencies ${\rm Im}(\gamma)$ (d) as a function of $\beta$.} \label{Fig: mu27_Abs_instability}
\end{figure*}

\subsection{HMRI and H-SMRI for all wavelengths }

From the purely theoretical perspective, it is appealing to allow all wavenumbers in our analysis by relaxing the restriction of the minimum wavenumber $k_{z,min}=2\pi/L_z$, which has been imposed due to a finite length of the cylinders. To perform the comparative study, in Fig. \ref{Fig:all_wavelength} we plot the growth rate maximized over all wavenumbers i.e., $k_z \geq 0$ (solid) and over $k_z\geq k_{z,min}$ (dashed) for varying $\beta$ and $\mu$. As expected, the domain of instability is larger in the first case. In Fig. \ref{Fig:all_wavelength} (a), for fixed $\mu=0.27$ and $\beta=0$, we observe that the marginal stability curves for both $k_z\ge 0$ and $k_z\ge k_{z,min}$ effectively overlap with a very little difference at larger $Lu$. For fixed $\beta=0$, similar trend is observed for $\mu=0.3$ and $\mu=0.35$ as seen in Figs. \ref{Fig:all_wavelength}(b) and \ref{Fig:all_wavelength}(c), respectively. 

For $\mu=0.27$ and $\beta=2$, while the growth rate curves overlap notably, the domain of instability is significantly larger for $k_z\ge 0$ than that for $k_z\ge k_{z,min}$ (Fig. \ref{Fig:all_wavelength}a). This indicates that for $\mu=0.27$, larger $\beta$ yields unstable modes for wavenumbers much smaller than $k_{z,min}$ at larger $Lu$. For $\mu=0.3$ and $\beta=2$, the modification from the SMRI to H-SMRI modes occurs predominantly near the critical $Rm$ (Fig. \ref{Fig:all_wavelength}b). For Keplerian $\mu=0.35$ and $\beta=2$, there is no difference between the stability curves for both $k_z\ge 0$ and $k_z\ge k_{z,min}$ as seen in Fig. \ref{Fig:all_wavelength}(c).
	
Figure \ref{Fig:all_wavelength} shows that for fixed $\beta=2$ the region of instability, or the extended HMRI branch, becomes smaller with increasing $\mu$ while the overlap between the marginal curves at $k_z\ge0$ and $k_z\ge k_{z,min}$ increases with $\mu$ irrespective of $\beta$. Note that the critical $Lu_c$ and $Rm_c$ lying on these curves almost coincide. In other words, the choice of the wavenumber range, $k_z\ge 0$ or $k_z\ge k_{z,min}$, over which to optimize the growth rate does not alter much the critical $Lu_c$ and $Rm_c$ independent of $\mu$ and $\beta$. Hence, the assumption made in this paper to allow at least one full wavelength in the TC device by setting $k_{z,min}=2\pi/L_z$ holds very well with regard to detecting SMRI in the DRESDYN experiment.

\section{Absolute and convective H-SMRI}
	
We have shown that, as distinct from SMRI with zero frequency, H-SMRI comes with non-zero frequency, ${\rm Im}(\gamma)$, due to the presence of the background azimuthal field together with axial one. This implies that like HMRI, H-SMRI represents an instability in the form of a travelling wave with a growing amplitude. For such propagating instabilities, one normally distinguishes between convective and absolute versions. In the convective instability, perturbations appear as travelling wave packets which tend to decay at any spatial point in the flow at large times while continue to grow in the reference frame co-moving with the group velocity of those packets. By contrast, an instability is absolute when perturbations with zero group velocity stay in the flow and grow without limit at every point \cite{Huerre_Monkewitz_1990, Chomaz_2005}. Absolute instability is experimentally more important
than the convective instability, because the latter may be rapidly carried out of the flow system that has a finite size before growing sufficiently strong for detection. On the other hand, the absolute instability remains within the flow, exhibiting sustained growth that can be detected in experiments.

The characteristically different nature of convective and absolute instabilities and its importance in laboratory experiments on magnetic dynamo instabilities were studied earlier \cite{Gailitis_1996, Stefani_1999}. Applying these concepts to HMRI, the differences between convective and absolute forms of HMRI in TC flow \cite{Priede_Gunter_2009} and particularly in the context of PROMISE experiments \cite{Stefani_Gerbeth_Gundrum_Etal_2009PhysRevE} were analyzed, showing that experimentally the onset criteria and growth properties of HMRI are in better agreement with the absolute instability results than with convective ones. Recently, Mishra et al. \cite{Mishra_etal2021}, studied the convective and absolute forms of azimuthal version of MRI (AMRI) and showed that the absolute instability is able to explain the characteristics of the flow as seen in the early PROMISE experiment \cite{Seilmayer_etal2014}.   

Up to now in this paper, we have studied SMRI, HMRI and H-SMRI in the form of convective instability, assuming the axial wavenumber $k_z$ to be a real number, as commonly done in the literature \cite{Ruediger_etal_2018_PhysRepo}. In this section, we investigate the absolute form of H-SMRI and compare it with the results of the convective instability analysis above. Absolute instability is determined by analytically extending the dispersion relation $\gamma(k_z)$ from the real
$k_z$-axis to the complex $k_z$-plane and finding its saddle points $k_{z,s}$ \cite{Gailitis1980,Huerre_Monkewitz_1990,Chomaz_2005}, at which the derivative of the complex eigenvalue becomes zero, 
\begin{equation}
\frac{\partial \gamma(k_z)}{\partial k_z}\Big|_{k_z=k_{z,s}}=0,~~~k_{z,s} \in \mathbb{C}.
\label{absolute}
\end{equation}
There exists an absolute instability in the flow if the real part of $\gamma$ at $k_z = k_{z,s}$ is positive, ${\rm Re}\gamma(k_{z,s}) > 0$, which is then the growth rate of the instability and $\omega \equiv {\rm Im}(\gamma(k_{z,s}))$ is the frequency. The condition (\ref{absolute}) means that the real group velocity of the absolute instability is zero at $k_{z,s}$, i.e., $\partial \omega/\partial {\rm Re}(k_z)|_{k_z=k_{z,s}}=0$, although it can still have a non-zero phase velocity. Following our previous work on convective/absolute AMRI \cite{Mishra_etal2021}, here we apply this method to H-SMRI, too. For SMRI, as we have seen below, the convective and absolute versions are the same, since the eigenmodes has a zero frequency (phase velocity) and therefore does not have a form of a travelling wave. For this purpose, we did similar calculations of the eigenvalues $\gamma$ with the above-described code and same boundary conditions, except assuming now the axial wavenumber to be complex.  

Figure \ref{Fig: mu27_Abs_instability} shows the instability growth areas ${\rm Re}(\gamma)>0$ in the complex $k_z$-plane at $Lu=5$, $Rm=25$ in (a) the SMRI and (b,c) H-SMRI regimes for $\mu=0.27, Pm=7.77\times10^{-6}$ (point C in Fig. \ref{Fig6: mu27_Pm_comparison}c) and varying $\beta$. In each of these three plots, we clearly see a saddle point (red cross), where the group velocity of the instability is zero. This saddle point is the absolute instability point. In these plots, we also mark with black stars those points of the convective instability at the real ${\rm Re}(k_z)$-axis where its growth rate is maximal. In Fig. \ref{Fig: mu27_Abs_instability}(a) we show the instability growth area for SMRI at $\beta=0$. The saddle point is located at $k_{z,s}=(2.21, 0)$, i.e., lies on the real ${\rm Re}(k_z)$-axis, hence coinciding with convective instability, and has the growth rate ${\rm Re}(\gamma(k_{z,s}))=(0.127, 0)$. On the other hand, an imposed azimuthal magnetic field  ($\beta \ne 0$) modifies SMRI with zero frequency to H-SMRI with non-zero frequency. This leads to the saddle point being shifted now in the upper half (${\rm Im}(k_z)>0$) of the $k_z$-plane and equal to $k_{z,s}=(2.21, 0.33)$ and $k_{z,s}=(2.25, 0.67)$, respectively, for $\beta=1$ and $2$ (Figs. \ref{Fig: mu27_Abs_instability}b and c). The absolute H-SMRI growth rates given by the value of ${\rm Re}(\gamma)$ at these saddle points, ${\rm Re}(\gamma(k_{z,s}))=0.12$ and ${\rm Re}(\gamma(k_{z,s}))=0.1$, are lower than the corresponding maximum growth rates of the convective H-SMRI, ${\rm Im}(\gamma)=0.13$ and ${\rm Im}(\gamma)=0.12$  at the same $\beta$.

Figures \ref{Fig: mu27_Abs_instability}(d) and (e) show, respectively, the growth rates and frequencies of the absolute (red) and convective (dashed-black) H-SMRI as a function of $\beta$. For the convective instability, the plotted curves are the growth rate maximized over all $k_z \geq k_{z, min}$ and the frequency corresponding to this maximum growth rate, while for the absolute instability the growth rates and frequencies are calculated at the saddle point. Since we are in the H-SMRI regime, the growth rates of both the absolute and convective instabilities drop with increasing $\beta$, with the former decreasing faster than the latter. Note that the frequencies of the absolute and convective H-SMRI are quite close to each other and both increase linearly with $\beta$, which appears to be a distinctive feature of H-SMRI that can be important in experiments to distinguish it from HMRI.

\section{Summary and conclusions}
	
In this paper, using 1D linear stability analysis, we studied three basic -- standard, helical and helically modified-standard -- types of axisymmetric MRI (for short SMRI, HMRI and H-SMRI) and the connections between them in a cylindrical Taylor-Couette flow of liquid sodium threaded by helical current-free magnetic field. We were motivated by upcoming experimental campaigns in frame of the DRESDYN project, aiming at detecting SMRI and its variants in new liquid sodium large-scale TC experiments. The present study was mainly intended as a first preparatory step towards carrying out these experiments. The theoretical results obtained here will form the basis and provide essential guidance for experimentally ascertaining (H-)SMRI for the first time and interpreting the experimental outcomes. For this reason, we focused our analysis on the ranges of characteristic parameters of the TC flow that are achievable in the DRESDYN-MRI machine, the most important ones for SMRI and H-SMRI being the Lundquist number ($Lu\leq 10$, defined in terms of the axial magnetic field) and the magnetic Reynolds number ($Rm\leq 40$). We also varied the strength of the azimuthal magnetic field relative to the axial one ($\beta$) and the ratio of angular velocities of the outer to inner cylinders ($\mu$) within allowable limits. Our main result was that SMRI and H-SMRI can in principle be detected in the DRESDYN-MRI experiment for the considered range of the cylinder rotation ratios $\mu \in [0.26,0.35]$, including the astrophysically most relevant and important Keplerian rotation profile with $\mu=0.35$. This is in contrast to previous studies of SMRI in the laboratory context where the instability had still remained elusive in the experiments usually adopting smaller TC devices. \emph{Thus, new DRESDYN experiments offer an unique possibility to observe and characterize SMRI in the laboratory for the first time.} The results obtained in this paper for the case of infinite cylinders are still preliminary; further detailed linear and nonlinear studies taking into account the endcaps in a finite length TC device are required to better understand the specific features of SMRI emerging in the experiment. In this regard, it should be mentioned that recent nonlinear simulations of SMRI have already been performed in the context of related Princeton MRI experiments \cite{Wei_etal2016,Winarto_etal2020}, but for Reynolds numbers orders of magnitude lower than those required for the onset of SMRI in real experiments. A follow-up study of the nonlinear saturation and dynamics of H-SMRI in the TC flow for the same ranges of the main parameters ($\mu, \beta, Lu, Re, Rm$) that are relevant to DRESDYN experiments will be presented elsewhere. Specifically, in the nonlinear regime, we will trace the transition from the saturated HMRI to H-SMRI and ultimately to SMRI with monotonically decreasing $\beta$ and increasing $Lu$ and $Rm$. In this way, we will examine how this transition occurs in the nonlinear regime compared to that in the linear one studied here. This will be based on our previous study  of the nonlinear dynamics of pure HMRI \cite{Mamatsashvili_etal2018} using a similar numerical technique.

We analyzed in detail the helically modified SMRI (H-SMRI) which results from the modification of SMRI by the azimuthal magnetic field. It was shown that the background azimuthal field has a stabilizing influence on SMRI and therefore H-SMRI generally has a lower growth rate and occurs at higher Lundquist, $Lu$, and magnetic Reynolds, $Rm$, numbers than SMRI. Specifically, given the experimental constraints on the rotation rates of the cylinders (measured by $\mu$ and $Re$) and on the strength of the imposed axial and azimuthal magnetic fields (measured by $Lu$ and $\beta$), we clearly identified the regions of SMRI and H-SMRI in the parameter $(Lu,Rm)$-plane at different $\mu$ and $\beta$. Unlike SMRI, the frequency of H-SMRI is non-zero (overstability) and increases linearly with the background azimuthal field ($\beta$ parameter). This is an important property of H-SMRI which can utilized for its experimental detection. We also studied HMRI in the context of DRESDYN, which occurs instead at much lower $Lu$ and $Rm$, and showed that there is a continuous and monotonous transition from HMRI to H-SMRI as the parameters change. This transition in the present 1D stability analysis is consistent with the results of the local WKB analysis of Ref. \cite{Kirillov_Stefani_2010ApJ}, where the transition was shown to occur smoothly through the exchange of instabilities between the two modes though an exceptional point of the spectrum. 

We considered quasi-Keplerian rotation in the given Taylor-Couette flow, because of its significance for astrophysical disks. It is well known that essential HMRI does not exist for Keplerian rotation, so only SMRI and H-SMRI can operate in this regime. We characterized the onset properties and growth rates of SMRI and H-SMRI in the $(Lu,Rm)$-plane for different $\beta$. We showed that azimuthal magnetic field have a stabilizing effect on SMRI, that is, H-SMRI always have smaller growth rate and unstable region located at higher $Lu$ and $Rm$ the larger is the azimuthal field over the axial one. As a result, sufficiently strong azimuthal magnetic field can eventually render Keplerian flow stable against H-SMRI in a given range of $Lu$ and $Rm$, while for $0 \leq \beta \lesssim 4$, SMRI and H-SMRI at Keplerian rotation still fall in the range of these parameters accessible in DRESDYN.

Since H-SMRI is an overstability with nonzero frequency, we studied its absolute and convective forms. From an experimental viewpoint the absolute instability is more relevant, since it has zero group velocity and hence stays in the flow, exhibiting a sustained growth, while the convective instability is eventually carried out of the device. We compared the growth rates and frequencies of absolute and convective H-SMRI and showed that in both cases the growth rates decrease with $\beta$, with the absolute instability decreasing faster than the convective one. The  corresponding frequencies are close to each other and increase linearly with $\beta$, which can be a useful diagnostics for the detection of SMRI in the experiments.

Finally, we would like to briefly discuss the possible non-modal growth of MRI in the considered magnetized TC flow configuration, which could be important for interpreting outcomes of the planned MRI experiments. It is well-known from the theory of hydrodynamic shear flows that even in the spectrally stable case (i.e., in the absence of exponentially growing unstable modes), such flows can exhibit large non-modal, or transient growth of perturbations as a result of non-orthogonality of eigenmodes due to flow shear in the classical modal approach \cite{Reddy_etal1993, Trefethen_etal1993, Schmid_Henningson2001, Trefethen_Embree2005, Schmid2007}. This non-modal growth has an important consequence -- it can trigger subcritical transition to turbulence in spectrally stable smooth shear (e.g., plane Couette, pipe Poiseuille, etc.) flows, as usually observed in numerical simulations and experiments \cite[e.g.,][]{Grossmann2000,Eckhardt_etal2007,Schmid2007,Mamatsashvili_etal2016,Kerswell2018}.

Non-modal effects are naturally at work in a differentially rotating TC flow, since it is a special case of shear flows \cite[e.g.,][]{Meseguer2002,Maretzke_etal2014}, and therefore inevitably affect MRI that is driven by the flow shear, despite the fact that it is a modal instability  \cite{Mamatsashvili_etal2013, Squire_Bhattacharjee2014a, Squire_Bhattacharjee2014b, Gogichaishvili_etal2017, Gogichaishvili_etal2018, Mamatsashvili_Stefani2016, Mamatsashvili_Stefani2017, Meduri_etal2019}. Using the local flow model, it was shown in those papers that over short, of the order of dynamical (shear) times, the non-modal growth of MRI can be larger than its modal exponential growth with the latter dominating only at asymptotically large times. Thus, the non-modal growth of MRI over short and intermediate time-scales can be actually of more relevance for its nonlinear outcome than the modal growth. In the present case of MRI in the magnetized TC flow, the distinction between modal and non-modal dynamics can be especially important near the marginal stability curves in the ($Lu, Rm$)-plane (Figs. 2f and 4f), where the flow may still exhibit subcritical transition to a sustained nonlinear MRI state (turbulence) due to the non-modal amplification of finite amplitude perturbations regardless of being (marginally) MRI-stable according to the modal approach. To the best of our knowledge, MRI in this global TC flow has not been yet studied from this non-modal viewpoint. This, in its own right, represents a very interesting and important line of research, since the non-modal MRI growth can largely define its nonlinear outcome, especially in the marginal modal stability regime that can be approached experimentally. 
\\
\\
\begin{acknowledgments}
We thank Prof. R. Hollerbach for providing the linear 1D code used in this paper and for his kind assistance in testing it for the present problem. We also thank anonymous Referees for useful suggestions that improved the presentation of our results. This work received funding from the European Union's Horizon 2020 research and innovation programme under the ERC Advanced Grant Agreement No. 787544.
\end{acknowledgments}

\bibliography{ref.bib}

\end{document}